\documentclass[11pt]{article}
\usepackage{jheppub}
\usepackage[T1]{fontenc} 

\usepackage{amssymb}
\usepackage{amsfonts}
\usepackage{amsbsy}
\usepackage{amsmath}
\usepackage{amsthm}
\usepackage{graphicx}
\usepackage{epstopdf}
\usepackage{multirow}
\usepackage{latexsym}
\usepackage{array}

\input cyracc.def 
\newfont{\twelvecyr}{wncyr10 at 12pt}

\def\Z{\mathbb{Z}}
\def\F{\mathbb{F}}

\def\cL{\mathcal{L}}
\def\cO{\mathcal{O}}
\def\cM{\mathcal{M}}

\def\P{\mathbb{P}}
\def\bP{\mathbb{P}}

\def\n3a{t}

\def\b2{S}

\def\ge{{\mathfrak{e}}}
\def\gso{{\mathfrak{so}}}
\def\gsu{{\mathfrak{su}}}
\def\gsp{{\mathfrak{sp}}}
\def\gf{{\mathfrak{f}}}
\def\gg{{\mathfrak{g}}}

\title{$\P^1$-bundle bases and
the prevalence of non-Higgsable structure in 4D F-theory models}

\author[1]{James Halverson}
\author[]{and}
\author[2]{Washington Taylor}

\affiliation[1]{Kavli Institute for Theoretical Physics\\ University of California, Santa Barbara\\ Santa Barbara, CA 93106, USA}
\affiliation[2]{Center for Theoretical Physics\\
Department of Physics\\
Massachusetts Institute of Technology\\
77 Massachusetts Avenue\\
Cambridge, MA 02139, USA}

\emailAdd{{\tt jim} {\rm at} {\tt kitp.ucsb.edu}}
\emailAdd{{\tt wati} {\rm at} {\tt mit.edu}}

\preprint{NSF-KITP-15-068, MIT-CTP-4677}

\abstract{
  We explore a large class of F-theory compactifications to four
  dimensions.  We find evidence that gauge groups that cannot be
  Higgsed without breaking supersymmetry, often accompanied by
  associated matter fields, are a ubiquitous feature in the landscape
  of ${\cal N} = 1$ 4D F-theory constructions.  In particular, we
  study 4D F-theory models that arise from compactification on
  threefold bases that are $\P^1$ bundles over certain toric surfaces.
  These bases are one natural analogue to the minimal
  models for base surfaces for 6D F-theory compactifications.  Of the
  roughly 100,000 bases that we study, only 80 are weak Fano bases in
  which there are no automatic singularities on the associated
  elliptic Calabi-Yau fourfolds, and 98.3\% of the bases have
  geometrically non-Higgsable gauge factors.  The $\P^1$-bundle
  threefold bases we analyze contain a wide range of distinct surface
  topologies that support geometrically non-Higgsable clusters.  Many
  of the bases that we consider contain $SU(3)\times SU(2)$
  seven-brane clusters for generic values of deformation moduli; we
  analyze the relative frequency of this combination relative to the
  other four possible two-factor non-Higgsable product groups, as well
  as various other features such as geometrically non-Higgsable
  candidates for dark matter structure and phenomenological
  ($SU(2)$-charged) Higgs fields.  }

\begin{document}

\maketitle

\flushbottom

\section{Introduction}

F-theory \cite{Vafa-F-theory, Morrison-Vafa-I, Morrison-Vafa-II}
compactifications to four and six dimensions provide a broad
perspective on the landscape of four- and six-dimensional string vacua.
In six dimensions, F-theory may provide a construction for 
vacua in
all branches of the space of
consistent supersymmetric theories of gravity coupled to tensor
fields, gauge fields, and matter \cite{universality, KMT-I, KMT-II}.
In four dimensions, on the other hand, F-theory as currently
formulated only captures one part of the apparently vast landscape of
${\cal N} = 1$ string vacua; for many heterotic vacua, infinite
families of IIA flux vacua (see {\it e.g.} \cite{dgkt}), and large
classes of $G_2$ M-theory compactifications (see {\it e.g.}
\cite{Acharya1,Acharya2} for phenomenologically oriented work
and \cite{Halverson-Morrison} for recent progress on global compactifications), there is no known duality to F-theory, and
in many cases it seems likely that such a duality cannot be found
without a substantial extension of the framework of F-theory.

F-theory may nonetheless be the region of the ${\cal N} = 1$ landscape
that is currently most amenable to study away from weakly coupled regimes,
since supersymmetry and algebraic geometry together give strong
control over nonperturbative aspects of the theory that are not
accessible from other approaches.  F-theory compactifications can be
thought of as compactifications of type IIB string theory where the
axiodilaton $\tau = C_0 + i e^{-\phi}$ varies over a compact space
$B$, where $B$ is a complex two-dimensional (three-dimensional)
manifold $B_2$ ($B_3$) for an F-theory compactification to six (four)
dimensions. Since $e^\phi$ determines the type IIB string coupling
$g_s$, F-theory is more general than weakly coupled type IIB string
theory in that it allows for the possibility of varying and/or strong
coupling.

In recent years it has been found that nearly all six-dimensional
supersymmetric F-theory compactifications give rise to non-Higgsable
gauge groups and matter in the resulting low-energy 6D supergravity
theory.  These non-Higgsable structures arise from curved base
geometries that force nonperturbative seven-brane configurations
carrying nonabelian gauge groups to arise on certain divisors or
combinations of divisors in the base.  These gauge groups are not
broken in any supersymmetric vacuum that can be reached without
changing the topology of the base.  A small number of basic
irreducible ``non-Higgsable clusters'' \cite{clusters} can be used to
systematically study the space of elliptic Calabi-Yau threefolds and
6D F-theory models in their maximally Higgsed phase.  In addition to
single-factor non-Higgsable gauge groups, which can arise in F-theory
compactifications with smooth heterotic duals, F-theory can also give rise to
non-Higgsable product groups with matter charged under multiple gauge
factors.  Similar non-Higgsable clusters arise at the level of
geometry in 4D F-theory models \cite{ghst, 4D-NHC}.

The purpose of this paper is to initiate a systematic study of the
kinds of non-Higgsable structures and bases that  arise in concrete
elliptic
Calabi-Yau fourfolds and 4D F-theory constructions.  In particular, we
investigate here a specific class of compact toric threefold bases for
4D F-theory compactifications.  The geometries we consider here are of
interest for two reasons.  First, they contain a broad class of
topologically distinct complex surface types whose local geometry as
divisors in the F-theory base gives rise to geometrically
non-Higgsable gauge and matter content.  Second, considering generic
elliptic fibrations over the bases we construct here provides a view
of a broad class of 4D F-theory vacua and gives some initial evidence
that in 4D as well as in 6D, the vast majority of supersymmetric F-theory vacua 
contain non-Higgsable gauge groups and matter. The construction used
here gives rise to a class of $109,158$ threefold bases $B$, of
which 98.3\% (all but $1,824$) have non-Higgsable gauge groups at
generic points in  moduli space.

The structure of this paper is as follows: In  Section
\ref{sec:review} we review some of the basics of F-theory
constructions to six and four dimensions, and the notion of
non-Higgsable clusters in F-theory geometry.  In Section
\ref{sec:classification}, we describe the systematic classification of
all threefold bases that can be constructed as $\P^1$ bundles over
toric base surfaces $\b2$ that themselves support elliptic Calabi-Yau
threefolds. In Section \ref{sec:Hodge}, we consider the Hodge
structure of the generic elliptic fourfolds over the toric $\P^1$ bundle
threefold bases and explore how this class of fourfolds fits into the
known set of possible Hodge numbers for elliptic fourfolds.
In Section \ref{sec:4D-NHC}, we describe various aspects of the set of
non-Higgsable clusters that arise for the bases we have constructed,
and in Section \ref{sec:conclusions} we make some concluding remarks.

\section{Review of F-theory, Weierstrass models and non-Higgsable
  clusters} 
\label{sec:review}

\subsection{Seven-branes and non-Higgsable clusters}

In one description of F-theory, the IIB axiodilaton $\tau$ is conveniently
encoded as the complex structure modulus of an elliptic curve
(complex torus) that is
fibered over the compact spatial dimensions.  When the total space $X$
of the resulting elliptic fibration $\pi:X\rightarrow B$ is
Calabi-Yau, supersymmetry is preserved in the non-compact spacetime.
One of the main features of F-theory is that the compactification
manifold $B$ can have curvature and a nontrivial canonical class $K$,
unlike Calabi-Yau compactifications where the canonical class of the
compactification manifold is trivial (up to torsion).  When $B$ has
nontrivial (Ricci) curvature, for the total space $X$ to be Calabi-Yau
seven-branes must be added that wrap complex codimension one cycles in
$B$.  Specifically, seven-branes are located on loci in the base over
which the elliptic fiber parameterized by $\tau$ becomes singular.
For simple seven-brane configurations, the total space $X$ of the
elliptic fibration can still be smooth even when the fibration
structure is singular.  When multiple seven-branes come together over
a common (complex) codimension one locus, however, the total space $X$
of the fibration itself becomes singular.  Such a situation is
necessary for a nonabelian gauge group to arise in the low-energy
supergravity theory of the F-theory compactification.

For some simple choices of the base manifold $B$, such as $\P^2$ or
$\P^3$ for 6D or 4D F-theory compactifications, respectively, the
generic elliptic fibration $X$ over $B$ is smooth, and the
seven-branes can be separated so that there is no nonabelian gauge
group in the low-energy theory.  Nonabelian gauge groups can be
``tuned'' over such a base by piling up multiple seven-branes on a
common codimension one locus, but in 6D, and in 4D in
the absence of flux, the resulting
nonabelian gauge groups can always be broken
by varying the moduli of the elliptic fibration. In the low-energy
theory this corresponds to a Higgsing transition where a vacuum
expectation value is given to a matter field that is charged under the
nonabelian gauge group.  Such charged matter fields are encoded in the
moduli of the holomorphic $\tau$ function describing the elliptic
fibration.  

For many choices of base geometry $B$, however, the curvature of the
base is so strong that multiple seven-branes must pile up in certain
places, giving nonabelian gauge groups in the low-energy theory that
cannot be removed by any simple deformation of the geometry
 that preserves the structure of the base as well as the
Calabi-Yau property and supersymmetry.  A nonabelian gauge group
arising from such a seven-brane configuration in a compact space $B$
is said to be \emph{geometrically non-Higgsable}, meaning that there
is no holomorphic deformation of the $\tau$ function for an
elliptically fibered Calabi-Yau threefold $X$ over the base $B$ that
breaks the gauge symmetry on the seven-branes.  The geometrically
non-Higgsable gauge symmetry can be enhanced on subloci in the moduli
space, but it can never be broken by a geometric deformation.  Whether
or not this phenomenon occurs depends on the topology of $B$; it
cannot occur for F-theory compactifications to eight dimensions, but
it appears quite generally for F-theory compactifications to six
dimensions, and there is growing evidence (to which this paper
represents a contribution) that this is also the case in four
dimensions.

One way to understand the appearance of non-Higgsable gauge groups and
matter is from the local geometry of codimension one\footnote{Note that in
general when we refer to codimension $k$ loci, we mean complex
codimension $k$ loci in the base manifold.} algebraic cycles,
or {\it divisors}, on the base $B$.  When a divisor in $B$ is embedded
with a sufficiently negative normal bundle, this produces so much
curvature in the local geometry that seven-branes must pile up on the
divisor to maintain the Calabi-Yau nature of the total space $X$.  A
systematic analysis of circumstances under which a set of divisors in
a threefold $B$ must carry non-Higgsable gauge groups and matter is
given in \cite{4D-NHC} in terms of the geometry of the divisors and
their normal bundles.   In particular, the normal bundle of a divisor
is simply a complex line bundle over the divisor, and the presence of
a non-Higgsable gauge group can be determined by the absence of
sections for related line bundles over the divisor.

\subsection{Weierstrass models and non-Higgsable clusters}
\label{sec:Weierstrass}

To be more explicit and to discuss specific models, it is helpful to
briefly review the relevant mathematics of Calabi-Yau elliptic
fibrations. We start the discussion from the physical point of view,
in terms of seven-branes of type IIB string theory.
We consider a
Calabi-Yau
elliptic fibration $\pi:X\rightarrow B$
that  has a global section and a
representation as a Weierstrass model
\footnote{By a theorem of Nakayama \cite{Nakayama}, any elliptic
  Calabi-Yau threefold with section can be described through a
  Weierstrass model.  This is not proven for fourfolds, but holds 
at least for
  the class of examples we consider.}
\begin{equation}
y^2 = x^3 + f\, xz^4 + g\, z^6
\end{equation}
where $f$ and $g$ are sections $f\in\Gamma(\cO(-4K))$, $g\in
\Gamma(\cO(-6K))$, $K$
is the canonical class of the base $B$, $(x,y,z)$ are
homogeneous coordinates in the weighted projective space 
$\bP(2,3,1)$ and $z=0$ is the
section. One often passes to the patch where $z=1$, giving the common
form $y^2 = x^3 + f\, x + g$.  For fixed $f$ and $g$ the Weierstrass equation represents
a particular Calabi-Yau elliptic fibration $X$. More generally,
a
family ${\cal  M}$ of Calabi-Yau elliptic fibrations
is parameterized by a
continuous set of choices for $f$ and $g$. 
In the bulk of  the moduli space ${\cal  M}$,
varying $f$ and $g$ varies
the complex structure of a generic member of the family. We henceforth refer
to $\cM$ as the Weierstrass moduli space.  Note that ${\cal M}$ includes not only the complex
structure moduli space for the Calabi-Yau threefold associated with
the generic elliptic fibration over $B$, but also strata associated
with the moduli spaces of other threefolds that arise when
higher-order singularities, corresponding to enhanced gauge groups,
are tuned on some divisors in the base.

The elliptic fiber becomes singular over the discriminant locus
\begin{equation} \Delta \equiv 4 f^3 + 27 g^2=0,
\end{equation} which defines a divisor in $B$. An $SL(2,\Z)$
monodromy is induced on $\tau$ by following any closed loop around the locus
$\Delta = 0$.  This implies
in the physical language of F-theory
that there   are seven-branes
on $\Delta=0$ that source the
axiodilaton. As $f$ and $g$ are varied the structure
of seven-branes may change, giving rise to different low-energy
physics. For example, obtaining non-abelian gauge symmetry along
a seven-brane configuration on a locus described by
$z=0$
in a local coordinate system requires that $\Delta$ is of the form
\begin{equation}\Delta = z^N\, \tilde \Delta\end{equation}
with $N\geq 2$.

Given this language for describing the elliptic fibration, a
geometrically non-Higgsable seven-brane configuration arises when
$\Delta = z^N \tilde{\Delta}$ with
 $N > 2$ for any allowed choice of $f$ and $g$.  In such a situation,
  there is no variation of $f$ and $g$ that removes this
  factorization, which means that there is no symmetry breaking flat
  direction in the Weierstrass moduli space. If $B$ is toric, then $f$
  and $g$ can be described as polynomials in homogeneous coordinates,
  and it is a straightforward exercise to classify all possible
  monomials that might appear in them. If all the monomials contain a
  nonzero power of some homogeneous coordinate $z$, then so do $f$ and
  $g$, and therefore so does the discriminant. In this way, given a
  toric variety, it is straightforward to determine whether an
  F-theory compactification on a toric base $B$ contains geometrically
  non-Higgsable seven-branes on toric divisors (see
  \cite{mt-toric, Anderson-Taylor} for explicit monomial analyses of
  toric base manifolds for F-theory fibrations to 6D and 4D
  respectively, in this context). This method is the one
  we use.

  Let us be concrete about how the conditions that $f, g$ vanish to
  given orders on a divisor $D$ in a threefold base can be expressed
  particularly simply in terms of sections of line bundles on
  $D$. Though these can be computed easily in the toric case, we keep
  the discussion general since these methods apply even when $B$ and
  $D$ are not toric.  If $D$ is defined in terms of a local
  coordinate $\zeta$
  via $\zeta = 0$, then we expand $f$ and $g$
\begin{equation}
f =  f_0 + f_1 \zeta + \cdots \qquad \qquad g = g_0 + g_1 \zeta + \cdots
\end{equation}
in a power series around $\zeta = 0$.
As described in \cite{Anderson-Taylor, 4D-NHC}, the coefficient
functions $f_i, g_i$ are naturally described as global sections of line
bundles over $D$
\begin{eqnarray*}
f_i & \in & \Gamma ({\cal O}(-4K_D) \otimes N_{D/B}^{4-i})  \label{eq:fi}\\
g_i & \in & \Gamma ({\cal O}(-6K_D)\otimes N_{D/B}^{6-i})  \,. \label{eq:gi}
\end{eqnarray*}
When the divisors defining these line bundles are not effective (that is,
when those bundles do not have global sections and accordingly some 
$f_i$ or $g_i$ do not exist), the
corresponding terms in the expansion of $f, g$ must vanish, giving
rise to non-Higgsable structure when the resulting orders of vanishing
are sufficiently large.

\begin{table}
\begin{center}
\begin{tabular}{|c |c |c |c |c |c |}
\hline
Type &
ord ($f$) &
ord ($g$) &
ord ($\Delta$) &
singularity & nonabelian symmetry algebra\\ \hline \hline
$I_0$&$\geq $ 0 & $\geq $ 0 & 0 & none & none \\
$I_n$ &0 & 0 & $n \geq 2$ & $A_{n-1}$ & $\gsu(n)$  or $\gsp(\lfloor
n/2\rfloor)$\\
$II$ & $\geq 1$ & 1 & 2 & none & none \\
$III$ &1 & $\geq 2$ &3 & $A_1$ & $\gsu(2)$ \\
$IV$ & $\geq 2$ & 2 & 4 & $A_2$ & $\gsu(3)$  or $\gsu(2)$\\
$I_0^*$&
$\geq 2$ & $\geq 3$ & $6$ &$D_{4}$ & $\gso(8)$ or $\gso(7)$ or $\gg_2$ \\
$I_n^*$&
2 & 3 & $n \geq 7$ & $D_{n -2}$ & $\gso(2n-4)$  or $\gso(2n -5)$ \\
$IV^*$& $\geq 3$ & 4 & 8 & $E_6$ & $\ge_6$  or $\gf_4$\\
$III^*$&3 & $\geq 5$ & 9 & $E_7$ & $\ge_7$ \\
$II^*$& $\geq 4$ & 5 & 10 & $E_8$ & $\ge_8$ \\
\hline
non-min &$\geq 4$ & $\geq6$ & $\geq12$ & \multicolumn{2}{c|}{does not
appear for supersymmetric vacua} \\
\hline
\end{tabular}
\end{center}
\caption[x]{\footnotesize  Table of
codimension one
singularity types for elliptic
fibrations and associated nonabelian symmetry algebras.
In cases where the algebra is not determined uniquely by the degrees
of vanishing of $f, g$,
the precise gauge algebra is fixed by monodromy conditions that can be
identified from the form of the Weierstrass model.
}
\label{t:Kodaira}
\end{table}

The types of singularities that can arise in an elliptic fibration,
and the Dynkin diagrams associated with the corresponding low-energy
gauge groups, are determined by the famous Kodaira classification; the
singularity types and associated gauge group factors can be determined
by the Tate algorithm \cite{Morrison-Vafa-II} according to the orders
of vanishing of $f$ and $g$, augmented by more detailed structure
associated with the effects of outer monodromy that can give rise to
non-simply-laced groups in 6D and 4D.  The standard table of
singularity types is reproduced in Table~\ref{t:Kodaira}.  When there
is a singularity associated with a nonabelian gauge group, the
Weierstrass model of the elliptically fibered Calabi-Yau $X$ is
singular, corresponding to a limit in which K\"ahler moduli associated
with cycles on $X$ that correspond to the nonabelian generators are
taken to vanish; these moduli may be turned on in a related M-theory
compactification to one lower space-time dimension. Alternatively,
singularities can also be studied by deformation of a local geometry
(which sometimes results from deformation of a global geometry); see
\cite{ghs1,ghs2,ghs3} for recent work on the connection between
deformation, singularities, and string junctions.  When the nonabelian
gauge content of the theory is enhanced by tuning a non-generic
seven-brane configuration supporting additional gauge generators, the
elliptic Calabi-Yau over $B$ changes topology.  Taken in the opposite
direction, the gauge group can be broken by turning on moduli in the
Weierstrass model.  In the low-energy theory, this corresponds to a
Higgsing of the additional nonabelian symmetry by giving vacuum
expectation values to charged matter fields associated with the tuned
Weierstrass moduli; see \cite{ghs2} for the explicit connection between
such Higgsing processes and string junctions representing the
massive W-bosons of the broken theory.  When the singularity of the elliptic fibration
becomes too strong, in particular when the order of vanishing of $f,
g$ on a codimension two locus reaches $(4, 6)$, the theory becomes a
superconformal field theory coupled to gravity \cite{Seiberg-SCFT}.
At this point, it is often possible to turn on new moduli and enter a
new branch of the F-theory moduli space.  Geometrically, this
corresponds to blowing up the codimension two locus on the base $B$,
giving a new base variety $B'$ \cite{Seiberg-Witten, Morrison-Vafa-II}.

\subsection{Non-Higgsable clusters for 6D F-theory vacua}
\label{sec:6D} 

A complete classification of non-Higgsable clusters for 6D F-theory
vacua was given in \cite{clusters}, and derived from an alternative
point of view in \cite{4D-NHC}.  We review a few of the salient
points here and describe some features of non-Higgsable clusters for 6D
F-theory models that are relevant to the main points of this paper.

For 6D F-theory compactifications, we are interested in the case where
the base variety used for the compactification is a complex surface $B_2 = \b2$.
In this situation, the codimension one loci that support seven-branes
are complex curves $C$.  The only situation in which a curve $C
\subset B$ can support a non-Higgsable cluster is when $C$ has genus
zero, {\it i.e.} is a rational curve with topology $S^2$; the complex
structure in fact makes it $\P^1$.  The normal
bundle of such a curve $N= {\cal O} (n)$ is characterized by a single
integer $n$, which also gives the self-intersection of the curve $C$
within $B$, $C \cdot C = n$.  Non-Higgsable clusters arise when there
are effective divisors that are rational curves
with sufficiently negative normal bundle that the self-intersection of
the curve satisfies
$n\leq -3$.   Isolated curves of self-intersection $-3, -4, -5, -6, -7,
-8, -12$ carry non-Higgsable gauge groups with Lie algebras $\gsu_3,
\gso_8, \gf_4, \ge_6, \ge_7, \ge_7, \ge_8$.  Curves of
self-intersection $-9, -10, -11$ always contain points where $f, g$
vanish to orders (4, 6), so that F-theory models on bases containing
such curves give superconformal field theories coupled to gravity.
Blowing up these points can lead to well-behaved supergravity models
with an additional tensor field.
In addition to the non-Higgsable clusters associated with isolated
curves of negative self-intersection, there are three configurations
of multiple curves that give rise to non-Higgsable product groups.  Chains of
curves of self-intersection $(-3, -2), (-3, -2, -2)$ give rise to
non-Higgsable gauge algebras $\gg_2 \oplus \gsu_2$, and the chain
$(-2, -3, -2)$ gives a non-Higgsable $\gsu_2 \oplus \gso_7 \oplus
\gsu_2$.

An interesting feature of non-Higgsable clusters  in 6D is that they
can carry only very specific types of charged matter.  The
non-Higgsable cluster on a curve of self-intersection $-7$, for
example, carries a gauge algebra $\ge_7$ and a half-hypermultiplet of
matter in the real $\frac{1}{2} {\bf 56}$ representation.  That this
matter cannot be given a vacuum expectation value to Higgs the $\ge_7$
can be seen from the fact that the D-term constraint cannot cancel
for a single such field.  Another way to understand this is from the
fact that no gauge-invariant holomorphic function can be formed from a
single such field \cite{Luty-Taylor}.

In six dimensions, F-theory geometry is tightly coupled to the physics
of the corresponding low-energy supergravity theory.  All physical Weierstrass
moduli in six-dimensional F-theory models are massless, and there is a
precise correspondence between the moduli space in the low-energy
theory and the complex structure moduli space  associated with a
Calabi-Yau threefold on each branch of the  space of F-theory vacua.  Thus,
for six-dimensional F-theory models, gauge groups that are
geometrically non-Higgsable are also physically non-Higgsable.
Furthermore, all branches of the moduli space of F-theory vacua are
connected, suggesting that there is a unique consistent ${\cal N} = 1$
supersymmetric quantum theory of gravity in six dimensions.

The gauge groups associated with non-Higgsable clusters in 6D theory
models are ``generic'' in two distinct senses.  The first sense in
which this type of structure is generic is the fact that when there is
a non-Higgsable structure over a given base $B$, the resulting gauge
group and matter are present in every 6D theory over that base,
independent of the choice of Weierstrass moduli. While a larger gauge
group can often be tuned at special
subloci of the Weierstrass moduli space, the only way to reduce the size of the gauge group
in this situation is through a tensionless string transition to a
simpler base geometry, corresponding to blowing down one or more $-1$
curves in the base. 

\begin{figure}
\begin{center}
\includegraphics[width=8cm]{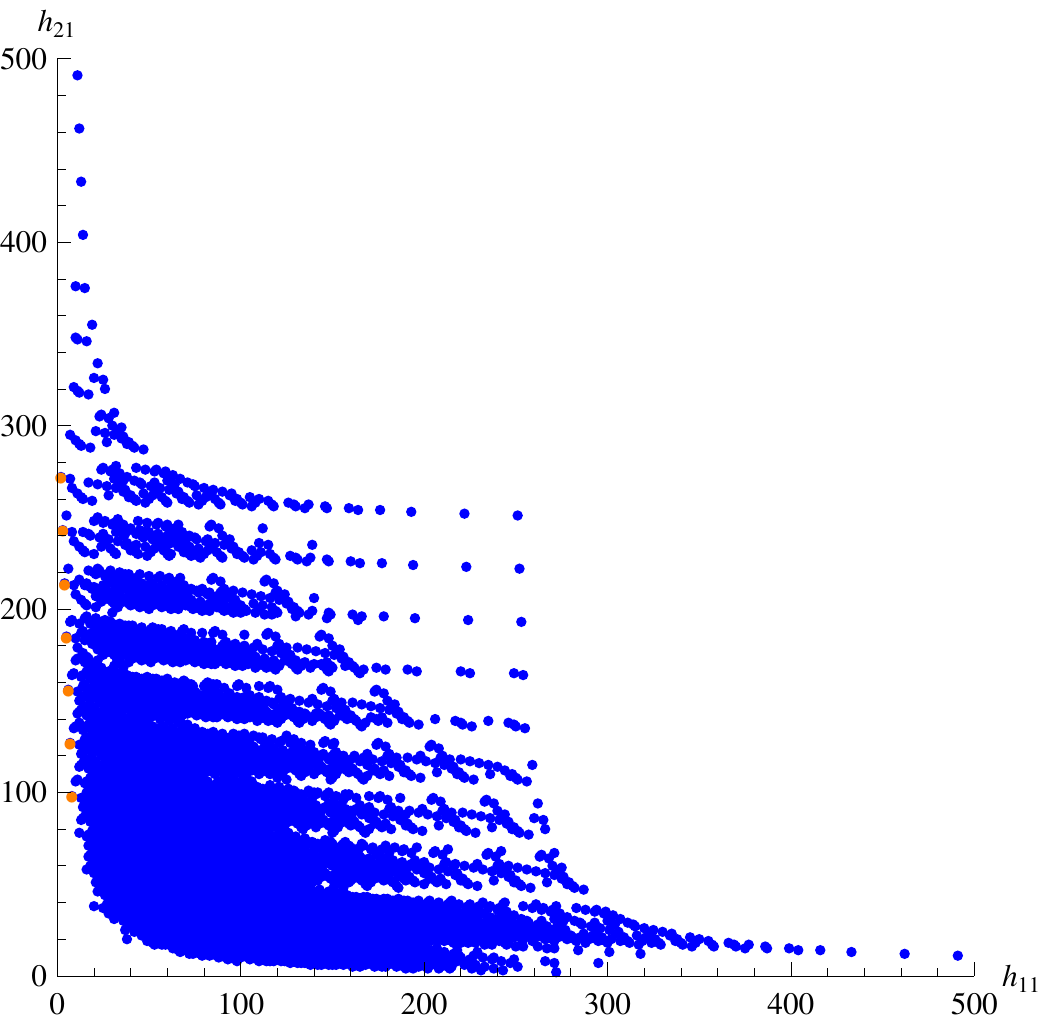}
\end{center}
\caption[x]{\footnotesize Hodge numbers for generic elliptic
  Calabi-Yau manifolds fibered over the 61,539 toric base surfaces
  that support elliptically fibered Calabi-Yau's \cite{mt-toric,
    Hodge}.  Only seven Hodge numbers (on the left in orange)
  correspond to elliptic fibrations over {\it weak Fano}
bases that do not give rise
  to non-Higgsable clusters.}
\label{f:6D-Hodge}
\end{figure}

The second sense in which non-Higgsable clusters are generic in 6D
F-theory models is that almost all base manifolds $B$ that support
elliptically fibered Calabi-Yau threefolds give rise to non-Higgsable
clusters. While of course a measure on the space of vacua is difficult
to define, the set of possible base manifolds $B$ is a finite set, so
a reasonable (if coarse) measure is to simply look at the fraction of
the finite number of bases that support NHC's, or even more roughly to
consider the fraction of the set of possible Hodge numbers for generic
elliptic fibrations over $B$ that are associated with bases forcing
non-Higgsable structure. By either measure, virtually all allowed
bases $B$ have negative self-intersection curves that give
non-Higgsable clusters. We summarize briefly the extent to which this
is understood at a quantitative level.  From the minimal model program
for surfaces \cite{bhpv} and the work of Grassi \cite{Grassi}, it is
known that all twofold bases that support nontrivial elliptic
fibrations can be formed by blowing up points on $\P^2$ or on the
Hirzebruch surfaces $\F_m$. This approach has been used to
systematically explore the space of possible bases $B$ and to
enumerate various classes of elliptically fibered Calabi-Yau
threefolds. In \cite{mt-toric, Hodge, Martini-WT}, the set of all
toric and ``semi-toric'' bases $B$ that support an elliptically
fibered Calabi-Yau threefold and the Hodge numbers of the
corresponding generic elliptic fibrations were analyzed, and
in \cite{Wang-WT}
a systematic classification was given for all bases (including those
without toric or semi-toric structure) that support elliptically
fibered CY threefolds with small $h^{1, 1}$ or large $h^{2, 1}$.
Rigorous upper bounds on the set of CY threefolds with large $h^{2,
  1}$
\cite{Hodge, Johnson-WT}, and the close correspondence between these analyses and the set
of Calabi-Yau threefolds constructed through toric methods in
\cite{Kreuzer-Skarke}, indicate that the set of possible base surfaces $B$ is not
only finite but also fairly well understood, at least at a coarse
level of detail.  Of the over 60,000 toric bases constructed in
\cite{mt-toric}, only 16 lack non-Higgsable clusters, and of the over
130,000 semi-toric bases found in \cite{Martini-WT} only 27 lack
non-Higgsable clusters.  Indeed, the only base surfaces that do not
have at least one curve of self-intersection $-3$ or below are the
generalized del Pezzo surfaces, which number on the order of several
hundred. 
(Generalized del Pezzo surfaces can be defined as non-singular
projective surfaces with the properties that the canonical class $K$ satisfies $K \cdot K
> 0$
and that the surface is
{\it weak Fano}, {\it i.e.} $-K \cdot C \geq 0$ 
for every
effective curve $C$ in the surface.)
For all these surfaces, the generic Calabi-Yau elliptic
fibration has a set of Hodge numbers $(2 + T, 272 -29 T)$, where $T =
0, \ldots, 9$ corresponds to the number of tensor multiplets in the 6D
theory resulting from compactification on a generalized del Pezzo
surface formed from blowing up $\P^2$ at $T$ points. Thus, of all
possible generic elliptic Calabi-Yau manifolds over all possible
bases, there are only 10 possible Hodge number combinations associated
with bases that do not admit non-Higgsable clusters. In
Figure~\ref{f:6D-Hodge}, we have graphed the Hodge numbers associated
with generic elliptic fibrations over all toric base surfaces; of
these, only seven Hodge number pairs, on the left-hand side of the
graph, correspond to F-theory compactifications that are free of
non-Higgsable gauge groups.

\subsection{Non-Higgsable clusters for 4D F-theory vacua} 
\label{sec:4D}

At the level of Calabi-Yau geometry, there is a close parallel between
the structure of F-theory compactifications to 4D and 6D.  For a 4D
compactification, we have an elliptically fibered Calabi-Yau fourfold
$X= X_4$ over a complex threefold base $B$.  As in 6D, divisors $D \subset
B$ with a  ``sufficiently negative'' ({\it i.e.,} far from effective)
normal bundle must support a
geometrically non-Higgsable gauge group.  A general approach to
analyzing the geometry of non-Higgsable clusters for 4D F-theory
models in terms of the geometry of surfaces and normal bundles was
worked out in \cite{Anderson-Taylor, 4D-NHC}.  While similar in
general outline to the 6D story, the set of possible non-Higgsable
clusters in 4D is rather more complicated than in 6D.  In addition,
for 4D F-theory compactifications there are  complications in
the relationship between the underlying geometry and the low-energy
physics.  For one thing, flux (G-flux) in F-theory compactifications
can produce a superpotential that lifts some or all of the geometric moduli of
the theory.  This can in principle drive the model to special subloci
of the moduli space with enhanced gauge symmetry.  Furthermore,
instantons and other structure on the worldvolume of the seven-branes
can break otherwise non-Higgsable gauge groups
(see {\it e.g.} \cite{Anderson-Taylor} for an explicit example), so that in 4D there is
not necessarily a precise correspondence between non-Higgsable
geometry and the gauge group and matter content in the low-energy
theory.  These are important issues, but a necessary prerequisite to a
systematic study of the role of non-Higgsable clusters in low-energy
theory must begin with the purely geometric aspect of the question.
Thus, we focus here only on the geometric structure of non-Higgsable
clusters, leaving further study of the role of fluxes and seven-brane
worldvolume degrees of freedom to further work.
While in principle all aspects of non-Higgsable clusters explored in
the later part of this paper are purely based on geometry, and all
such clusters should be referred to as ``geometrically non-Higgsable''
to be precise, we will often drop the adjective and leave it as
understood in the remainder of this paper.

As found in \cite{4D-NHC}, there are strong local constraints on the
types of gauge groups that can appear in 4D (geometric) non-Higgsable
clusters, and on the products of two gauge groups that can appear with
jointly charged matter, either in an isolated cluster or as part of a
larger cluster.  The nonabelian gauge algebras that can appear in a 4D
non-Higgsable cluster are \cite{ghst, 4D-NHC}
\begin{equation}
 \gsu_2, \gsu_3, \gg_2, \gso_7,  \gso_8, \gf_4, \ge_6, \ge_7, \ge_8 \,.
\end{equation}
The only possible gauge algebra combinations that can appear in a
product of two groups with jointly charged matter are
\begin{equation}
\begin{array}{ccccc} 
\gsu_2 \oplus \gsu_2, &\hspace*{0.1in} & \gsu_2 \oplus \gsu_3,
&\hspace*{0.1in} & \gsu_3 \oplus \gsu_3, \\
\gg_2 \oplus \gsu_2, &\hspace*{0.1in} & 
\gso_7 \oplus \gsu_2
&\hspace*{0.1in} & 
\end{array}
\label{eq:pairings}
\end{equation}
Unlike in 6D, however, where only rational curves can support a
non-Higgsable gauge group factor, the possible divisor geometries that
can support non-Higgsable gauge group factors are quite varied in four
dimensions.  One of the goals of this work is to explore some of the
range of possible divisor geometries that can support non-Higgsable
gauge groups.  Non-Higgsable clusters in four dimensions can also have
much more complicated structure than for 6D theories.  As shown in
\cite{4D-NHC}, non-Higgsable clusters can have gauge groups that are
combined into ``quiver'' diagrams\footnote{A quiver diagram for a
  gauge theory is a graph in which the nodes represent simple
gauge group factors
  and edges represent bifundamental type matter between
  pairs of gauge factors.} that exhibit branching, loops, and
long chains.  We also explore some of this structure in the following
sections.

\section{Classification of $\P^1$ bundle bases} 
\label{sec:classification}

The threefold bases $B$
that we consider here are defined as $\P^1$
bundles over the set of surfaces $S$ classified in \cite{mt-toric},
which are those smooth toric surfaces
that themselves support smooth elliptically fibered Calabi-Yau threefolds. 
Threefolds with $\P^1$
bundle structure have been considered previously in the context
of heterotic/F-theory duality
(see, {\it e.g.}, \cite{fmw}). 
In particular, in \cite{Anderson-Taylor}, a systematic analysis was
given of all F-theory/heterotic dual
constructions where the 
F-theory model is a $\P^1$ bundle and the dual Calabi-Yau on the
heterotic side is smooth, which occurs when the surface $S$ is a
generalized del Pezzo surface.
While in principle the F-theory models that result from the threefold
bases we consider here also have heterotic duals, the compactification
manifolds on the dual heterotic side are generally highly singular.
The  simplest classes of  $\P^1$-bundle threefold bases, where the
surface $S$ is taken to be $\P^2$ or a Hirzebruch surface $\F_m$, were
also studied in the context of F-theory in \cite{Klemm-lry, Mohri,
  Berglund-Mayr, Grimm-Taylor}. 

The bases in the class that we consider here are of interest for several reasons.  Firstly,
the $\P^1$ bundle threefold base over a surface $S$ has a divisor
(actually two divisors, corresponding to sections of the $\P^1$
bundle) with the geometry of $S$.  Thus, this class of threefold bases
provides a rich set of examples of distinct divisor topologies and
normal bundles.  In fact, the sections of each of the many threefold
bases we consider here correspond to distinct divisor/normal bundle
geometries for each distinct threefold base.  The second reason that
the bases in this class are of interest is because they are one
natural analog of the Hirzebruch surfaces $\F_m$ (which are $\P^1$
bundles over $\P^1$), which form most of the ``minimal model'' bases for the
set of 6D F-theory compactifications.  Many of the threefold bases
that we consider here are similarly, in a certain sense, minimal model
bases for 4D F-theory compactifications.  As we discuss further below,
a more systematic and rigorous analysis along these lines would
involve the framework of Mori theory \cite{Mori}, and will be left for
further work.  Finally, the geometries in the class we consider here
are interesting because they give a rich sample of threefolds that can
act as bases for elliptically fibered CY fourfolds, which goes well
beyond the set of such bases that have been studied previously.  This
gives us a class of examples with which to explore what kinds of
non-Higgsable structures may be typical in the 4D F-theory landscape.
Because the class of bases explored here is rather specialized, the
lessons taken in this latter context must be taken as only suggestive;
nonetheless, some general conclusions that we can draw may be
sufficiently robust that they will persist over larger distributions
of threefold bases that can be studied using other methods.

\subsection{Twists and allowed bundles} 
\label{sec:twists}

To define a threefold $B$ that is a $\P^1$ bundle with section over a
surface $S$ we must choose a class $T \in H_2(S,\Z)$.  From one
perspective, $T$ can be thought of as a ``twist'' corresponding to the
first Chern class of a line bundle ${\cal L}$ over $S$ so that $B$ is
the projectivization $B=\P (\cal L\oplus {\cal O}_S)$.  Alternatively,
$\cO_S(\pm T)$ are the normal bundles of the two sections.  Note that
choosing one line bundle $\cL$ is sufficient for a general
construction since $\P(\cL_1 \oplus \cL_2) = \bP((\cL_1 \otimes
\cL_2^*)\oplus \cO_S)=:\P(\cL\oplus \cO_S)$.

In the
language of toric geometry, if $v_i =(x_i, y_i), i = 1, \ldots, N$ give the
rays that define the toric fan for $S$,
then the fan for  $B$ can be defined using the set of rays
\cite{Anderson-Taylor}
\begin{eqnarray}
w_i & = &  (x_i, y_i, t_i), \; 1 \leq i \leq  N
\label{eq:wi}\\
w_{N+ 1, N+ 2} =
s_\pm & = &  (0, 0, \mp 1) \label{eq:ws}\,,
\end{eqnarray}
where the coordinates $t_i$ determine the twist $T$ by $T=\sum_i t_i d_i$
with $d_i\in H_2(S,\Z)$ the toric divisors in $S$ associated to the vertices
$v_i$. There are two linear relations amongst the $d_i$, so
that the number of independent parameters in the twist $T$ is $N-2 =
h^{1, 1} (S)$.

As discussed above,
an F-theory model can be defined over the base $B$ when there exists
an elliptic fibration (with section) over $B$ that has a Calabi-Yau
resolution, and a Weierstrass model\footnote{As mentioned above, unlike the case of threefolds, it has not been proven that every such
 fourfold admits a Weierstrass model, 
but one exists for all the elliptic fibrations over
toric
 bases $B$ that we consider here.}
\begin{equation}
 y^2 = x^3+ f x+ g \,.
\label{eq:Weierstrass}
\end{equation} 
For $B$ to admit an elliptic fibration with a Calabi-Yau resolution,
there cannot be any divisors  on $B$ where the order of
vanishing of $f, g$ is generically $(4, 6)$.  We also assume that we
are in a supergravity phase of the theory that does not include
superconformal sectors, so that there are no curves
with $(4, 6)$ vanishing.   These conditions impose stringent
conditions on the geometry of $B$, and in particular
place constraints on the set
of allowed twists $T$ for any given base surface $S$.
Note that some weaker singularities may arise that cannot be resolved
to a total space that is a Calabi-Yau fourfold, where consistent
F-theory supergravity models may still be possible.  For example, such
singularities may include codimension 3 loci where $f, g$ vanish to
orders $(4, 6)$ \cite{Morrison-codimension-3}, 
or certain other mild local singularity types that
are weaker than $(4, 6)$ vanishing and yet cannot be resolved to
smooth Calabi-Yau spaces.
We allow for such milder singularity types in our analysis, leaving a
more detailed study of the relevant physics to future work.

In \cite{Anderson-Taylor},  the F-theory conditions on the structure
of a $\P^1$ bundle were shown to correspond to constraints on the
components $\tilde{t}_i = T \cdot d_i$ that limit the set of possible
$T$ for a given surface $S$ to a finite set.  When the
self-intersection of the curve $d_i$ is $d_i \cdot d_i = - n$, then
the constraints on the corresponding twist component are
\begin{eqnarray}
n = 0: & \hspace*{0.1in} &  |T \cdot d_i | \leq 12 \label{eq:t-0}\\
n = 1: & \hspace*{0.1in} &  |T \cdot d_i | \leq 6\label{eq:t-1}\\
6 \geq
n  \geq 2: & \hspace*{0.1in} &  |T \cdot d_i | \leq 1 \,.\label{eq:t-2}
\\
n  \geq  7: & \hspace*{0.1in} &  T \cdot d_i  = 0 \,.\label{eq:t-big}
\end{eqnarray}
While these constraints in principle allow only a finite number of
possible twists $T$ for any given surface $S$, these constraints
are only necessary and not sufficient.  For any $T$ satisfying these
constraints, a more detailed check must be carried out to determine if
the resulting base $B$ is free of $(4, 6)$ divisors or curves.
A general framework for carrying out such an analysis for an arbitrary
base $B$ was developed in \cite{4D-NHC}.  
For toric bases such as those considered here, the order of vanishing
of $f, g$ on each of the toric divisors and curves can be directly
computed using toric methods, or can be determined using the more
general methods of \cite{4D-NHC}.

As the complexity of the surface $S$ grows, the number of solutions
$T$ to the constraints (\ref{eq:t-0}--\ref{eq:t-big}) can grow
exponentially, but most such solutions will have problems such
as $(4, 6)$ singularities on curves.  To reduce the combinatorial
problem of identifying all allowed twists $T$ over a given toric base
surface
$S$ to a tractable computation, stronger constraints are needed.  We
have used methods related to those of \cite{Anderson-Taylor, 4D-NHC}
to develop a set of stronger constraints that enable a complete
classification of the twists $T$ for any given surface $S$ in a
relatively modest computational time.  The details of these stronger
constraints are described in Appendix~\ref{sec:method-1}
and Appendix~\ref{sec:method-2}.

\subsection{Classification of bases} 

We have used the constraints described in the Appendices to carry out
a systematic analysis and enumeration of all bases $B$ that have the
form of a $\P^1$ bundle over any of the 61,539 toric base surfaces $S$
identified in \cite{mt-toric}.  This gives a total of 109,158
threefolds $B$ that can support elliptic Calabi-Yau fourfolds, and
thus can act as compactification spaces for F-theory to four
dimensions.  This set of threefolds includes the 4,962 bases $B$ that
were constructed in \cite{Anderson-Taylor} as $\P^1$ bundles over the
subset of 16 surfaces $S$ that are generalized del Pezzo surfaces,
corresponding to heterotic duals on elliptic Calabi-Yau manifolds with
generically smooth Weierstrass models.
Note that we have only counted as distinct bases $B$ that have
distinct topology, independent of the $\P^1$ bundle structure.  So,
for example, the $\P^1$ bundle with twist $(t_1, t_2, t_3, t_4) =(0, 0, 0, 1)$ over the base
$\F_0 = \P^1 \times \P^1$ is considered to be the same threefold
base $B$ as $\P^1
\times \F_1$.

Note that in this classification we have not included any bases that
have curves on which $f, g$ are forced to vanish to orders $(4, 6)$.
In particular, we do not consider base threefolds that contain a
divisor $D$ that supports a non-Higgsable $E_8$ if $g$ vanishes to
order $6$ on any curve $C \subset D$; such a curve can arise from a
$II^*$-$I_1$ collision where the $E_8$ seven-brane intersects the
residual discriminant.  These would be the analogue of bases
containing $-9, -10,$ and $-11$ curves for 6D F-theory models.  While
in principle these base threefolds describe superconformal field
theories coupled to gravity, and could give (generally non-toric)
acceptable bases for 4D supergravity models without $(4, 6)$ curves
after blowing up the offending curves $C$, this analysis is more
complicated than the corresponding story in six dimensions, and we do
not explore these bases further in this paper.
The technical details of the procedure used to rule out these bases
are described in Appendix~\ref{sec:appendix-e8}.

In the remainder of this paper we provide some details of
the structure of the 109,158 threefolds
that we have constructed.

\subsection{Distribution of $\P^1$ bundle bases} 

We begin by characterizing the range of bundles that are allowed for
the different base surfaces $S$.

The distribution according to $h^{1, 1}(S)$ of toric base surfaces $S$
that support elliptically fibered CY threefolds is described in
\cite{mt-toric}.  The maximum number of distinct such bases peaks
around $h^{1, 1}(S) \cong 25$ and then drops rapidly, with a
relatively small number of bases having $h^{1, 1}(S) > 80$, and the
  maximum value being $h^{1, 1}(S) = T + 1 = 194.$

Only a subset of these surfaces admit $\P^1$ bundles over them that form
valid bases $B$ that can directly support elliptic CY fourfolds.
In particular, if the surface $S$ contains a $-12$ curve, then there
is no good corresponding $B$.  This can be seen as follows.  From the
analysis of twists in \S\ref{sec:twists}, for a $-12$ curve $d_i$, we
must have $T \cdot d_i = 0$.  This means that in the local geometry of
any $\P^1$ bundle $B$ over $S$, the $\P^1$ bundle over the curve $d_i$
is a divisor $D$ in $B$ with the geometry of $\F_0$, and normal bundle
$N_{D/B}=\cO_D(-12X)$, where  $X, Y$ are the standard basis for the cone of effective
curves on $D$ with $X^2=Y^2=0$ and $X\cdot_D Y=1$.  There is a type $II^{*}$
($E_8$) singularity on $D$ and, using the language of \cite{4D-NHC} we compute
\begin{equation}
g_5 \in \Gamma(\cO_D(-6K_D)\otimes N_{D/B}) = \Gamma(\cO_D(12Y)) 
\end{equation} 
from which we see that $g_5=0$ defines a nontrivial curve that is
itself a divisor in the surface $D$.  
Along this curve, $f$ and $g$ vanish to orders $(4,
6)$. Similarly, any surface $S$ with a curve of $-9$ or below gives a
divisor $D$ carrying an $E_8$ singularity where the curve $g_5=0$ in
$D$ is a $(4,6)$ locus; thus, we need not consider any surfaces $S$
with curves of self-intersection $-9$ or below.

On the other hand, if $S$ does not admit any curves of
self-intersection $-9$ or below, then there is at least one $\P^1$
bundle $B$ over $S$ that can support a good elliptic Calabi-Yau
fourfold, namely the trivial bundle $B =\P^1 \times S$.  This gives us
a subset of 24,483 surfaces  $S$ of the $61,539$ found in
\cite{mt-toric} over which to construct bases $B$.
Of these, the largest has $h^{1, 1}= 72$, and is described by a chain
of toric curves containing 12 copies of the basic sequence $-8, -1,
-2, -3, -2, -1, -8, \ldots$ terminated by a $0$ curve, with the
next-to-last $-8$ curve on each end replaced by a $-7$.

\begin{figure}
\begin{center}
\includegraphics[width=10cm]{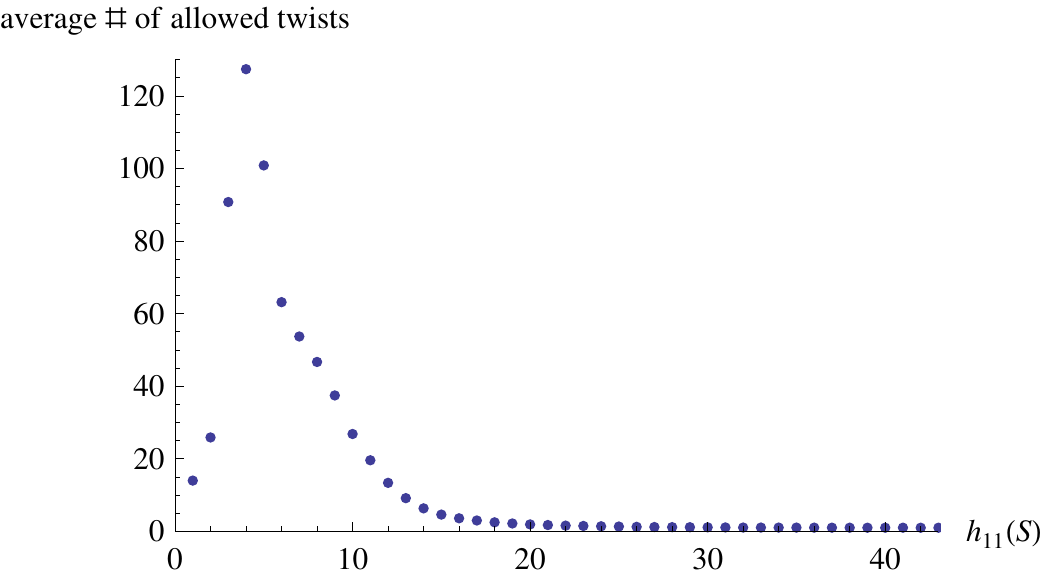}
\end{center}
\caption[x]{\footnotesize  
The average number of distinct $\P^1$
bundles $B$ over a toric base surface $S$ in the set found in \cite{mt-toric},
as a function
  of $h^{1, 1}(\b2)$.}
\label{f:twists-average}
\end{figure}

The average number of allowed twists over each base $S$ is graphed as
a function of $h^{1, 1}(S)$ in Figure~\ref{f:twists-average}.
Note that for $h^{1, 1}> 43$, the only $\P^1$ bundle
bases $B$ that support elliptic CY fourfolds are trivial bundles $\P^1
\times S$.
The largest number of distinct twists possible over a given surface
$S$ is 1119, which occurs for the  generalized del Pezzo surface
with a toric description in
terms of curves of self-intersection $((-1, -1, -2, -1, -2, -1,
-1))$.\footnote{We generally use
the notation $((n_1, \ldots, n_N))$ to denote the self-intersections
of the divisors $d_i$ in the toric
base surface $S$ for the $\P^1$ fibration of a base $B$.}
This surface can be constructed by blowing up the del Pezzo surface
$dP_3$ at the intersection point between a pair of $-1$ curves, and
the set of base threefolds constructed as $\P^1$ bundles over it was
also analyzed in \cite{Anderson-Taylor}.

\subsection{Divisors supporting non-Higgsable gauge groups} 

One particularly interesting aspect of the class of bases we are
considering here is that they give a broad set of distinct local
structures for divisors and normal bundles that can arise in 3D
F-theory bases.  Indeed, essentially every base in the class we have
constructed represents two distinct combinations of divisor and normal
bundle, corresponding to the two sections of the $\P^1$ bundle and
$\pm$ the twist, so that we have roughly 200,000 distinct local
divisors and normal bundles that can arise in 3D bases.  (A very small
number of bases formed from nontrivial $\P^1$ bundles have a symmetry
under exchanging the two sections so that the local structure on both
sides is the same, so the total number of local geometries is slightly
smaller than $2 \times 109,158 = 218,316$.)  Each
base $B$ also contains a set of divisors $D_i$ that each have the
topology of a Hirzebruch surface, associated with the restriction of
the $\P^1$ bundle to the toric curves $d_i$ in the surface $S$.  Because the
range of geometries is much larger for the divisors associated with
sections, we focus more on those here.

It is interesting to consider the subset of the local divisor + normal
bundle geometries that support non-Higgsable gauge groups.  In total
we find that only 368 of the 109,158 bases have twists that produce
non-Higgsable gauge groups on the divisor associated with either
section. The divisor $D$ in each case is topologically described by
the base $\b2$ of the
$\P^1$ fibration.  The divisors that support non-Higgsable groups all
have $h^{1, 1}(D) \leq 14$; the three divisors with the largest value
of 14 have toric self-intersections
\begin{eqnarray*}
 &  & 
((-4, -1, -4, -1, -2, -3, -2, -1, -4, -1, -4, -1, -2, -3, -2, -1))\\
 & &((-4, -1, -4, -1, -4, -1, -2, -3, -2, -1, -2, -3, -2, -1, -4, -1))
  \\
& & ((0, -3, -2, -1, -4, -1, -4, -1, -4, -1, -4, -1, -4, -1, -2, -3)) \,.
\end{eqnarray*}
Note that none of these divisors contains any curves of
self-intersection below $-4$.  There are some bases/divisors that
support NHC's and
that contain $- 5$ curves, beginning with
$\F_5$, but no divisors supporting a
non-Higgsable group arise in our data set that have curves of
self-intersection $-6$ or below.
This suggests that no divisor that
supports a non-Higgsable gauge group can have any curves of
self-intersection below $-5$, which would dramatically restrict the set
of possible divisors supporting non-Higgsable groups.

\begin{figure}
\begin{center}
\includegraphics[width=10cm]{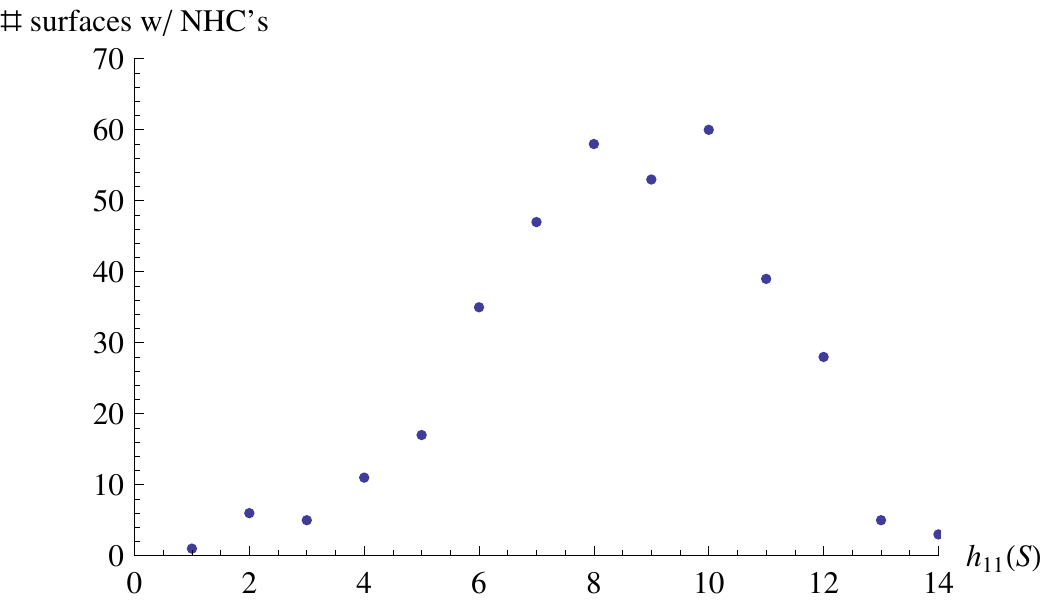}
\end{center}
\caption[x]{\footnotesize  Number of bases $\b2$ that support non-Higgsable
  groups on a section of the $\P^1$ bundle over $\b2$, as a function
  of $h^{1, 1}(\b2)$.}
\label{f:divisors-Hodge}
\end{figure}

The number of bases $B$ that support non-Higgsable groups on one or
both sections is graphed as a function of $h^{1, 1}(\b2)$ in
Figure~\ref{f:divisors-Hodge}.

There are a total of 8355 distinct bases $B$ that have non-Higgsable
clusters on one or both of the sections, of which 508 have
non-Higgsable clusters on both sections.  Thus, 
in the
set of bases we have constructed
there are close to
9000 distinct divisor + normal bundle combinations
that support a non-Higgsable gauge group.

\begin{figure}
\begin{center}
\includegraphics[width=10cm]{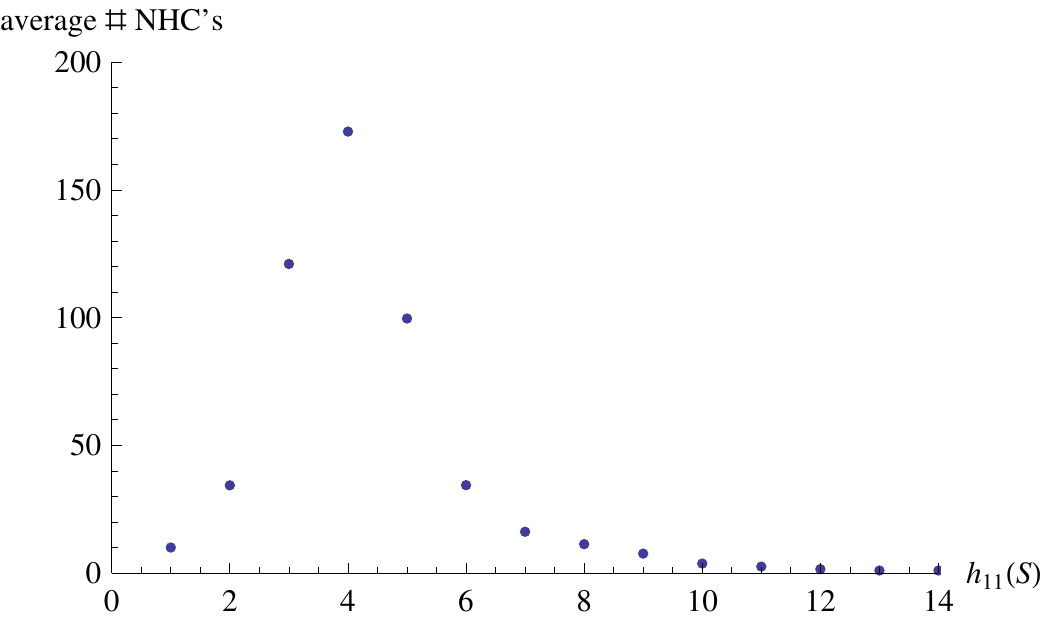}
\end{center}
\caption[x]{\footnotesize  Average number of different normal bundles
  compatible with a non-Higgsable group on a section divisor,
as a function
  of $h^{1, 1}(\b2)$}
\label{f:twists-Hodge}
\end{figure}

The average number of possible normal bundles per divisor
for sections that support a non-Higgsable gauge group factor is
graphed as a function of $h^{1, 1}(D)$ in Figure~\ref{f:twists-Hodge}

\subsection{The prevalence of non-Higgsable clusters}

One basic question regarding threefold bases for 4D F-theory models is
the extent to which non-Higgsable clusters are generic features in the
sense that they arise for a large fraction of threefold bases.  As
described in \S\ref{sec:6D}, virtually all base surfaces for 6D
F-theory compactifications give rise to non-Higgsable gauge groups,
with the only exceptions being the {\it weak-Fano} generalized del
Pezzo surfaces.  It is natural to ask whether a similar story holds in
four dimensions.  While the story for base threefolds is more
complicated, the bases we study here provide some evidence that the
general picture may be rather similar, at least for toric threefold
bases.

We discuss some simple aspects of this question here, leaving further
analysis for later work.  We begin by discussing 
weak Fano threefolds;
we then describe some explicit results
from the classification of $\P^1$-bundle bases that we have carried
out, and then examine some general features and analyses that may help
both to explain the explicit results found here and lead towards
further understanding of the generality of geometrically non-Higgsable
structure in 4D models.

\subsubsection{Weak Fano threefolds}

In any dimension, a weak Fano variety satisfies the condition that $-K
\cdot C \geq 0$ for any effective curve $C$ ({\it i.e.}, $-K$ is {\it
  nef}).\footnote{A weak Fano variety also has an anti-canonical class
  $-K$ that is {\it big}, but $-K$ is big for any toric variety
  \cite{aipsv}, so we focus attention in this paper on the nef
  condition.}
For surfaces, this condition implies that there are no irreducible
effective
curves of
self-intersection $-3$ or below.  Thus, for 6D F-theory
compactifications, weak Fano bases are precisely those that have no
non-Higgsable gauge group factors.

For toric threefolds $B$, the weak Fano condition has a similar but
slightly different connection to the singularity structure of an
elliptically fibered Calabi-Yau fourfold over $B$.  
In particular,
the condition that $-K\cdot C \geq 0$  for all effective $C$ is
equivalent to the condition that there are no codimension one, two, or
three loci in $B$ where $f$ or $g$ are forced to vanish.
Mathematically this is encoded in the statement that a divisor on a
smooth toric variety is {\it basepoint free} if and only if it is nef
(see \cite{La}, Proposition 1.5).
At a heuristic level this can be understood from the fact that when    
$-K \cdot C < 0$, it means that there is no representative of the
  curve class $C$ that is disjoint from or transversely intersecting
  $-K$, since such a representative would have a vanishing or positive
  intersection number.  Thus, $-K \cdot C < 0$ implies that $C$ is
    contained within the locus $-K$, and therefore it is also contained
    in $-4K$ and $-6K$, so that $f$ and $g$ would both
    need to vanish on $C$.  We see from this that any base that is not
    weak Fano must have at least one curve on which $f$ and $g$ are
    forced to vanish.  In the other direction, if
$-K \cdot C \geq 0$ for all effective curves, then we expect that all
    curves can be put in general position so that they are not
    contained within $-K$.  In the situation where the base is toric
    this expectation is realized and there is no locus on which $f, g$
    are forced to vanish when $B$ is weak Fano.  When $B$ is not
    toric, however, this condition is not always correct.  In any
    case, for toric bases such as those we consider here, the weak
    Fano condition is equivalent to the condition that there are no
effective    curves $C$ in $B$ where $f, g$ necessarily vanish.

Note, however, that unlike for 6D compactifications, for 4D
compactifications the fact that a base is not weak Fano does not
necessarily imply the existence of a divisor on which $f, g$ must
vanish to at least orders $1, 2$, which is the condition needed for
the presence of a geometrically non-Higgsable gauge group factor.  In
fact, there are threefold bases that are not weak Fano in which there
are divisors where $(f, g, \Delta)$ vanish only to orders $(1, 1,
2)$. These exhibit a Kodaira type II singularity with no gauge group,
which were studied via a local deformation in \cite{ghs2}; the fact
that such singularities have no gauge group can be understood from the
different vanishing cycles of the colliding branes. There are also
threefold bases that are not weak Fano where $f, g$ vanish only on
curves.  We explore next the cases where this occurs in the bases we
have constructed.

\subsubsection{Explicit computation of
base threefolds without non-Higgsable groups}

Of the 109,158 threefold bases that we have constructed, all but 1,824
have a geometrically non-Higgsable gauge factor on at least one
divisor.  As mentioned above, however, there is a richer structure for those
bases without non-Higgsable gauge factors than there is in
six-dimensional compactifications.  In particular, even in the
absence of non-Higgsable gauge groups, a base can have $(f, g,
\Delta)$ vanishing to order $(1, 1, 2)$ on one or more divisors, or to
higher orders on one or more curves.  

Of the 1824 threefold bases without non-Higgsable gauge groups, 1788
have no generic vanishing of $f, g$ on any divisor.  Of the other 36,
there are 35 that have a divisor where $(f, g)$ vanish to orders $(1,
1)$.  The simplest of these is a $\P^1$ bundle base $B$ constructed
from a toric generalized del Pezzo surface $S$ with curves of
self-intersection $((1, -2, -1, -2, -2, 0))$ and twist $T = d_3 + d_4 +
2d_5 + 3d_6$.  The resulting threefold has $f, g$ both vanishing to
order one on $\Sigma_-$, the divisor associated with the toric ray
$s_-= (0, 0, 1)$ in (\ref{eq:ws}). There is also one case in which $f,
g$ vanish to orders $(1, 0)$ on a toric divisor, given by the
threefold base associated with a $\P^1$ bundle over the toric surface
having self-intersections $((-2, -2, -1, -2, -2, -1, -2, -2, -1))$ and
twist $d_3 + d_4 + d_5 + 2d_6 + d_7$.

Of the 1788 bases that have no generic vanishing of $f, g$ on a
divisor, all but 80 have the property that $f, g$ vanish on some
curves in the base, often to relatively high orders.  A typical
example is given by the $\P^1$ bundle over $\F_1$, the toric surface
with self-intersections $((1, 0, -1, 0))$, with twist $T =2d_4$.  In
this case, there is no generic vanishing of $f, g$ on any toric
divisor but $f, g$ vanish to orders $(2, 3)$ on the curve defined by
$D_3 \cap \Sigma_-$.  An interesting open question is the physical
nature of F-theory compactifications on bases of this type.  While
there is no non-Higgsable gauge group at the level of geometry, the
higher-order codimension two singularity suggests the presence of some
kind of ``matter curve''.  It is possible that these specific
singularities may just be a generalization of the kind of cusp
singularity that appears at points in the discriminant locus for 6D
compactifications, without matter from $7-7$ strings, but it is also
possible that these singularities have further physical significance,
for example as uncharged matter.  This is an interesting question for
further investigation. 

\subsubsection{General features of bases without non-Higgsable
  structure}

One simple feature of the $\P^1$-bundle bases that lack non-Higgsable
gauge groups is that they are all constructed from surfaces $S$ that
have no curves of self-intersection $-4$ or below.  This is relatively
easy to understand.  Over a curve $d_i$ of self-intersection $d_i
\cdot d_i = n =-4$ or below, from (\ref{eq:t-2}) the twist must
satisfy $|T \cdot d_i | \leq 1$.  If $T \cdot d_i = 0$, then the local
geometry is simply that of $\P^1 \times S$ in the vicinity of the
curve, so there must be a non-Higgsable cluster from the standard 6D
analysis.  If $T \cdot d_i = 1$ (or equivalently $-1$), then the local
geometry of $D_i$ is that of $\F_1$ with a normal bundle of
$\cO_{D_i}(n F)$, where $X, F$ are the standard basis of effective
divisor curves on $\F_1$ with $ X \cdot X = -1, X \cdot F = 1, F \cdot
F = 0, -K_{D_i} = 2X + 3F$.  In this case from (\ref{eq:g}) we have
\begin{equation}
f_0 \in \Gamma ({\cal O}(8 X + 12F+4nF)) \,.
\label{eq:f0-example}
\end{equation}
The divisor $8 X + 12F + 4nF$ in $D_i$ is not effective unless $n \geq -3$.  
Similarly,
\begin{equation}
g_0 \in \Gamma ({\cal O}(12 X + 18F+6nF)) \,,
\end{equation}
\begin{equation}
g_1 \in \Gamma ({\cal O}(12 X + 18F+5nF)) \,.
\end{equation}
In fact these bundles do not have global sections if $n\leq -4$, in which
case $f_0$, $g_0$ and $g_1$ do not exist and there
must be a non-Higgsable gauge group on $D_i$.

We can also consider explicitly the consequences of imposing the weak
Fano condition discussed above on the threefold base $B$.  The weak
Fano condition states that $-K \cdot C \geq 0$ for any curve $C$ in
the base $B$.  For the toric $\P^1$-bundle bases that we are
considering, the effective holomorphic curves $C$ are either curves
that can be described as intersections of the form $\Sigma_\pm \cdot
D_i$, or of the form $D_i \cdot D_{i + 1}$ (using the cyclic
convention $N + 1 \rightarrow 1$).  There are linear relations between
these curves; in particular, each of the curves $D_i \cdot D_{i + 1}$
is simply the fiber $F$ of the $\P^1$ bundle, and the curves
$\Sigma_-\cdot D_i$ are essentially the curve classes $d_i \subset
\Sigma_-$, which have two relations in homology, while the curves
$\Sigma_+ \cdot D_j$ are linear combinations of the $\Sigma_-\cdot
D_i$ with the fiber $F$.  It is convenient to keep all of these
classes separate, however, as they act as an effective (though
redundant) set of generators for the Mori cone (see {\it e.g.}
\cite{Cox-vr}).  To check the weak Fano condition, therefore, it
suffixes to check that $-K \cdot C \geq 0$ for each of the curve
classes just mentioned.  Since all of these curves may be represented
in homology as the intersection of toric divisors, we can assess this
condition by computing the triple intersection numbers of toric
divisors.  The divisors associated with sections satisfy the relation
\begin{equation}
\Sigma_+ = \Sigma_-+ \sum_{i}t_iD_i \,.
\end{equation}
The 3D cones of the fan are all associated with nonzero triple
intersection numbers
\begin{equation}
\Sigma_\pm \cdot D_i \cdot D_{i + 1} = 1
\end{equation}
where $i + 1$ is again
taken cyclically on the base.  
From the structure of the base we inherit the intersections
\begin{equation}
\Sigma_\pm \cdot D_i \cdot D_i = n_i \,.
\end{equation}
We have $\Sigma_+ \cdot \Sigma_-= 0$, and
\begin{equation}
\Sigma_+ \cdot \Sigma_+ \cdot D_i = \Sigma_+ \cdot (\sum_{j}t_jD_j) 
\cdot D_i = T \cdot d_i \,.
\end{equation}
with intersections in $\b2$ implied in the last expression and henceforth.
Similarly, we have $\Sigma_-\cdot \Sigma_-\cdot D_i = -T \cdot d_i$.

With this information we can now check the weak Fano condition.
Using  the fact that
the anti-canonical class of $B$ is $-K = \Sigma_+ + \Sigma_-+
\sum_{j}D_j$ we compute
\begin{equation}
-K \cdot \Sigma_\pm \cdot D_i
= 2+n_i  \pm T \cdot d_i \,.
\label{eq:fanocalc}
\end{equation}
If this expression is negative the weak Fano condition is violated,
which (using equation (\ref{eq:t-0})-(\ref{eq:t-big})) occurs
whenever we have
a -3 curve in the base, or a -2 curve $d$ with a nonzero twist $T
\cdot d \neq 0$, or a $-1$ curve $d$ with a twist $| T \cdot d | \geq
2$, or more generally a curve in the base of self-intersection $m$
with a twist $| T \cdot d | \geq 3 + m$. We also compute
\begin{equation}
-K \cdot D_i \cdot D_{i+1} = 2
\end{equation}
and thus any violation of the weak Fano condition must come from
curves of the form $\Sigma_{\pm}\cdot D_i$.
We therefore conclude that there is a curve $C$
with $-K\cdot C <0$ if and only there is a curve $d_i$ with
\eqref{eq:fanocalc} negative, and we use this to check whether or
not $B$ is almost Fano.

Analyzing the allowed threefold bases $B$ in terms of the base
surfaces $\b2$ and twists $T$ (according to the condition just
derived) shows that there are precisely 80 cases that are weak Fano.
These are the cases where there is no vanishing of $(f, g)$ on any
curve in the threefold base $B$.  In the cases that are not weak Fano,
we can carry out an explicit local analysis that demonstrates the
appearance of curves where $(f, g)$ vanish.  For example, consider the
case mentioned above of a $\P^1$ bundle on $\F_1$ with twist $2d_4$.  In this
case, the section $\Sigma_-$ is a complex surface $\F_1$ embedded into
$B$ with normal bundle $N = -2F$.  As in (\ref{eq:f0-example}),
\begin{equation}
f_k \in \Gamma ({\cal O}_{\F_1}(8 X + 12F-2 (4-k)F)) \,.
\end{equation}
For $k = 0$, the divisor associated with the line bundle is $c =8 X +
4F$, which satisfies $c \cdot X = -4$, so $f_0$ vanishes on $X$.
Similarly for $f_1, g_0, g_1, g_2$, so we see that in general this
type of weak Fano base with a divisor $\F_1$ having normal bundle $N
= \cO_{\F_1}(-2F)$ must have a curve with orders of vanishing of $(f, g)$ of at
least $(2, 3)$.  Similar arguments can be made in other cases where
the base is not weak Fano.

Thus, we see that there are essentially 3 distinct classes of bases.
For the weak Fano bases, there are no curves in the
base on which $f, g$ must vanish.  For a somewhat larger class of
bases, which are not weak Fano, there are no
non-Higgsable gauge groups, but $f, g$ vanish on some curves or
possibly divisors in the base.  And for the majority of threefold
bases, there are divisors that carry non-Higgsable gauge groups.

While there is no complete classification of weak
Fano threefolds\footnote{Note, however, that
  there are only about $100$ Fano threefolds, which instead satisfy
  the strict inequality $-K\cdot C>0$; we consider these further in
the following section.}, we expect that this set of bases forms a very small
fraction of the full set of allowed threefold bases for F-theory
compactifications to 4D.  We also expect that the class of bases
without non-Higgsable gauge group factors may be relatively small.  In
the rest of this paper we give some further evidence for these
conclusions. 

\section{Hodge numbers  of elliptic CY fourfolds over $\P^1$-bundle bases}  
\label{sec:Hodge}

For 6D F-theory models, the Hodge numbers of the generic elliptic
Calabi-Yau threefold over each base provide a convenient
parameterization of the set of models that gives a simple birds-eye
view of the space of allowed theories.  Figure~\ref{f:6D-Hodge}, for
example, shows the Hodge numbers for the generic elliptic CY
threefolds over the full set of toric bases $\b2$.  From the analysis
of \cite{mt-toric, Hodge, Martini-WT, Johnson-WT, Wang-WT}, we know
that the outline of the set of all possible models is clearly captured
in this diagram, with the set of bases that do not give rise to
non-Higgsable clusters a small subset confined to the far left of the
diagram.

While for 6D models, at least for large $h^{2, 1}(X_3)$, rigorous bounds
are known for the set of possibilities, the state of knowledge for
Calabi-Yau fourfolds is much weaker.  Nonetheless, from what is known
of the set of toric constructions for CY fourfolds, it seems at least
based on our current limited understanding that the
story for elliptic CY fourfolds may be not be too wildly different from that for CY
threefolds.  To get some initial sense of how the models we consider
here fit into the broader landscape of possibilities for elliptic
Calabi-Yau fourfolds, we describe here some rough aspects of the set
of Hodge numbers for elliptically fibered Calabi-Yau fourfolds
produced from generic Weierstrass models over the base threefolds we
have constructed.

Systematic constructions of a large set of Calabi-Yau fourfolds (which
may or may not be elliptically fibered) were carried out using toric
and related methods in \cite{Klemm-lry, Kreuzer-Skarke-4D, lsw-4D,
  krs-4D}.  The distribution of Hodge numbers
$h^{1, 1},h^{3, 1}$ of these fourfolds has a striking similarity to
the ``shield'' pattern of Hodge numbers for CY threefolds identified
in the analysis of Kreuzer and Skarke \cite{Kreuzer-Skarke}; as we now
understand it, this shield pattern for toric hypersurface
CY threefolds defines the outer boundary of the
set of Hodge numbers for {\it all} elliptic CY threefolds, whether
toric or not.  In the absence of examples to the contrary, it is
natural to expect that the Hodge numbers of all elliptic CY fourfolds
may similarly be contained within the shield-shaped pattern of known
fourfold constructions.
In this section we explore how the bases we have
constructed fit into the larger class of known Calabi-Yau fourfolds
using estimates of the Hodge numbers based on the base of the
fibration.

\subsection{Computing Hodge numbers of fourfolds from base geometry}

For a Calabi-Yau threefold that is a generic elliptic fibration over a
complex surface base $S$, the Hodge numbers of the CY threefold can be
determined from the geometry of the base, as described in \cite{Hodge,
  Martini-WT}.  In principle,
a similar approach can be taken to computing the Hodge
numbers of a generic elliptic Calabi-Yau fourfold $X$ over a toric
threefold base, though some aspects of such a computation are not yet
fully understood.  We focus here on the Hodge numbers $h^{1, 1}(X),h^{3,
  1} (X)$.  
A more complete description of the computation of Hodge numbers from
the base geometry will be given elsewhere \cite{Hodge-btw}; here we
simply summarize some aspects of that story relevant for the
computations here.

We begin with $h^{1, 1}(X)$, for which the analysis is the simplest.
From the Shioda-Tate-Wazir relation \cite{stw,
  Morrison-Vafa-I},
\begin{equation}
h^{1, 1}(X) =h^{1, 1} (B) + {\rm rank} (G) + 1 \,,
\label{eq:h11}
\end{equation}
where $G$ is the (non-Higgsable) gauge group associated with the
F-theory compactification on the generic elliptic fibration.  This
relation asserts that the set of divisors on $X$ is spanned by
divisors on $B$ lifted to divisors on $X$, ``vertical'' (sometimes
called Cartan) divisors
arising from the resolution of codimension one singularities
associated with nonabelian components of $G$, and ``horizontal''
linearly independent sections.  The contributions to $h^{1, 1}(X)$
from $h^{1, 1}(B)$ and the nonabelian part of $G$ are easy to compute
from the geometry of $B$, since $G$ for the generic elliptic fibration
is simply the non-Higgsable nonabelian gauge group in $G$.  For most
bases we expect only a single section for the generic elliptic
fibration, so there is no contribution to rank $G$ from abelian
factors.  In general, the rank of the abelian gauge group is
determined by the Mordell-Weil group of the fibration, with the number
of $U(1)$ factors given by $r = k-1$, where $k$ is the number of
linearly independent global sections of the fibration.  For toric base
surfaces for 6D compactifications, $h^{1, 1}(X_3)$ can be computed
independently using anomaly cancellation conditions, and in all cases
matches (\ref{eq:h11}) with no contribution from the Mordell-Weil
group, so that there are no non-Higgsable $U(1)$ factors over any
toric base surface for elliptic Calabi-Yau threefolds.  The situation
for fourfolds is less clear.  We do not have an independent approach
analogous to the anomaly cancellation mechanism of 6D theories to
verify the absence of non-Higgsable abelian factors, though in all
cases where explicit checks have been made there are no such
contributions.  In 6D, a very small class of non-toric bases has been
identified that support non-Higgsable $U(1)$ factors \cite{Martini-WT,
  Morrison-Park-Taylor}.  While similar non-toric bases with
non-Higgsable abelian factors are to be expected for 4D models, we
have no reason to expect that this can happen over toric bases.  In
our computation of $h^{1, 1}(X)$ for the elliptic fourfolds over base
threefolds $B$ we assume the absence of a nontrivial Mordell-Weil
group, and simply compute using the non-Higgsable nonabelian gauge
group.  We leave open the possibility that the resulting Hodge numbers
may be slightly off, if non-Higgsable $U(1)$ factors arise in certain
cases.

We now turn to $h^{3, 1} (X)$.  This corresponds to the number of
complex structure moduli for the elliptic Calabi-Yau fourfold $X$.  A
simple estimate for $h^{3, 1}$ can be determined by counting the
allowed monomials in the Weierstrass model.  The number of such
monomials $W_f$ in $f$ is simply the number of vectors $q \in N^*$
such that $\langle q, w_i \rangle \geq -4$ for all rays $w_i \in N$ in
the fan defining $B$ as a toric variety, and similarly for $W_g$ with
$\langle q, w_i \rangle \geq -6$.  As in the 6D case \cite{mt-toric},
there are several additional terms that must be included.  First,
there is an overparameterization of the Weierstrass model by the
number $w_{\rm aut}= 3 + w_{\rm polar}$ of automorphisms, where 3 is
the dimension of the base variety and the number of universal toric
automorphisms, and $w_{\rm polar}$ is the set of additional
automorphisms, which are in one-to-one correspondence with lattice
points in the interior of a codimension one face of the polar polytope
defined through $\langle q, w_i \rangle \geq -1$
\cite{cox-homogeneous}.  Second, there are additional degrees of
freedom that must be added, associated with each combination of a ray
$w_i$ that lies on the interior of a one-dimensional face of the
convex polytope constructed from all the $w_j$'s, and a monomial $q$
that lies on the interior of the dual one-dimensional face of the dual
polar polytope in $N^*$.  As described in more detail in
\cite{Hodge-btw}, this term follows from the combinatorial analysis of
Batyrev \cite{Batyrev} in cases where there is a simple reflexive
polytope that can be built from a $\P (2, 3, 1)$ fibration over the
toric surface $S$, and we assume that it holds more generally in other
circumstances as well.  We denote the number of these additional
degrees of freedom by $W_{11}$.  Including these additional terms, and
subtracting one for the universal natural scalar hypermultiplet, we
have
\begin{equation}
h^{3, 1}(X) = W_f + W_g-w_{\rm aut} + W_{11}  -1\,.
\label{eq:h31}
\end{equation}
As an example of a situation where the extra term $W_{11}$ becomes
relevant, consider the $\P^1$ bundle over $\F_0$ with twist $2 X$.  In
this case we have $w_-= (0, 0, 1) =(w_1 + w_3)/2$, where $w_1 = (0, 1,
0)$ and $w_3 = (0, -1, 2)$, so $w_-$ is in the interior of a 1D face
of the convex polytope. It is easy to check that there is one monomial
in the interior of the dual 1D face of the polar polytope, so in this
case we have $W_{11} = 1$.

There are a number of subtleties in the computation of the Hodge
numbers of $X$ that force us to treat the formulae (\ref{eq:h11}) and
(\ref{eq:h31}) as only approximate ``Hodge numbers''.  One issue is
that unlike for Calabi-Yau threefolds, where it is known that there is
a geometric resolution of all the relevant singularities \cite{Roan},
for fourfolds there are situations where singularities can arise that
cannot be resolved.  In such cases there is no obvious geometric
meaning to the Hodge numbers $h^{1, 1},h^{3, 1}$.  In
\cite{Klemm-lry}, the approach taken was to use Vafa's formulae
\cite{Vafa-Hodge} in terms of the chiral ring of Landau-Ginsburg
models, with the idea that the Hodge numbers should make sense in this
context even in the absence of a geometric resolution.  In various
simple cases the computation we get here using base geometry matches
with that analysis.  As mentioned above, we are assuming that there
are no generically nontrivial Mordell-Weil groups over any of the
bases we consider.  There are also issues in applying (\ref{eq:h31})
when there is not a direct simple construction of the fourfold using a
reflexive polytope, as discussed further in \cite{Hodge-btw}, though
we expect the formula to still be valid in those cases.  In any case,
for the purposes of this paper we can simply treat (\ref{eq:h11}) and
(\ref{eq:h31}) as ``approximate'' computations of the Hodge numbers of
the generic elliptic Calabi-Yau fourfold (or its analogue when there
is no geometric resolution), as a rough way of characterizing the
distribution of bases in the context of the larger set of possible
Calabi-Yau fourfold Hodge numbers.
In the remainder of this paper we simply use these estimates as
``Hodge numbers'' with no further apology, keeping in mind the various
issues just raised.

\subsection{(Approximate) Hodge numbers of
elliptic CY's over $\P^1$-bundle bases}

\begin{figure}[h]
  \centering
  \includegraphics[scale=1]{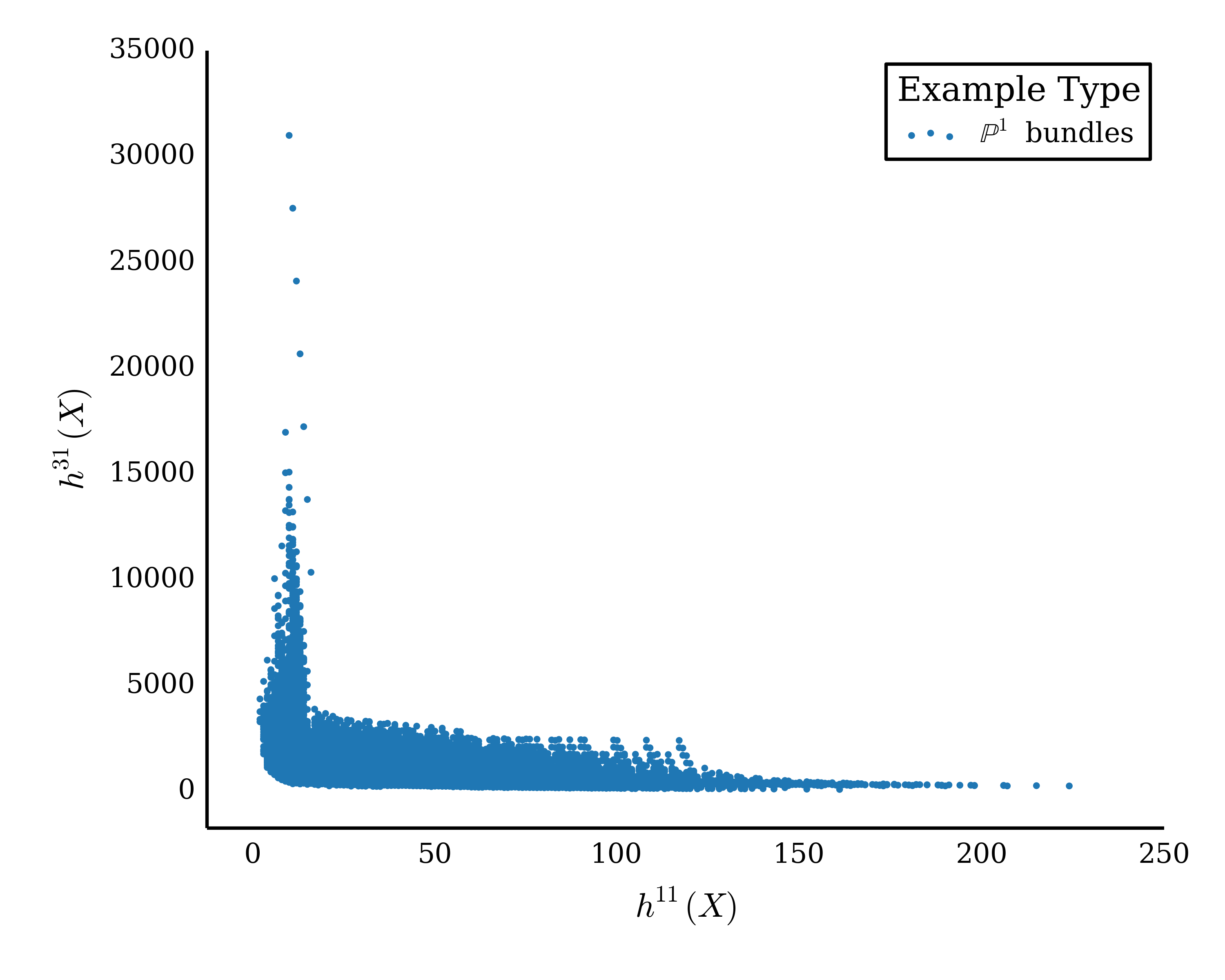}
  \caption{Plotted are the Hodge numbers for each example in our data
    set, as estimated from the geometry of each base.}
\label{f:Hodge-4D}
\end{figure}

\begin{figure}[h]
  \centering
  \includegraphics[scale=1]{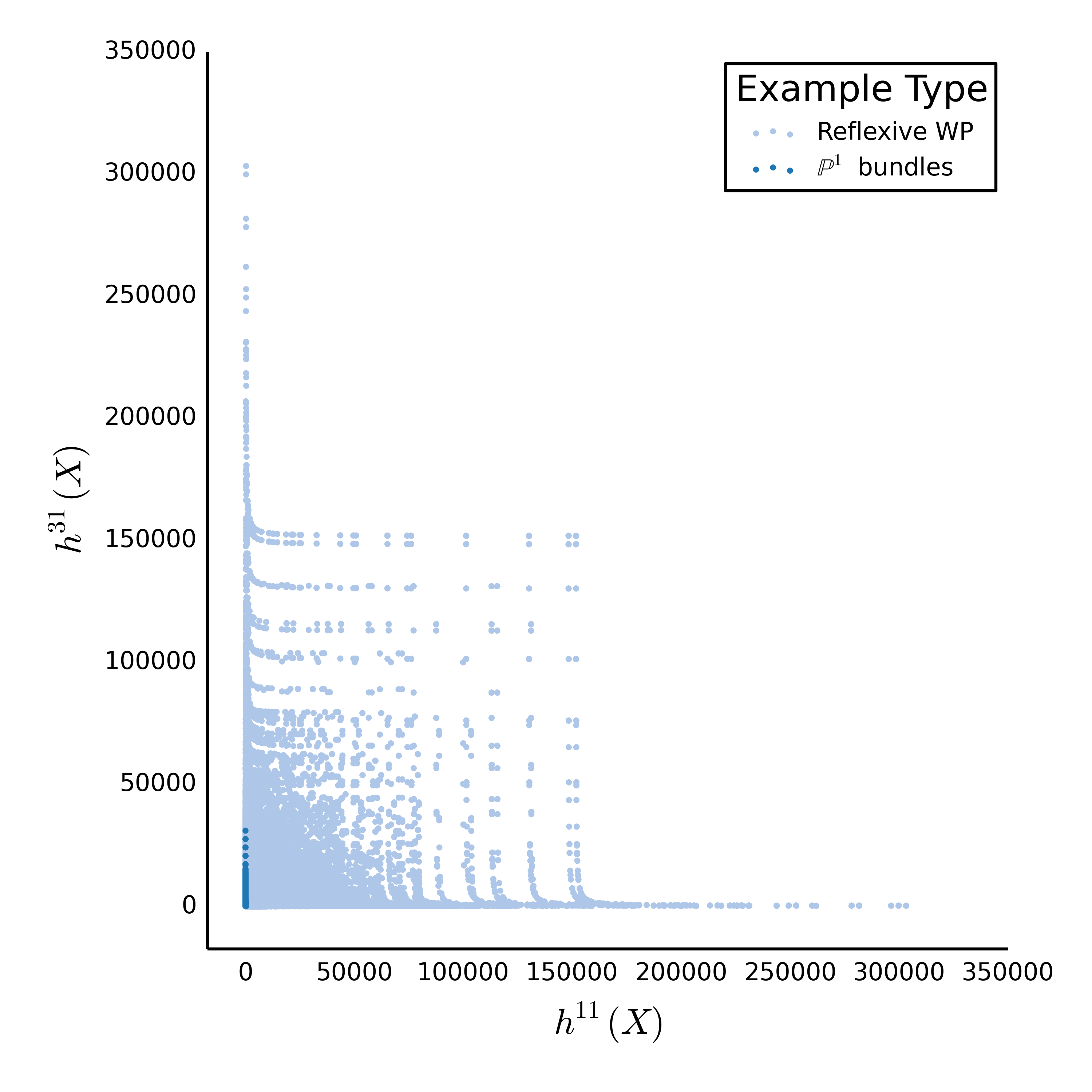}
  \caption{Hodge numbers of the examples in our data set plotted in
    the context of the set of Hodge numbers from Calabi-Yau fourfolds
    described through reflexive polytopes
as hypersurfaces
in 5D weighted projective space \cite{Kreuzer-Skarke-4D}.}
\label{f:Hodge-4D-ks}
\end{figure}

Using the approach described in the preceding section, we have
computed the approximate ``Hodge numbers'' of the generic elliptic
Calabi-Yau fourfold over each base in our list in terms of the
geometric data of the toric base $B$.  The distribution of Hodge
numbers for our bases are shown in Figure~\ref{f:Hodge-4D}, and
compared in Figure~\ref{f:Hodge-4D-ks} to the Hodge numbers computed
in \cite{Kreuzer-Skarke-4D} for the full set of Calabi-Yau fourfolds
realized as hypersurfaces in weighted projective space using the
Batyrev reflexive polytope approach.  Note that though the Calabi-Yau
fourfolds from \cite{Kreuzer-Skarke-4D} are not necessarily
elliptically fibered, recent evidence suggests that a large fraction
of Calabi-Yau threefolds are in fact elliptically fibered \cite{Hodge,
  Johnson-WT}, particularly at large Hodge numbers, and along similar
lines a recent analysis of complete intersection Calabi-Yau fourfolds
\cite{Gray-hl-2} shows that over 99.5\% are actually elliptically
fibered.

In general, the bases we have constructed give
fourfolds with relatively small Hodge numbers, in particular with
small $h^{1, 1}(X)$.  As we describe in the remainder of this section,
we take this to mean that our data set gives a higher than typical fraction of 4D
F-theory models {\it without} non-Higgsable gauge groups, so that from
this set we are actually getting a smaller fraction of models with
non-Higgsable gauge factors, and in the full set of threefold bases we
might expect non-Higgsable clusters over much more than 98\% of the
possible bases.

\subsection{Hodge numbers and the hierarchy of non-Higgsable groups}

There is a natural hierarchy in the complexity of bases, associated
with the maximal Kodaira singularity type that is forced to arise
for a generic Weierstrass model over any given
base $B$.  Lowest in this hierarchy are the Fano and weak Fano bases,
which support no generic Kodaira singularities of any codimension, for
threefold bases as well as for surfaces.  These  bases generally give
rise to elliptic Calabi-Yau manifolds that have
relatively small Hodge numbers; in particular the Picard number $h^{1,
  1}(X)$ is generally quite low for Fano and weak Fano bases.  As the
Kodaira singularity types increase, the Hodge numbers, particularly
$h^{1, 1}(X)$ increase.  We illustrate this hierarchy first for the
well-understood case of elliptic Calabi-Yau threefolds, and then
compare with the results for our threefold base constructions for
elliptic Calabi-Yau fourfolds.

\begin{figure}
\centering
\includegraphics[scale=1]{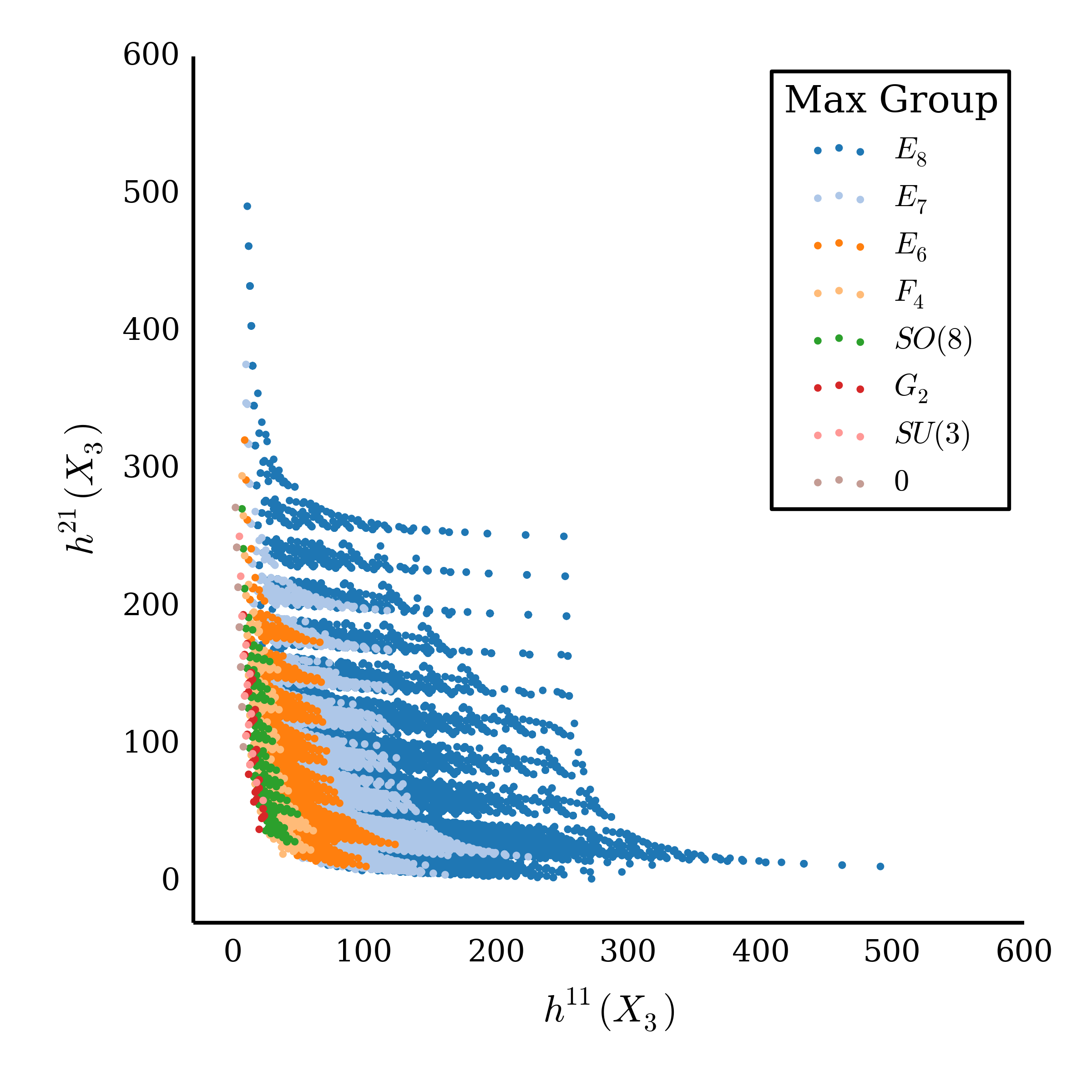}
\caption[x]{\footnotesize  Distribution of Hodge numbers for elliptic
  Calabi-Yau threefolds over bases that have a given maximum Kodaira
  singularity type.  When different bases give the same Hodge numbers,
the color shown is the minimum across bases of the maximum for each
base; {\it i.e.}, all numbers associated with weak Fano bases that
produce no singularities are shown, all numbers associated with bases
that produce at most type IV singularities but are not weak Fano are
shown, etc.}
\label{f:hierarchy-6D}
\end{figure}

A plot of the set of Hodge numbers that can arise for generic
elliptically fibered CY threefolds
over a given toric base with each possible maximal Kodaira
singularity type is shown in Figure~\ref{f:hierarchy-6D}.  We focus
primarily on toric bases for simplicity, though consideration of
non-toric bases as in \cite{Martini-WT, Johnson-WT, Wang-WT} shows
that essentially the same pattern holds for non-toric bases.  Clearly,
the Hodge numbers of threefolds over bases that support no singularity
({\it i.e.}, Fano and weak Fano bases) or small Kodaira singularity
types ({\it e.g.} type III, IV, $I_0^*$) are relatively small.  The
absolute numbers of threefolds that have various automatic Kodaira
singularity types --- {\it i.e.}, non-Higgsable clusters --- are
increasingly large as the singularity type increases.  As mentioned
earlier, there are only 10 Hodge number pairs associated with generic
elliptic fibrations over bases that do not support Kodaira
singularities.  Of the 7524 distinct Hodge number pairs that are
realized by generic elliptic fibrations over toric bases, 7 are
associated with such weak Fano bases.  Only another 50 Hodge number
pairs are associated with bases giving only type III singularities
({\it i.e.} with only non-Higgsable $SU(3)$ factors).  And indeed,
only 426 of the Hodge number pairs (less than 10\%) are associated
with bases that generate Kodaira singularities that are not worse than
$I_0^*$. 5296 of the Hodge number pairs are associated with $E_8$
non-Higgsable gauge groups.\footnote{Note that this set of toric data
  includes bases that have $-9, -10,$ and $-11$ curves, which lead to
  bases that are not strictly toric.  In our fourfold analysis in this
paper we have not included such bases so we miss a large number of
theories with geometrically non-Higgsable $E_8$ gauge groups, as
discussed further below.}
The upshot of this analysis is that there is a clear picture of the
distribution of bases for 6D F-theory compactifications.  There is a
very small fraction of bases that do not give rise to non-Higgsable
gauge groups in the low-energy theory.  The majority of bases give
rise to non-Higgsable gauge groups, with the rank of the groups
generally increasing with the Hodge numbers of the elliptic threefold
(and correspondingly with the numbers of fields in
the low-energy supergravity theories).  

\begin{figure}
  \centering
  \includegraphics[scale=1]{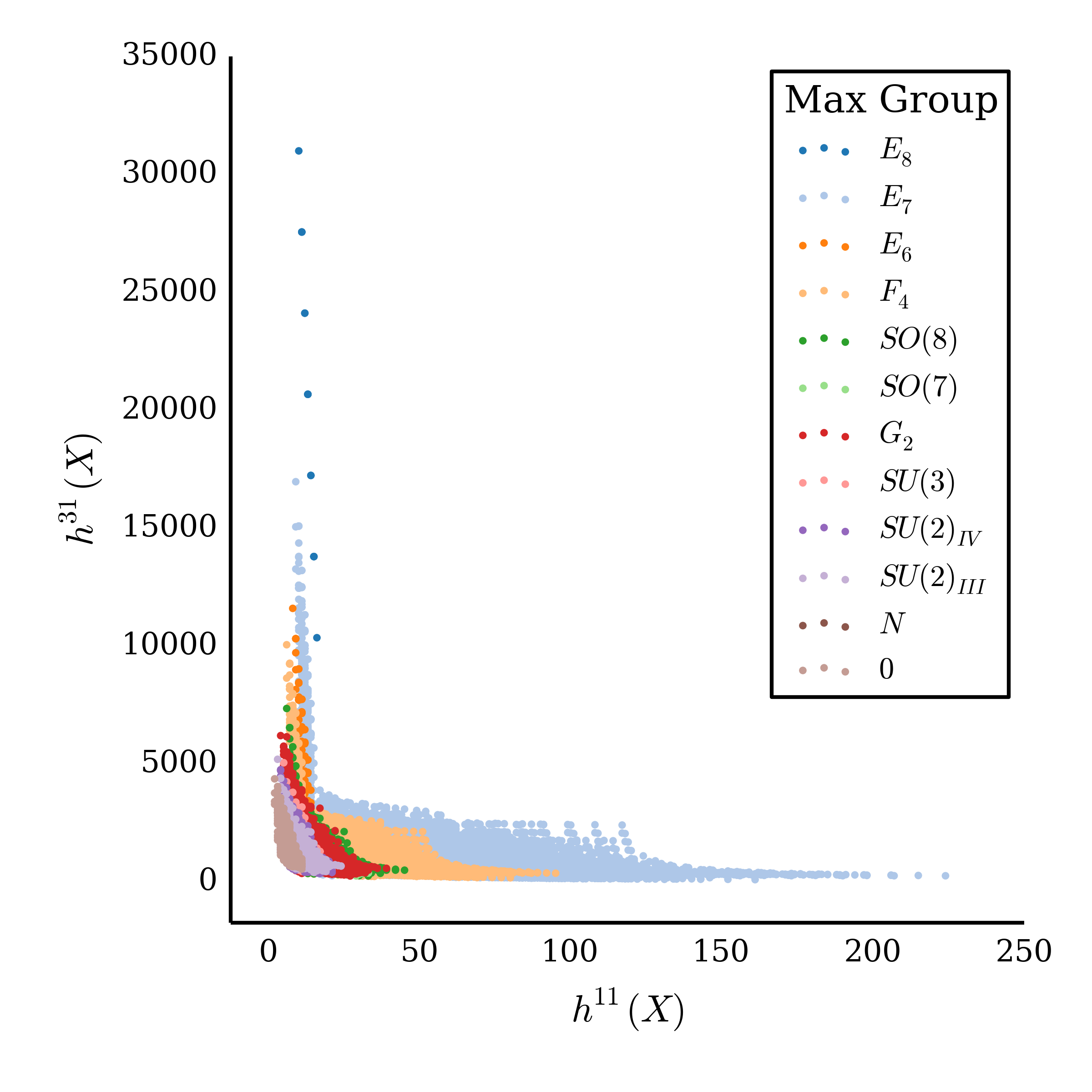}
\caption[x]{\footnotesize  Distribution of Hodge numbers for elliptic
  Calabi-Yau fourfolds over bases that have a given maximum Kodaira
  singularity type.  When different bases give the same Hodge numbers,
the color shown is the minimum across bases of the maximum for each
base, as in the corresponding 6D plot.}
\label{f:hierarchy-4D}
\end{figure}

We now consider the distribution of Kodaira singularity types for the
$\P^1$-bundle threefold bases that we have constructed.  A graph of
the Hodge numbers, again coded by maximum Kodaira singularity type, is shown
in Figure~\ref{f:hierarchy-4D}.  Again, we see the pattern that the
bases that generate smaller Kodaira singularity types are less
numerous and produce fourfolds with smaller Hodge numbers.  As
mentioned above, we have only included strictly toric $\P^1$-bundle
bases that support $E_8$ non-Higgsable gauge factors when the $E_8$
factor arises on a divisor with no codimension two $(4, 6)$
singularities, unlike in the 6D plot, Figure~\ref{f:hierarchy-6D}.
Thus, we see very few examples with $E_8$ non-Higgsable groups in the
4D plot.  This
is presumably an artifact of the class of bases we are considering; we
expect that as for 6D, bases that produce at least one non-Higgsable
$E_8$ will dominate the distribution for larger Hodge numbers.  In
particular, we note that when we include $\P^1$-bundle constructions
over toric bases where there is an $E_8$ on a divisor with a $(4, 6)$
codimension two singularity on a curve within the $E_8$ divisor,
analogous to a $-9, -10,$ or $-11$ curve for a twofold base, we
find an additional 68,528 threefolds that support a non-Higgsable
$E_8$.  Unlike in the 6D case, however, computing the Hodge numbers of
the threefold that results after blowing up the $(4, 6)$ curves is not
straightforward, so we have not included these bases in our analysis.  This
does suggest, however, that the domination by $E_8$ non-Higgsable
groups at large Hodge numbers for 4D theories will parallel the
structure for 6D theories.

\begin{figure}[h]
  \centering
  \includegraphics[scale=1]{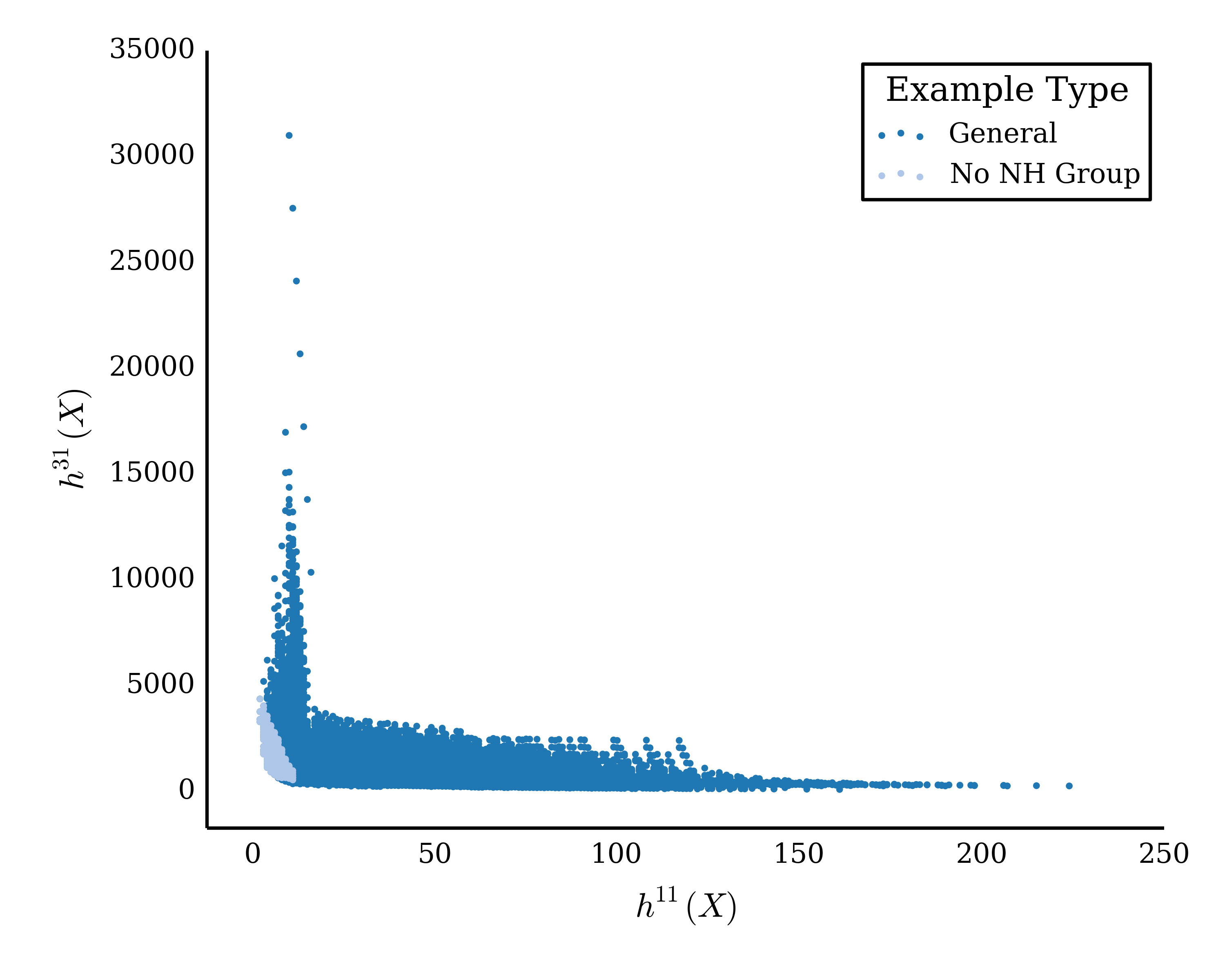}
  \caption{Plotted are the Hodge numbers for each example in our data
    set, where examples without a non-Higgsable gauge group are
    denoted in light blue.}
\label{fig:hodge gen and smooth}
\end{figure}

Considering some specific types of bases, we begin with Fano threefold
bases.  It is known that there are 105 Fano varieties of dimension
three \cite{3D-Fano}.  Of these 18 are toric, and were considered as
F-theory bases in \cite{Klemm-lry}.  Of these 18, 10 are included in
our list of $\P^1$-bundle bases, and are included in the 80 weak-Fano
bases that produce no generic singularities.  As mentioned above,
these are a subset of the 1824 bases that produce no non-Higgsable
gauge groups (Figure~\ref{fig:hodge gen and smooth}).  The largest
value of $h^{1, 1}(X)$ for a Fano base in our set is 6 (for $\P^1
\times dP_3$), and the largest value for a weak Fano base is $h^{1,
  1}(X) = 9$ (for $\P^1 \times gdP_6$, where $gdP_6$ is the
generalized del Pezzo surface of degree 6 given by the toric surface
$S$ with self-intersections $((-2, -2, -1, -2, -2, -1, -2, -2, -1))$).
For bases that do not support any non-Higgsable gauge group, the
maximum of $h^{1, 1}(X)$ is 12, for bases that support at most an
$SU(2)$ the maximum is 25, etc.  The maximum values of $h^{1, 1}(X)$
where $X$ has each possible maximum Kodaira singularity type are
listed in Table~\ref{t:maximum-Kodaira}.  The picture for threefold
bases for 4D compactifications thus closely matches that of surface
bases for 6D compactifications.  In particular, the set of bases that
do not give rise to non-Higgsable gauge factors is relatively small
and confined to the region of small $h^{1, 1}(X)$.  We expect that
when a much broader class of elliptic Calabi-Yau fourfolds is
considered, this pattern will persist, and the large number of
fourfolds at larger $h^{1, 1}(X)$ that are not given through a
$\P^1$-bundle construction will be dominated by those with
non-Higgsable gauge groups.

\begin{table}
\begin{center}
\begin{tabular}{|c |c |c |c |c |c | c | c | c | c | c |}
\hline
 $G$ & none & $SU(2)$ & $SU(3)$ & $G_2$ & $SO(7)$ &
$SO(8)$ & $F_4$ & $E_6$ & $E_7$ & $E_8$\\
sing.\ type & $\leq II$ & III & IV & $I_0^*$& $I_0^*$& $I_0^*$&
$IV^*$ & $IV^*$ & $III^*$ & $II^*$\\
max $h^{1, 1}(X)$ & 12 & 25 & 25 & 40 & 40 & 46 & 96 & 96 & 225 
&  $492^*$\\
\hline
\end{tabular}
\end{center}
\caption[x]{\footnotesize Table of maximum value of Hodge number
  $h^{1, 1}(X)$ for any base $B$ in our dataset
  that supports Kodaira singularities that are no worse than a given
  type, with associated non-Higgsable
gauge group factor $G$.
The starred value $492$ for $E_8$ refers to the maximum of $h^{1,
  1}(X)$ for a slightly broader class of $\P^1$-bundle base
constructions where we allow $(4, 6)$ curves on a divisor supporting
an $E_8$; the base with this value is $\P^1 \times S_{491, 11}$, where
$S$ is the base surface that supports the elliptically fibered
Calabi-Yau threefold $X_3$ with largest known $h^{1, 1}(X_3)$
(see  {\it e.g.} \cite{mt-toric, Hodge}).
}
\label{t:maximum-Kodaira}
\end{table}

\subsection{Minimal models}

Some further discussion, and comparison to the better-understood
scenario in 6D F-theory models, may be helpful in clarifying the role
of the bases we construct in this paper in the context of the larger
set of all 4D F-theory compactifications.

We begin with a brief summary of the situation in 6D, expanding on the
review of
\S\ref{sec:6D}.  For 6D F-theory models, all possible base surfaces
(including non-toric bases, excluding only the trivial case of the
Enriques surface) are constructed as blow-ups of the minimal
bases $\P^2$ and $\F_m, m = 0, \ldots, 8, 12$.  These blow-ups are
performed by blowing up points in the minimal bases, giving additional
divisors and increasing $h^{1, 1}(B_2)$ and $h^{1, 1}(X_3)$.  Any such
blow-up leaves fixed or decreases the value of the self-intersection
of any given curve, and thus cannot decrease the set of non-Higgsable
gauge groups.  Thus, the set of all F-theory models associated with
generic elliptic fibrations over any base is constructed by blowing up
points on the minimal bases, where each blow-up decreases $h^{2, 1}
(X_3)$, increases $h^{1, 1} (X_3)$, and either leaves invariant or
increases the non-Higgsable gauge content of the theory.  This matches
with the structure of Figure~\ref{f:hierarchy-6D}, where the del Pezzo
and generalized del Pezzo surfaces are realized by blow-ups of $P^2$,
which has $(h^{1, 1},h^{2, 1}) = (2, 272)$ for a generic elliptic
fibration, and further blow-ups on $-2$ curves or below produce bases
with larger $h^{1, 1}(X_3)$ and non-Higgsable gauge groups.  As
another illustration, the base $\F_8$ has Hodge numbers (10, 376) and
a $-8$ curve supporting an $E_7$ non-Higgsable factor.  Blow-ups of
this base on points that do not lie on the $-8$ curve produce a new
set of bases that support a maximal $E_7$ gauge group ---
seen in the family of (light blue) points that sit down and to the
right from the point $(10, 376)$ in Figure~\ref{f:hierarchy-6D}.  This pattern
of $E_7$ structures was also noted in \cite{Candelas-cs}.  In
particular, this picture makes it clear that the only bases that
support elliptically fibered Calabi-Yau threefolds without
non-Higgsable gauge groups are those that come from a limited set of
blow-ups on the only minimal bases that do not support non-Higgsable
gauge groups ({\it i.e.}, those without curves of self-intersection
$-3$ or below, which are $\P^2, \F_1, $ and $\F_2$.

We expect that
a similar story will hold in 4D, though there are many
additional complications.  The analogue of the minimal model story for
surfaces is the Mori program for threefolds \cite{Mori}, which is
rather more complicated.  We leave a more thorough treatment of this
story for future work, and make only a few general heuristic comments
on the structure of what is expected as motivation to set the context
for the bases we have constructed here.  Roughly speaking, the minimal
threefolds for the Mori program should be distributed in three general
classes: first, Fano bases; second, bases that are $\P^1$ bundles over
a base surface $S$; and third, bases that are surface $S$ bundles over
$\P^1$.  The Fano bases are a relatively small set including spaces
like $\P^3$.  The second is essentially the class we are studying
here, with various
further restrictions such as that $S$ and $B$ are toric.
And the third set is another interesting class of constructions that
will be described further elsewhere. Note that the third class
includes a large number of constructions with much larger $h^{3, 1}
(X)$ than those considered here.  For example, the Calabi-Yau fourfold
with the largest known $h^{3, 1} = 303,148$, is an $S$ bundle over
$\P^1$ with $S= S_{251, 251}$ with $S_{251, 251}$ the base surface for
the elliptic CY threefold with Hodge numbers $(251, 251)$.  The third
class produces divisors with less interesting non-Higgsable structure,
however, since when we restrict to the toric context all divisors are
either Hirzebruch surfaces $\F_m$ or the fiber surface $S$ with a
trivial normal bundle.  For a full analysis of the Mori theory story,
one should include the possibility of singularities in the bases.
Such singularities can be associated with superconformal field
theories coupled to the supergravity theory \cite{Seiberg-SCFT,
  SCFT-1, SCFT-2}.  In the present work we focus on smooth toric bases
$B$ constructed as $\P^1$ bundles over smooth surfaces $S$ in the
class constructed in \cite{mt-toric}.

In this context, we can analyze our $\P^1$-bundle bases 
and consider
which of them are ``minimal'' in the simple sense that there is  no
divisor on $B$ that can be blown down ({\it i.e.} shrunk to a point)
to give another smooth toric
base $B'$.  Such bases are not truly minimal in the sense of Mori
theory, since we do not consider flips, flops, or blowing down to
singular spaces.  But they do give a simple picture of how  some of
the bases we consider may form a set of
smooth minimal toric base threefolds from which a wider range of
threefold bases, with larger non-Higgsable gauge groups, can be
constructed by sequential blow-ups.  The criterion that 
one of the
$\P^1$-bundle bases in our set can be blown down on a divisor is very
simple.  The divisor $\Sigma_-$ can only be blown down to a point in
the case where the base surface is $S= \P^2$ and the twist is $T =
H$.  $\Sigma_-$ can be blown down to a curve only if the base is
$S=\F_m$ and the twist is $T = 2 X + kY$, where $X \cdot X = -m$, $Y$
is the fiber of the Hirzebruch, and $k \leq 2m$ is an integer.
The story is identical for $\Sigma_+$ with opposite values for the
twist.  Otherwise, the only way of blowing down a divisor is for a
divisor $D_i$ to be associated with a ray $w_i = w_{i -1} + w_{i + 1}$
(with $i$ cyclic as usual), which occurs only when $T \cdot d_i = 0$.
In such a case, $D_i$ can be blown down to a curve.

\begin{figure}[h]
  \centering
  \includegraphics[scale=1]{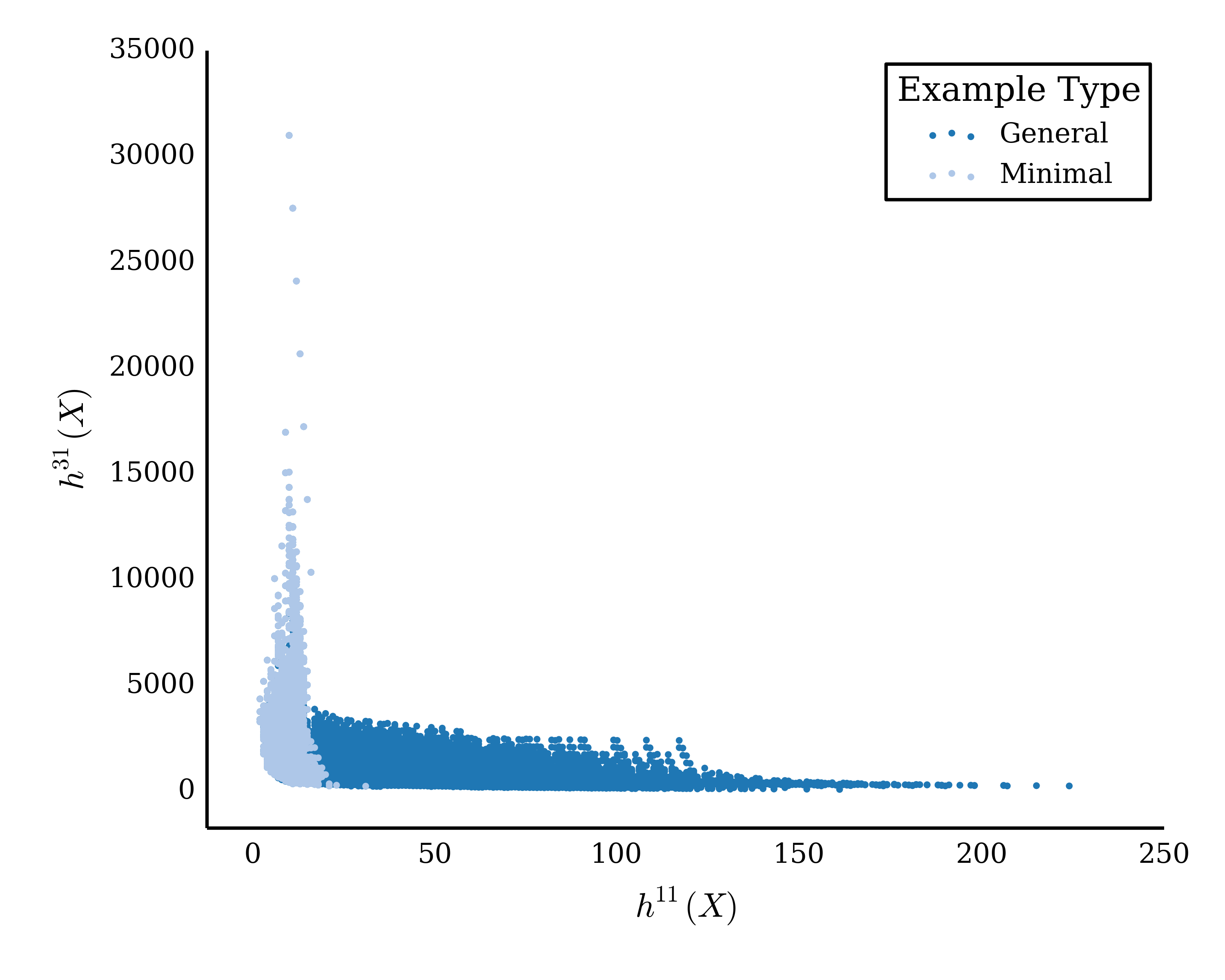}
  \caption{Plotted are the Hodge numbers for each example in our
  data, where examples that are ``minimal'', in the sense that there
  is no divisor that can be blown down to give another smooth toric
  base, are denoted in light blue}
\label{f:Hodge-minimal}
\end{figure}

The set of Hodge numbers for generic elliptic fibrations over
threefold bases that are ``minimal'' in this minimal smooth toric
sense are plotted in Figure~\ref{f:Hodge-minimal}.  The point of this
plot is that, as for the Calabi-Yau threefold case with base surfaces, the minimal bases lie along the
left side of the plot, with very small values of $h^{1, 1}(X)$.  The
other bases, to the right, are formed from blow-ups of this set of
minimal bases.  We expect that more generally, in the full set of
elliptic CY fourfolds, there is a similar structure, with minimal
bases lying along the left side of the figure, and bases formed from
multiple blow-ups of the minimal bases, with increasing $h^{1, 1}$ and
increasing non-Higgsable gauge group content, going to the right.
Thus, this analysis reinforces the picture that we expect threefold
bases that do not generate non-Higgsable gauge groups to be a
relatively small set localized in the region with relatively small
$h^{1, 1}$.

\section{Non-Higgsable clusters} 
\label{sec:4D-NHC}

\begin{figure}[t]
   \centering
   \includegraphics[scale=1]{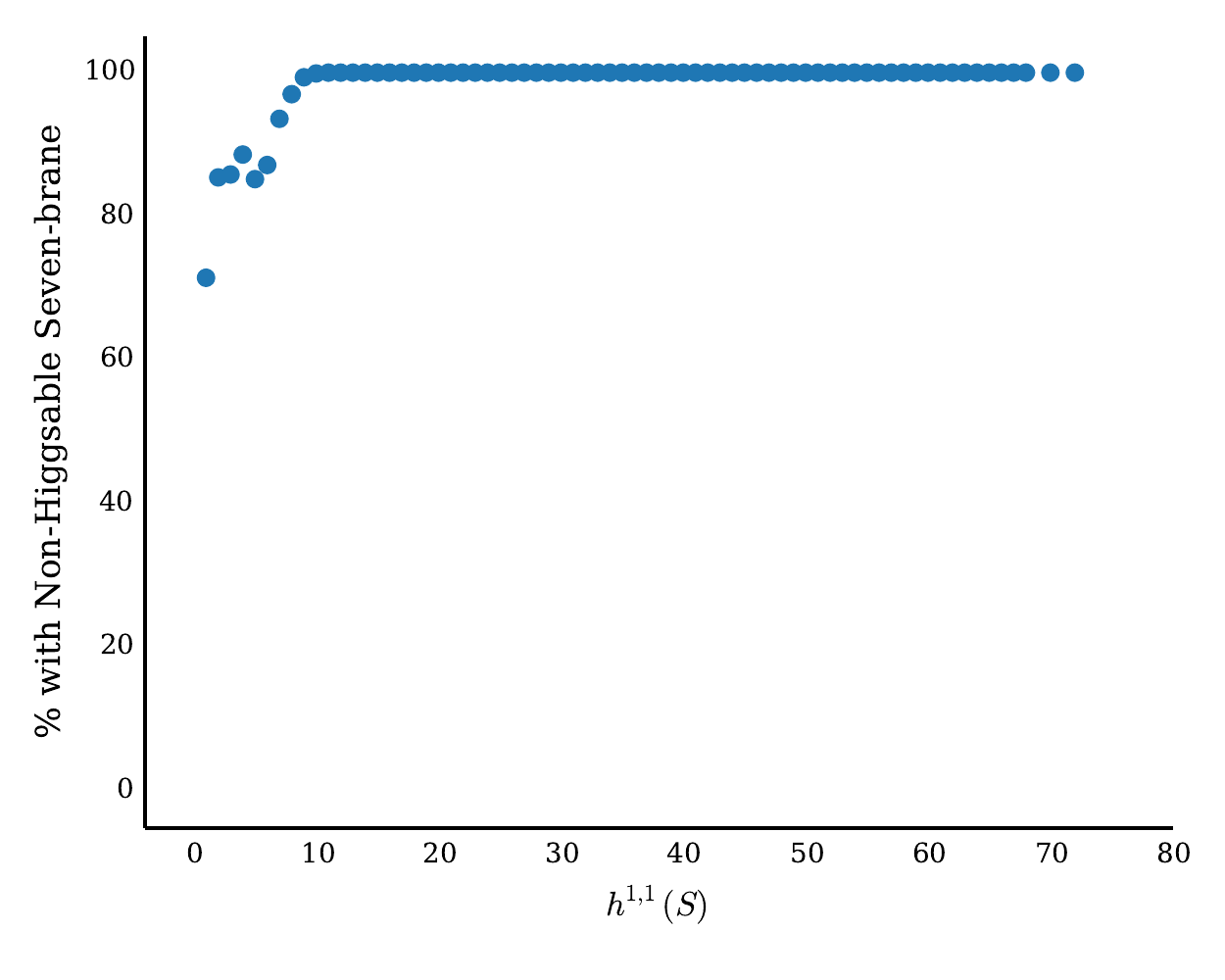}
   \caption{Percentage of bases $B$ with a non-Higgsable cluster as a
     function of $h^{11}(\b2)$.}
   \label{fig:percent NHC}
\end{figure}

\begin{figure}[t]
   \centering
   \includegraphics[scale=1]{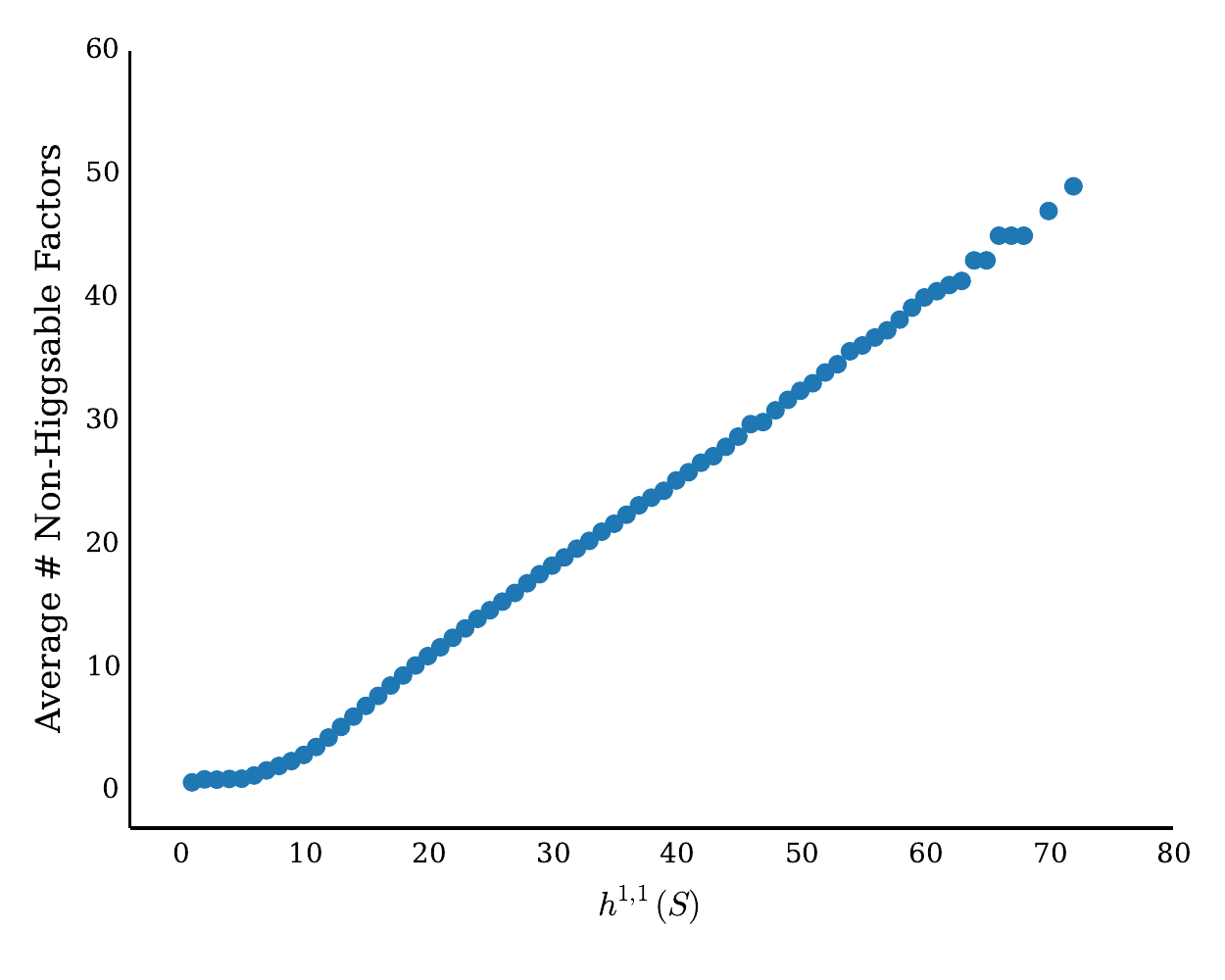}
   \caption{Average number of non-Higgsable gauge factors per example as a
     function of $h^{11}(\b2)$.}
   \label{fig:avg factors}
\end{figure}

In this section we study the structure of the non-Higgsable clusters
that arise in our examples. The most important conclusion that we
would like to draw from this set is that geometrically non-Higgsable
clusters seem to be generic features of 4D ${\cal N} = 1$ F-theory
vacua.  We have constructed $109,158$ bases, and $107,334$ of these
have non-Higgsable clusters; that is, $98.3\%$ of the examples exhibit
non-Higgsable gauge groups.  From Figure \ref{fig:percent NHC} it can
be seen that the likelihood of a non-Higgsable seven-brane
configuration appearing in our $\P^1$-bundle bases increases quickly
as a function of $h^{1,1}(\b2)$, plateauing near $100\%$ likelihood
for $h^{1,1}(\b2)\geq 10$; compare to the similar conclusion of Figure
\ref{fig:hodge gen and smooth}. Furthermore, from Figure \ref{fig:avg factors}
it can be seen that the \emph{number} of non-Higgsable gauge factors increases
with increasing $h^{1,1}(\b2)$, and the increase is approximately linear
for $h^{1,1}(\b2)\geq 10$.

From the analysis and arguments of
the previous section, the evidence we have so far suggests that for
more general bases supporting elliptic Calabi-Yau fourfolds, the
fraction of bases that give non-Higgsable clusters will be even
higher.  We do not have a rigorous argument for this conclusion, and
it is possible that there is some enormous class of elliptic
Calabi-Yau fourfolds that are unrelated to the known CY fourfolds from
toric and Landau-Ginsburg constructions. 
The hypothesis that non-Higgsable clusters are generically more
prevalent than in our restricted dataset seems
quite plausible, however, based on the parallel with the 6D story
and the fact that
the weak Fano bases and bases that have no non-Higgsable gauge
groups are in a subset with relatively small Hodge numbers among those
we have considered, combined with the observation that most of the set of known Calabi-Yau
fourfolds have larger values of $h^{1, 1} (X)$.  This conclusion is
also supported by the sense in which
many of the bases we have constructed here are essentially ``minimal''
smooth toric bases, which can be used to construct many other bases
with greater non-Higgsable content by consecutive blowing up.

Our goal in this section is to study
the detailed structure of the non-Higgsable clusters that arise in the
class of bases we have constructed here.  We consider the relative
frequency of appearance of the various allowed individual
non-Higgsable gauge group factors, as well as products, focusing on
some structures that may be relevant for producing semi-realistic
physics like that of the standard model.  One important issue to keep
in mind is that  the specific class of bases we have chosen to
consider here may have a strong influence on the distribution of
detailed aspects of the non-Higgsable clusters.  Some particular
aspects of this are as follows:

First, because we have only included strictly toric bases, we have
dropped all bases with $E_8$ non-Higgsable factors on divisors when
there is any curve on that divisor where there is a vanishing of $f,
g$ to orders $4, 6$.  As mentioned in the previous section, there are
an additional 68,528 $\P^1$ bundle bases that would be included in the
class under consideration here except for the appearance of such $(4,
6)$ curves.  As a consequence of this restriction, $E_8$ gauge group
factors are artificially suppressed in the set of bases we consider
here. 

Second, the global structure of the bases we have constructed is quite
specialized.  In particular, the structure of the set of divisors is
such that there is a ``ring'' of divisors $D_i$ associated with the
lifts of the curves $d_i$ in the base; each of the $D_i$ is a
Hirzebruch surface.  Then there are two sections $\Sigma_\pm$, each
with the topology of the base $S$.  This global structure has several
consequences.  Among other things, the distribution of gauge groups on
the sections $\Sigma_\pm$ will be taken from a broader range of local
divisor surface + normal bundle geometries, and may be more
characteristic of the full range of possibilities for general toric
threefold bases.  In addition, in terms of global structure there is
less opportunity for large structures with loops and branching in the
topology of the quiver associated with the non-Higgsable gauge group
factors in a given cluster.  And of course, we are restricting here to
toric bases, so that, for example, all the curves on which divisors
intersect are simply $\P^1$'s.

Despite these biases introduced by the choice of bases we
use here, we believe that there are some significant general lessons
that can be learned from the distribution of gauge groups and matter
found in the specific geometries we have considered.  In particular,
these bases provide a rich range of examples for more detailed
exploration of specific features of non-Higgsable clusters that may be
relevant both for phenomenological and more theoretical reasons.
Also, the general gist of our results, which is that there is a fairly
broad range of possibilities realized across the set of non-Higgsable
groups and products that are in principle allowed from F-theory,
should be valid in a broader class of bases.  In particular, we do not
find for example that the product of two gauge group factors $SU(3) \times
SU(2)$ is either hugely dominant or completely absent in the
distribution of possible non-Higgsable clusters, though it is somewhat
suppressed compared to other product groups such as $G_2 \times
SU(2)$, and individual factors $SU(3)$ and $SO(7)$ are less frequent
than the individual factors $SU(2)$ and $G_2$.  We expect these and
other such
general
patterns to persist over other classes of bases.

To compute the explicit non-Higgsable gauge factors associated with
each base we have used the straightforward toric approach in which the
Weierstrass functions $f, g$ are defined in terms of the sets of
monomials in the dual lattice, 
\begin{eqnarray}
{\cal F} & = &\{q \in N^*: \langle q, w_i
\rangle \geq -4 \;\forall i\}, \label{eq:f}\\
{\cal G}  & =  &
\{q \in N^*: \langle q, w_i
\rangle \geq -6  \;\forall i\}\,. \label{eq:g}
\end{eqnarray}
  The order of vanishing of $f$ on a
divisor $w$ is then given by ord$_wf = min_{q \in\cal F} \langle q, w
\rangle + 4$, and similarly for $g$.  The monodromy determining the
precise non-Higgsable
gauge algebra can also be read off directly from the set of monomials
as described in \cite{Anderson-Taylor, 4D-NHC}.

Note that in this section, in a more phenomenologically-motivated
spirit, we describe the non-Higgsable structure in terms of gauge
groups rather than gauge algebras, with the understanding that the
group is only explicitly determined from the orders of vanishing of
$f, g$ and monodromy at the level of the algebra, and further analysis
can give a quotient of the group by a discrete subgroup in specific cases.

\subsection{Single factors}
\label{sec:single factors}

In this subsection we consider the frequency with which any single
gauge factor $G_i$ appears in our examples. Figures~\ref{fig:num
  occurrences of each factor-general} and \ref{fig:num
  occurrences of each factor-section} display the number of overall
occurrences of each factor $G_i$ on a general divisor and on a
section, in each case as a
function of $h^{11}(\b2)$.  In Table
\ref{table:factor-percentages} we display the relative frequency of
each factor, computed by tabulating the number of times each factor
occurs in our data set and then computing for each the percentage of
the total. First we display the percentage of occurrence amongst all
factors, and then the percentage restricted to those factors that
appear on one of the sections.  As discussed earlier, we expect that
the latter may be more accurately demonstrative of the behavior that
occurs for broader sets of three-folds, since the sections exhibit a
broader set of topologies than the other divisors, which are all
Hirzebruch surfaces given by $\P^1$-bundles over base curves.  Note,
however, that the general pattern in the distribution is not highly
sensitive to this distinction.  In both cases the most frequently
appearing factors are $SU(2)$ and $G_2$.

\begin{table}[t]
\centering
\scalebox{.9}{\begin{tabular}{c|cccccccccc}
& $E_8$ & $E_7$ & $E_6$ & $F_4$ & $SO(8)$ & $SO(7)$ & $G_2$ & $SU(3)$
    &$SU(2)_{IV}$ & $SU(2)_{III}$  \\ \hline
Total Percentage & $.002$ & $9.93$ & $.028$ & $15.8$ & $.779$ & $6.09$ & $22.5$ & $.470$ & $18.2$ & $26.2$ \\
Section Percentage & $.204$ & $12.22$ & $2.43$ & $18.9$ & $4.12$ & $.154$ & $20.12$ & $4.27$ & $16.6$ & $21.0$ 
\end{tabular}}
\caption{Relative frequency of occurrence for each gauge factor, first
  for the total set of gauge factors, then for those arising on one of
  the sections.
Note that $SU(2)_{III}$ refers to an $SU(2)$ arising on a type III Kodaira
fiber, while $SU(3)_{IV}$ refers to an $SU(2)$ arising on a type IV
fiber with monodromy.}
\label{table:factor-percentages}
\end{table}

Looking at the results of this analysis, one aspect of the frequency
of the single factors can be thought of heuristically in terms of an
intuitively sensible principle.  Namely, if multiple groups may be
realized by the same Kodaira singularity type, but with different
outer monodromies, then the group associated with the largest
monodromy group action is the one that occurs most frequently.  In
particular, $F_4$ occurs more frequently than $E_6$ and both come from
Kodaira type $IV^*$, $G_2$ occurs more frequently than $SO(7)$ or
$SO(8)$ and all come from Kodaira type $I_0^*$, and $SU(2)_{IV}$
occurs more frequently than $SU(3)$ and both come from Kodaira type
$IV$. The reason that this is intuitively natural is because
\emph{not} having the largest monodromy group action requires that the
fibration satisfy additional conditions beyond that of the general
case.  Note that this means that one should expect
non-Higgsable $SU(3)$ to be relatively uncommon, because it can only
occur from a non-Higgsable cluster realized by a Kodaira type $IV$
fiber \emph{without} outer monodromy.

\begin{figure}[h]
  \centering
  \includegraphics[scale=1]{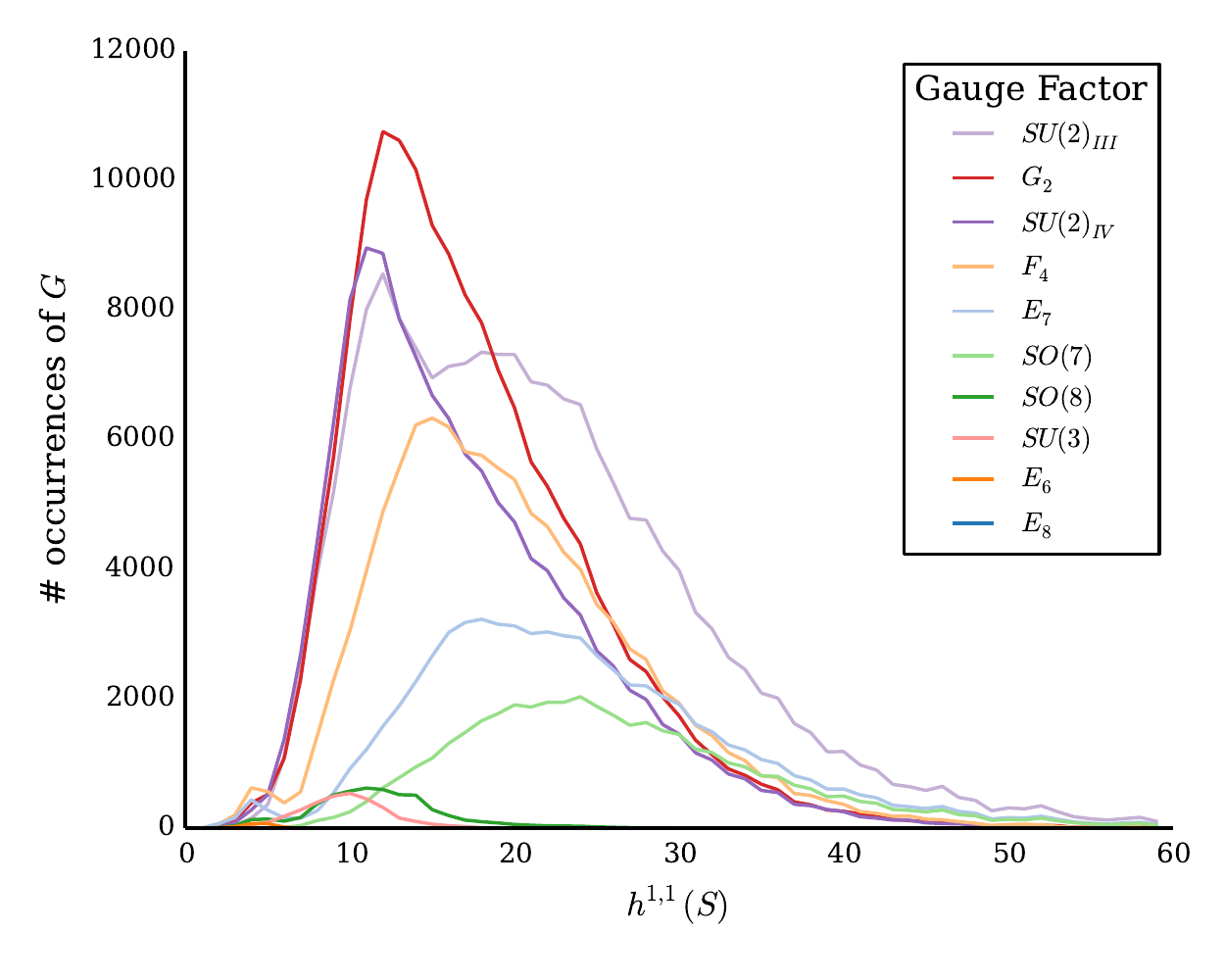}
  \caption{Displayed are the occurrences of a given
  gauge factor as a function of $h^{11}(\b2)$.}
  \label{fig:num occurrences of each factor-general}
\end{figure}
\begin{figure}[h]
  \centering
  \includegraphics[scale=1]{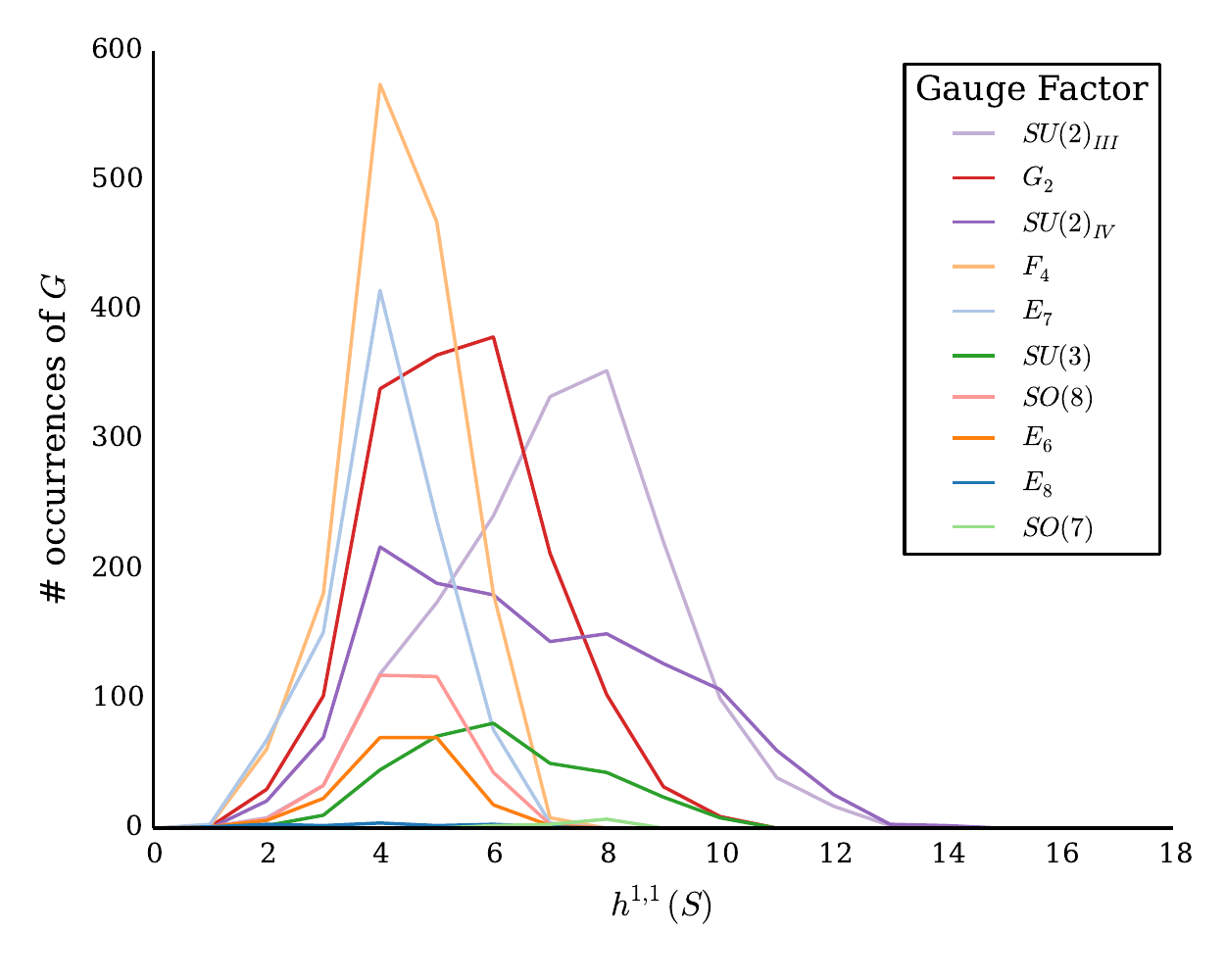}
  \caption{Displayed are the occurrences of a given
  gauge factor on one of the sections as a function of $h^{11}(\b2)$.}
  \label{fig:num occurrences of each factor-section}
\end{figure}

\subsection{Two-factor product subgroups}
We now briefly discuss the prevalence of non-Higgsable two-factor
products with jointly charged matter. By this we mean that the
non-Higgsable group contains a product of factors $G_1\times G_2$,
where the non-Higgsable seven-brane
configuration
carrying $G_1$ intersects the non-Higgsable seven-branes carrying
$G_2$.
In
fact, there are only five possibilities for two-factor products with
jointly charged matter, given in
\eqref{eq:pairings}.

\begin{figure}[h]
  \centering
  \includegraphics[scale=1]{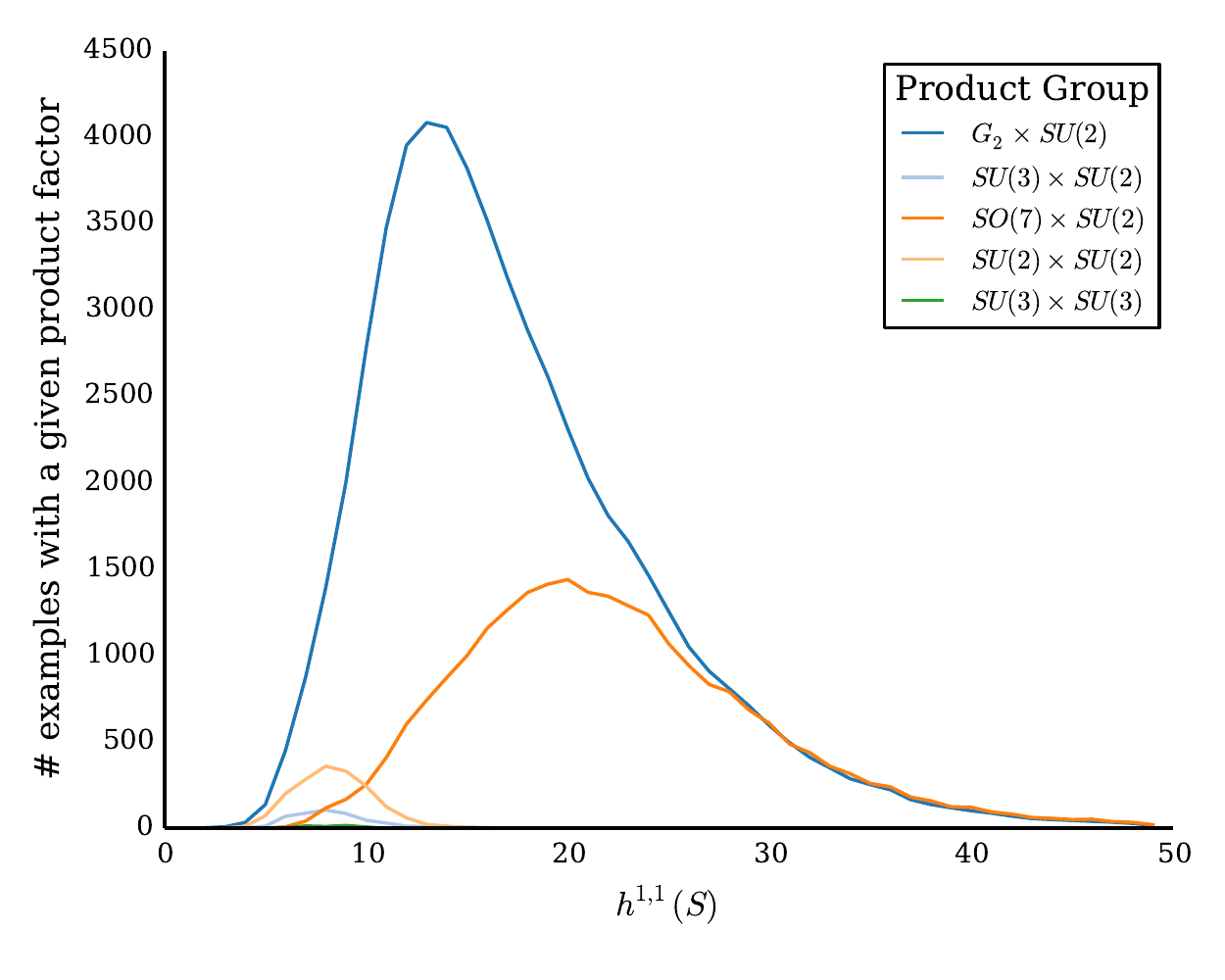}
  \caption{Displayed are the number of bases $B$ with a given
  product group as a function of $h^{11}(\b2)$.}
\label{f:bases-products}
\end{figure}
\begin{figure}[h]
  \centering
  \includegraphics[scale=1]{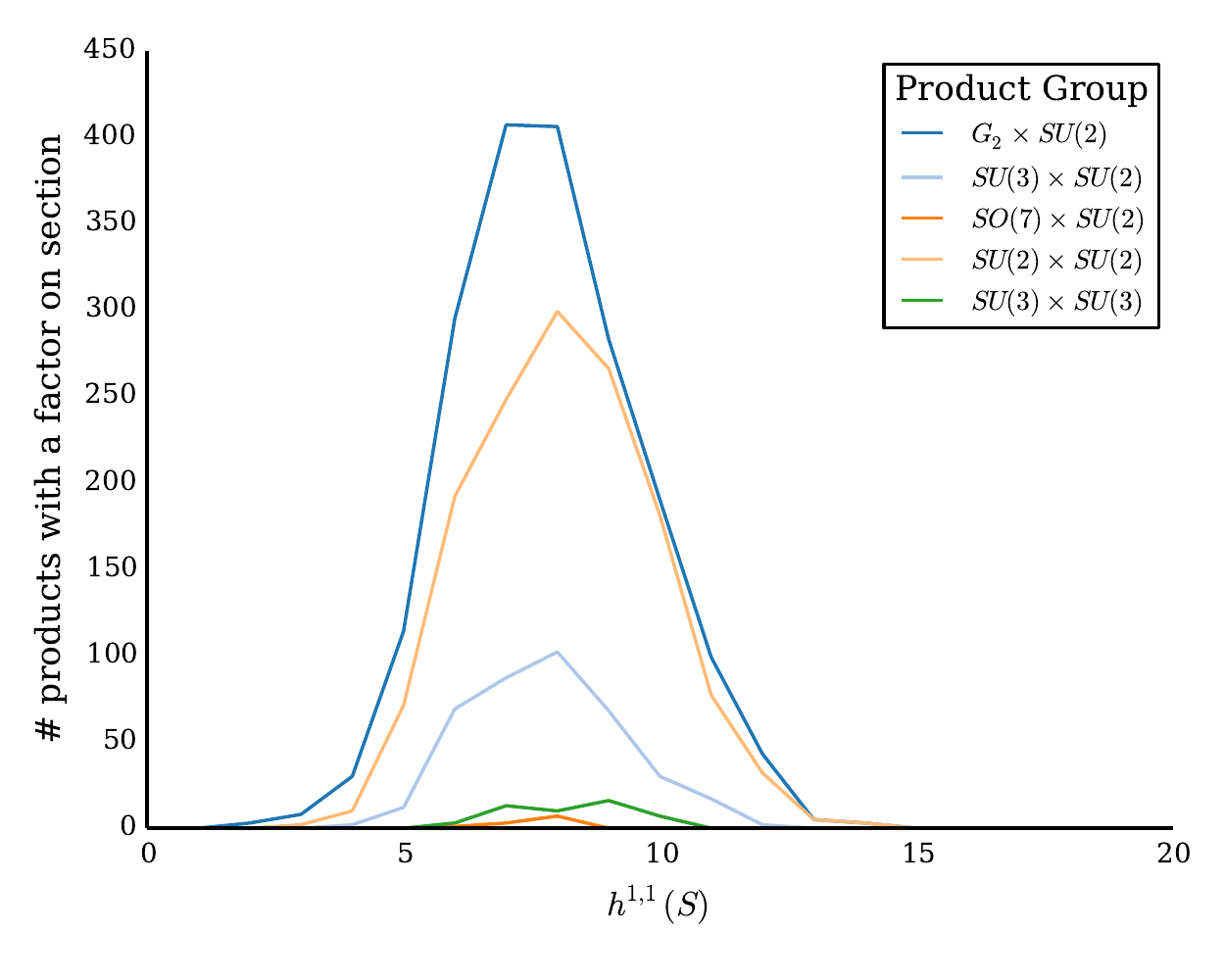}
  \caption{Displayed are the number of bases $B$ with a given
  product group as a function of $h^{11}(\b2)$, in the case that one of the
  factors in the product is on a section.}
\label{f:bases-products-sections}
\end{figure}

The prevalence of two-factor groups can again be understood according
to the principle discussed in section \ref{sec:single
  factors}. That principle is simply that for any given Kodaira fiber,
the mostly likely group to occur is the one with the largest outer
monodromy group action, since this one occurs generally, and the other
groups that may be realized by the same Kodaira fiber occur only if
additional conditions are satisfied. In particular, this means that
$SO(7)$ and $SU(3)$ are relatively unlikely groups compared to the
others, since they must satisfy additional conditions. This suggests that $G_2\times SU(2)$ and $SU(2)\times SU(2)$ should be much
more likely than $SU(3)\times SU(3)$, $SU(3)\times SU(2)$, or
$SO(7)\times SU(2)$, and in fact this is reflected in our data; see
Figures~\ref{f:bases-products} and
\ref{f:bases-products-sections}. Note by comparing the figures that
the prevalence of $G_2\times SU(2)$ and $SO(7)\times SU(2)$ relative
to other factors (including $SU(3)\times SU(2)$) goes down in
considering products where one of the factors sits on a section. As
discussed throughout the rest of the paper, we believe that the
analyses where one group factor sits on a section is more
demonstrative of the general 4D story, and thus particular attention
should be paid to figure \ref{f:bases-products-sections}.

\subsection{Cluster structure}

As described in \cite{4D-NHC}, an interesting feature of non-Higgsable
clusters in 4D F-theory models is that the associated quiver diagrams
can have nontrivial structure, with branchings and loops possible.
While the connectivity structure of divisors in the bases we consider
here is rather simple and does not support arbitrary branching and
looping structure, we explore briefly a few aspects of these features
of the non-Higgsable clusters that arise in the bases we have
constructed here.

First, we consider branching.
No branching is possible unless there is a non-Higgsable gauge factor
on at least one section.
For bases that have a non-Higgsable gauge group on one but not
both sections, the number of gauge groups on the divisors
$D_i$ associated
with curves on $\b2$ can range from 0 through 7.  The largest
branching occurs for the base $\b2$ having self-intersections
\begin{equation}
((0, -3, -2, -1, -4, -1, -4, -1, -4, -1, -4, -1, -4, -1, -2, -3))
\end{equation}
and twist
\begin{equation}
T = (t_1, \ldots, t_N) =
(0, 0, 1, 1, 0, -1, -1, -3, -2, -5, -3, -7, -4, -9, -5, -2) \,.
\end{equation}
This has a group $SU(2)$ on the  bottom section $\Sigma_+$, additional
$SU(2)$ factors on  vertical divisors $D_i$ with $i =2, 16$
(associated with the $-3$ curves $d_2, d_{16}$), and $G_2$ factors on
the vertical divisors $D_i$ with $i =5, 7, 9, 11, 13$
(associated with the $-4$ curves in $S$).  This is the only base with a
branching of degree seven.
A variety of bases support non-Higgsable clusters where a group factor
on a section has branching of degree 3, 4, 5, or 6; a simple example
of a case of branching of degree three
is described explicitly in \cite{4D-NHC}.

Considering bases that have a non-Higgsable gauge group on both
sections, there are only two examples that have more than one gauge
group on  divisors $D_i$, forming a closed loop in the quiver.
One example is on the base with self-intersections
\begin{equation}
((-3, -1, -2, -2, -1, -3, -1, -2, -2, -1))
\end{equation}
and twist
\begin{equation}
T =(0, 0,1, 2, 3, 0, -3, -2, -1, 0) \,.
\end{equation}
In this case there are four $SU(2)$ factors on the divisors
$\Sigma_\pm, D_1,$ and $D_6$, which intersect pairwise in the topology
of a closed loop.
In the other example, the base has self-intersections
\begin{equation}
((-1, -2, -2, -1, -4, -1, -4, -1, -2, -2, -1))
\end{equation}
and twist
\begin{equation}
T =(0, 0,0, 1, 1, 3, 2, 5, 3, 0, -3) \,.
\end{equation}
In this case there are $SU(2)$ factors on $\Sigma_\pm$ and $G_2$
factors on $D_5$ and $D_7$.

\subsection{Non-Higgsable QCD}
\label{sec:NH-QCD}

Motivated by the existence of an unbroken QCD sector in nature and the
ability of non-Higgsable clusters to avoid a large tuning problem in
Weierstrass moduli, with coauthors Grassi and Shaneson, we proposed
in \cite{ghst} studying the phenomenological
possibility of realizing $SU(3)_c$ from a non-Higgsable seven-brane
configuration. Such a configuration necessarily arises from a type
$IV$ Kodaira fiber, in which case the electroweak factor $SU(2)_L$
may arise from a type
$III$ fiber, a type $I_2$ fiber, or a type $IV$ fiber with outer
monodromy. The type $I_2$ case necessarily has a geometrically Higgsable
$SU(2)_L$, whereas the $III$ and $IVm$ realizations may be either
Higgsable or non-Higgsable depending on the base. Given current
knowledge, both Higgsable and non-Higgsable $SU(2)_L$ must be
considered because  the geometric non-Higgsability condition we are
using in F-theory is only a statement about one
sector of fields in a supersymmetric, ultraviolet theory.  The
possibility that the $SU(2)$ seen in the electroweak symmetry group in
nature is ``non-Higgsable'' in this sense is not
ruled out experimentally.  Indeed, supersymmetry breaking and/or
other effects could give rise to radiative electroweak symmetry
breaking in the infrared, as is common in many phenomenological
scenarios, even if there is an obstruction in the supersymmetric
ultraviolet theory to a Higgsing deformation, such as by a quartic
term arising from an F-term or D-term.
Realizing the $SU(3)$ or $SU(3) \times SU(2)$ nonabelian factors of
the standard model through non-Higgsable gauge groups that appear in
generic F-theory models over certain bases
presents an alternative to the well-studied GUT F-theory scenario
\cite{Donagi-Wijnholt,bhv, bhv2, Heckman-review, Weigand-review}.

To have matter that is jointly charged under the $SU(3)$ and $SU(2)$
factors, the divisors supporting these factors must intersect.  Along
the intersection there is an enhanced Kodaira singularity type that
can encode matter charged under both of the factors.  In
six-dimensional compactifications these codimension two singularities
always correspond to matter in the physical theory.  In 4D
constructions, however, the details of the matter content,
particularly the chirality of the matter, depend upon more detailed
considerations involving G-flux and other issues.  Only the
Lie-algebraic representation content is specified by the geometry of
the Weierstrass model.  We do not go deeply into this here (see
\cite{ghst} for further comments and references) but simply consider
any intersecting branes each carrying a nontrivial gauge group to have
``geometric matter''. In particular, the intersection of the divisors
carrying these gauge factors is a prerequisite for a realization of
the standard model, in which quarks are jointly charged under the two
factors.

\begin{figure}[t]
  \centering
  \includegraphics[scale=1]{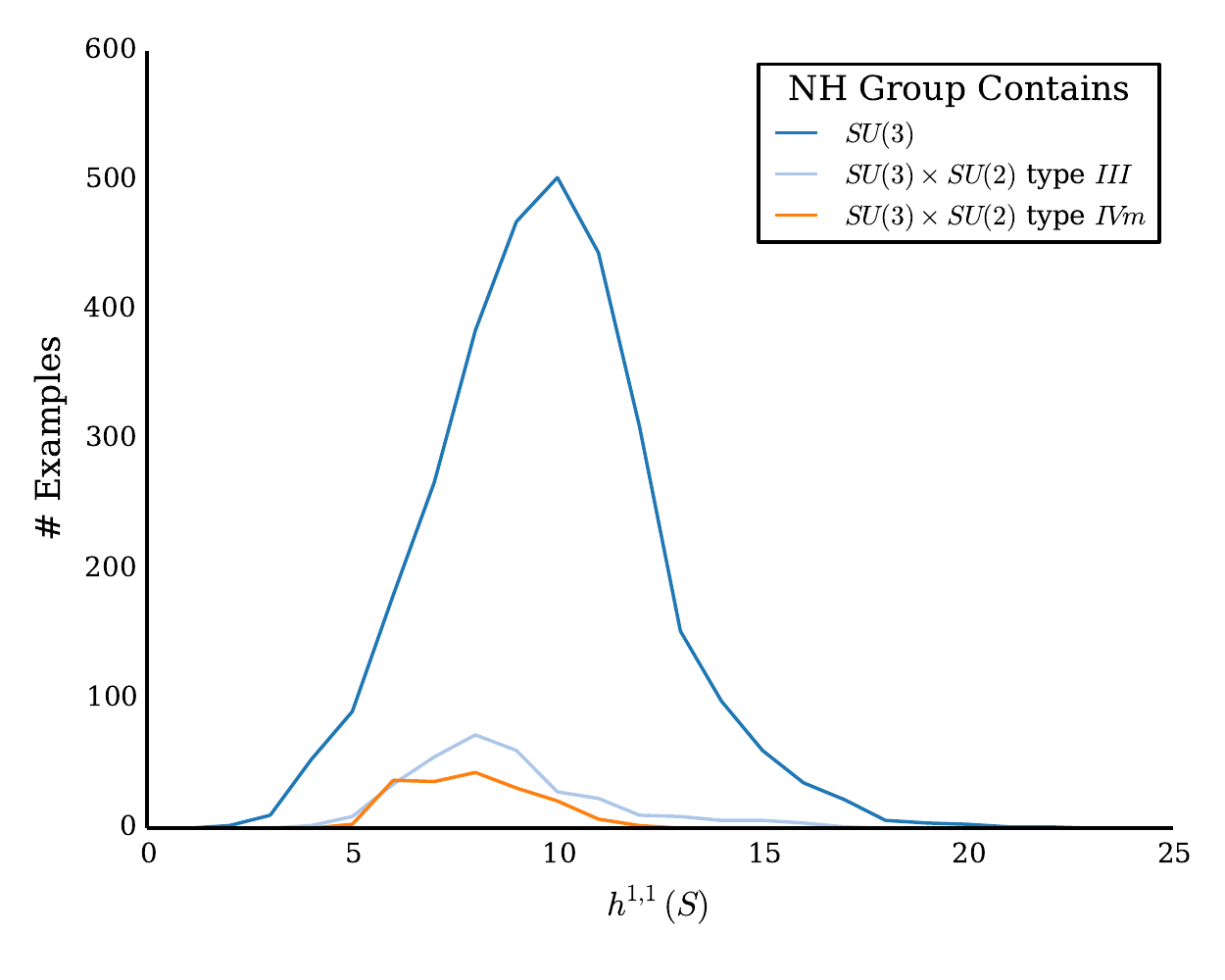}
  \caption{Number of bases $B$ that realize a scenario of Non-Higgsable QCD.}
  \label{plot:NHQCD}
\end{figure}

In this section we study the realizations of non-Higgsable QCD that occur
in our dataset, always with a non-Higgsable $SU(3)$ factor by definition, but
allowing all possibilities for the $SU(2)_L$ sector. The most coarse measure
of this phenomenon is presented in Figure \ref{plot:NHQCD},
which plots the number of bases $B$ that contain a non-Higgsable
$SU(3)$ factor, as well as the number containing a non-Higgsable
$SU(3)\times SU(2)$ factor on intersecting divisors, using either a
type $III$ or $IVm$ fiber for $SU(2)_L$. Note that this plot allows
for double counting: those examples with $SU(3)\times SU(2)$ are
included in the $SU(3)$ plot, and similarly some examples with
$SU(3)\times SU(2)$ may realize \emph{both} types of $SU(3)\times
SU(2)$ in a single example.  Note that while for many bases that have
an $SU(3)$ factor there will be some divisor that intersects the
$SU(3)$ divisor, on which it is possible to tune an $SU(2)$ through an
$I_2$ or other singularity, there can also be bases where this cannot
be done without introducing a $(4, 6)$ singularity on a curve.  Thus, the set of bases with some divisor having an
$SU(3)$ gives an upper bound and a general sense of the range of
possibilities, but is not a precise determination of a set of bases
that could admit an $SU(3) \times SU(2)$ with a Higgsable
$SU(2)$.

Let us be more specific about the counts. In our set of $109,158$
example bases $B$ there are $3,092$ examples that have a non-Higgsable
$SU(3)$ factor. Of those, $319$ have an intersecting $SU(3)\times
SU(2)$ factor (with geometric matter) of $IV$-$III$ type and $180$
have an $SU(3)\times SU(2)$ factor (with geometric matter) of
$IV$-$IVm$ type.  The \emph{total} number of bases $B$ in our set with
an $SU(3)\times SU(2)$ factor is $474$, which, from the other counts
just listed, implies that there are $319+180-474=25$ examples with
both a $IV$-$III$ and a $IV$-$IVm$. There are $313$ examples that have
a non-Higgsable $SU(3)\times SU(2)$ where the seven-branes that
support the $SU(3)$ and $SU(2)$ factors do not intersect any other
seven-branes. The base structure, twists, and gauge algebras for some
of those examples are listed in Table~\ref{tab:SU3SU2noexoticsnottb}
and Table~\ref{tab:SU3SU2noexoticsusestb}. The examples in
Table~\ref{tab:SU3SU2noexoticsnottb} all have neither the $SU(3)$
seven-branes nor the $SU(2)$ seven-branes on a section, and all of those
examples have hidden gauge sectors. On the other hand, the examples in
Table~\ref{tab:SU3SU2noexoticsusestb} have one of the $SU(3)$ or
$SU(2)$ seven-branes on a section, and these examples do not have
hidden gauge sectors. In other examples with non-Higgsable
$SU(3)\times SU(2)$, the $SU(3)$ and $SU(2)_{III}$ seven-branes intersect other
seven-branes that carry additional gauge factors not currently
observed in nature. Matter that is jointly charged with the extra
seven-branes may produce exotic states that in some cases may be
identified as exotic WIMPs; see section \ref{sec:DM}.

\begin{table}[t]
  \centering
  \scalebox{.8}{\begin{tabular}[t]{ccc}
      $d_i\cdot d_i$ for base curves $d_i$ & Twist $T\cdot d_i$ for each $d_i$ & Gauge Sectors \\ \hline
$(1, -5, -1, -2, -2, -3, -1, -3, -2, 0)$ & $(0, 1, -1, 1, 0, -1, 1, -1, 0, 0)$ & $G_2,SU(2)_{III}\times SU(3),SU(2)_{IV}$ \\
$(1, -5, -1, -2, -3, -1, -3, -2, -2, 0)$ & $(-1, 0, 0, 0, 1, -1, 1, 0, -1, 0)$ & $F_4,SU(2)_{IV},SU(3)\times SU(2)_{III}$ \\
$(1, -5, -1, -2, -3, -1, -3, -2, -2, 0)$ & $(2, 1, 0, 0, -1, 1, -1, 0, 1, 1)$ & $G_2,SU(2)_{IV},SU(3)\times SU(2)_{III}$ \\
$(0, -2, -2, -3, -1, -3, -2, -1, -6, -1, 0)$ & $(1, 1, 0, -1, 1, -1, 0, 0, 1, 0, 2)$ & $SU(2)_{III}\times SU(3),SU(2)_{IV},SO(8)$ \\
$(0, -2, -2, -3, -1, -3, -2, -1, -6, -1, 0)$ & $(0, 1, 0, -1, 1, -1, 0, 0, 1, -1, 2)$ & $SU(2)_{III}\times SU(3),SU(2)_{IV},F_4$ \\
$(0, -2, -3, -1, -3, -2, -2, -1, -6, -1, 0)$ & $(0, 0, 1, -1, 1, 0, -1, 1, -1, 0, 0)$ & $SU(2)_{IV},SU(3)\times SU(2)_{III},F_4$ \\
$(0, -3, -1, -2, -3, -1, -3, -2, -2, -1, -3)$ & $(0, 0, 0, 0, 1, -1, 1, 0, -1, 1, -1)$ & $SU(2)_{IV},SU(2)_{IV},SU(3)\times SU(2)_{III}$ \\
$(0, -3, -1, -2, -3, -1, -3, -2, -2, -1, -3)$ & $(1, 1, 0, 0, -1, 1, -1, 0, 1, 0, -1)$ & $SU(2)_{IV},SU(3)\times SU(2)_{III}$ \\
$(0, -4, -1, -2, -2, -3, -1, -3, -2, -1, -2)$ & $(0, 1, -1, 1, 0, -1, 1, -1, 0, 0, 0)$ & $SU(2)_{IV},SU(2)_{III}\times SU(3),SU(2)_{IV}$ \\
$(0, -4, -1, -2, -3, -1, -3, -2, -2, -1, -2)$ & $(-1, 0, 0, 0, 1, -1, 1, 0, -1, 0, 1)$ & $G_2,SU(2)_{IV},SU(3)\times SU(2)_{III}$ \\
$(0, -4, -1, -2, -3, -1, -3, -2, -2, -1, -2)$ & $(0, 0, 0, 0, 1, -1, 1, 0, -1, 1, -1)$ & $G_2,SU(2)_{IV},SU(3)\times SU(2)_{III}$ \\
$(0, -4, -1, -2, -3, -1, -3, -2, -2, -1, -2)$ & $(1, 1, 0, 0, -1, 1, -1, 0, 1, 0, 0)$ & $SU(2)_{III},SU(2)_{IV},SU(3)\times SU(2)_{III}$ \\
$(0, -5, -1, -2, -2, -3, -1, -3, -2, -1, -1)$ & $(0, 1, -1, 1, 0, -1, 1, -1, 0, 0, 0)$ & $G_2,SU(2)_{III}\times SU(3),SU(2)_{IV}$ \\
$(0, -5, -1, -2, -3, -1, -3, -2, -2, -1, -1)$ & $(-1, 0, 0, 0, 1, -1, 1, 0, -1, 0, 0)$ & $F_4,SU(2)_{IV},SU(3)\times SU(2)_{III}$ \\
$(0, -5, -1, -2, -3, -1, -3, -2, -2, -1, -1)$ & $(0, 0, 0, 0, 1, -1, 1, 0, -1, 1, -1)$ & $F_4,SU(2)_{IV},SU(3)\times SU(2)_{III}$ \\
$(0, -5, -1, -2, -3, -1, -3, -2, -2, -1, -1)$ & $(1, 1, 0, 0, -1, 1, -1, 0, 1, 0, 1)$ & $G_2,SU(2)_{IV},SU(3)\times SU(2)_{III}$ \\
$(0, -5, -1, -2, -3, -1, -3, -2, -2, -1, -1)$ & $(0, 1, 0, 0, -1, 1, -1, 0, 1, -1, 2)$ & $G_2,SU(2)_{IV},SU(3)\times SU(2)_{III}$ \\
$(0, -6, -1, -2, -2, -3, -1, -3, -2, -1, 0)$ & $(0, 1, -1, 1, 0, -1, 1, -1, 0, 0, 0)$ & $F_4,SU(2)_{III}\times SU(3),SU(2)_{IV}$ \\
$(0, -6, -1, -2, -3, -1, -3, -2, -2, -1, 0)$ & $(1, 1, 0, 0, -1, 1, -1, 0, 1, 0, 2)$ & $SO(8),SU(2)_{IV},SU(3)\times SU(2)_{III}$ \\
$(0, -6, -1, -2, -3, -1, -3, -2, -2, -1, 0)$ & $(0, 1, 0, 0, -1, 1, -1, 0, 1, -1, 2)$ & $SO(8),SU(2)_{IV},SU(3)\times SU(2)_{III}$ 
\end{tabular}}
  \caption{Examples with a non-Higgsable $SU(3)\times SU(2)$ sector
    that do not intersect other seven-branes, 
  where neither $SU(3)$ nor $SU(2)$ is on a section. These are all
the  examples with $h^{11}(S)< 10$.
The order of the gauge factors corresponds to the order of the curves
$d_i$, and in general the gauge factors lie on the curves with most
negative self intersections; {\it i.e.}, in the first example the
$G_2$ lies on $D_2$, $SU(2)_{III}$ on $D_5$, $SU(3)$ on $D_6$, etc.
}
  \label{tab:SU3SU2noexoticsnottb}
\end{table}

\begin{table}[t]
  \centering
  \scalebox{1}{\begin{tabular}[t]{ccc}
      $d_i\cdot d_i$ for base curves $d_i$ & Twist $T\cdot d_i$ for each $d_i$ & Gauge Sectors \\ \hline
$(2, -1, -1, -4, -1, -1)$ & $(5, 1, 1, -1, 1, 1)$ & $SU(3)\times SU(2)_{III}$ \\
$(2, -1, -2, -1, -4, 0)$ & $(5, 1, 0, 1, -1, 2)$ & $SU(3)\times SU(2)_{III}$ \\
$(1, -1, -2, -1, -4, -1, -1)$ & $(3, 1, 0, 1, -1, 0, 2)$ & $SU(3)\times SU(2)_{III}$ \\
$(1, -1, -2, -1, -4, -1, -1)$ & $(4, 1, 0, 1, -1, 1, 1)$ & $SU(3)\times SU(2)_{III}$ \\
$(1, -1, -2, -1, -4, -1, -1)$ & $(3, 2, -1, 1, -1, 1, 1)$ & $SU(3)\times SU(2)_{III}$ \\
$(1, -1, -2, -2, -1, -4, 0)$ & $(4, 1, 0, 0, 1, -1, 2)$ & $SU(3)\times SU(2)_{III}$ \\
$(1, -1, -2, -2, -1, -4, 0)$ & $(3, 2, -1, 0, 1, -1, 2)$ & $SU(3)\times SU(2)_{III}$ \\
$(0, -2, -1, -3, -1, -2, 0)$ & $(3, 0, 1, 0, 1, 0, 3)$ & $SU(2)_{IV}\times SU(3)$ \\
$(0, -2, -1, -3, -1, -2, 0)$ & $(-3, 1, -2, 1, -1, 0, -3)$ & $SU(2)_{III}\times SU(3)$ \\
$(0, -2, -2, -1, -3, -1, 0)$ & $(3, 0, 0, 1, 0, 1, 3)$ & $SU(2)_{IV}\times SU(3)$ \\
$(0, -2, -2, -1, -3, -1, 0)$ & $(3, 0, 0, 1, -1, 2, 2)$ & $SU(2)_{III}\times SU(3)$ \\
$(0, -2, -2, -1, -3, -1, 0)$ & $(-3, 0, 1, -2, 1, -1, -3)$ & $SU(2)_{III}\times SU(3)$ \\
$(1, -3, -1, -2, -2, -2, 0)$ & $(4, 0, 1, 0, 0, 0, 3)$ & $SU(2)_{IV}\times SU(3)$ \\
$(1, -3, -1, -2, -2, -2, 0)$ & $(-4, 1, -2, 1, 0, 0, -3)$ & $SU(2)_{III}\times SU(3)$
\end{tabular}}
\caption{Examples with a non-Higgsable $SU(3)\times SU(2)$ sector that do not intersect other seven-branes
  where either the $SU(3)$ or $SU(2)$ is on a section. These are all
  the  examples with $h^{11}(S)<6$.
  In each case the latter factor in the gauge group lies on a section
  $\Sigma_\pm$.  The first of these examples was also featured in
  \cite{ghst}.
}
  \label{tab:SU3SU2noexoticsusestb}
\end{table}

Some of the examples have a non-Higgsable $SU(2)\times SU(3)\times
SU(2)$, similar to a so-called left-right model, where there is
bifundamental matter charged under $SU(3)$ and each of the $SU(2)$
factors separately, but no bifundamental matter charged under both
$SU(2)$ factors. Though we
will not exhaustively categorize the possibilities, it is illustrative
to study this possibility in a simple subset.  One such subset is the
$25$ examples that have both $IV$-$III$ and $IV$-$IVm$ realizations of
$SU(3)\times SU(2)$; in fact these are typically left-right models. We
have displayed their algebraic structure in Table \ref{table:IV-III
  and IV-IVm}, where $SU(3)$ denotes non-Higgsable $SU(3)$, $SU(2)_{IV}$
denotes a non-Higgsable $SU(2)$ from a $IVm$ fiber, $SU(2)_{III}$ denotes a
non-Higgsable $SU(2)$ from a type $III$ fiber, and $II$ denotes a
singular seven-brane configuration
with a type $II$ fiber, which does not carry a
gauge group but might give rise to matter at intersections with other
branes. The first two entries represent the sections
$\Sigma_\pm$ of the $\P^1$
bundles, whereas the rest of the entries are for the divisors $D_i$
that are $\P^1$ bundles over the curves $d_i$ in the base
$\b2$. Seven-branes that occur on the sections may not intersect each
other, but intersect any seven-brane in the base, whereas seven-branes
on the divisors $D_i$ intersect one another only if their
respective base curves are adjacent.
\begin{table}
\centering
\begin{tabular}{c}
$(0,SU(3),0,0,SU(2)_{IV},0,II,SU(2)_{III},0,0)$\\
$(SU(3),0,0,SU(2)_{IV},0,II,SU(2)_{III},0,0,0)$\\
$(SU(3),0,0,SU(2)_{IV},0,II,SU(2)_{III},0,0,0,0)$\\
$(SU(3),0,0,SU(2)_{IV},0,II,SU(2)_{III},0,0,0,0)$\\
$(0,SU(3),0,SU(2)_{III},II,0,SU(2)_{IV},0,0,0,0)$\\
$(0,SU(3),0,SU(2)_{III},II,0,SU(2)_{IV},0,0,0,0)$\\
$(SU(3),0,SU(2)_{IV},0,0,0,0,SU(2)_{III},0,0,0,0)$\\
$(SU(3),0,SU(2)_{IV},0,0,0,0,0,SU(2)_{III},0,0,0)$\\
$(SU(3),0,SU(2)_{IV},0,0,0,0,0,SU(2)_{III},0,0,0)$\\
$(SU(3),0,0,SU(2)_{IV},0,0,0,0,0,0,SU(2)_{III},0)$\\
$(0,SU(3),0,SU(2)_{III},0,0,0,0,0,0,SU(2)_{IV},0)$\\
$(0,SU(3),0,SU(2)_{III},0,0,0,0,0,0,0,SU(2)_{IV})$\\
$(SU(3),0,SU(2)_{IV},0,II,SU(2)_{III},0,0,0,0,0,0)$\\
$(SU(3),0,SU(2)_{IV},0,II,SU(2)_{III},0,0,0,0,0,0)$\\
$(SU(3),0,SU(2)_{III},II,0,SU(2)_{IV},0,0,0,0,0,0)$\\
$(SU(3),0,SU(2)_{III},II,0,SU(2)_{IV},0,0,0,0,0,0)$\\
$(SU(3),0,SU(2)_{III},II,0,SU(2)_{IV},0,SU(2)_{IV},0,0,0,0,0)$\\
$(SU(3),0,SU(2)_{III},II,0,SU(2)_{IV},0,II,0,0,0,0,0)$\\
$(SU(3),0,SU(2)_{III},II,0,SU(2)_{IV},0,SU(2)_{III},0,0,0,0,0)$\\
$(SU(3),0,SU(2)_{III},II,0,SU(2)_{IV},0,0,SU(2)_{IV},0,0,0,0)$\\
$(SU(3),0,SU(2)_{III},II,0,SU(2)_{IV},0,0,II,0,0,0,0)$\\
$(SU(3),0,SU(2)_{III},II,0,SU(2)_{IV},0,0,SU(2)_{III},0,0,0,0)$\\
$(0,SU(3),SU(2)_{IV},0,0,SU(2)_{III},II,0,SU(2)_{IV},0,0,SU(2)_{III},II,0)$\\
$(SU(3),0,SU(3),0,0,SU(2)_{IV},0,II,SU(2)_{III},0,0,0,0,0)$\\
$(SU(3),0,SU(3),0,SU(2)_{IV},0,II,SU(2)_{III},0,0,0,0,0,0)$
\end{tabular}
\caption{Displayed is the algebra structure of the $25$ examples with both $IV$-$III$
and $IV$-$IVm$ realizations of $SU(3)\times SU(2)$. See the text for a discussion.}
\label{table:IV-III and IV-IVm}
\end{table}

There are a number of phenomena that can be seen easily in these $25$
examples, given the topological intersections just discussed. First,
note that there is always a non-Higgsable $SU(3)$ factor on one of the
sections, and the only $SU(2)$ factors arise from
divisors $D_i$. In these examples, any $SU(3)$ factor that is not on one
of the sections is not adjacent to any of the $SU(2)$ factors. Therefore,
the only $SU(3)\times SU(2)$ charged matter occurs at the intersection
between an $SU(3)$ factor on a section and an $SU(2)$ factor on one of
the divisors $D_i$.   Considering in particular the left-right symmetric
$SU(2)\times SU(3)\times SU(2)$ structure
just discussed, in all of these 25
examples the seven-brane carrying $SU(3)$ intersects at least two
seven-branes carrying $SU(2)$, none of which intersect each
other. When the $SU(3)$ intersects precisely two seven-branes carrying
$SU(2)$, the $SU(2)\times SU(3)\times SU(2)$ idea is realized. In
other examples the seven-brane with $SU(3)$ on a section intersects
three $SU(2)$'s, none of which intersect each other, in which case it
is the central vertex in a quiver that attaches to disjoint $SU(2)$
nodes.
In the last  two examples, the $SU(3)$ on the section intersects
two $SU(2)$ factors and one $SU(3)$ factor on the $D_i$, giving
another quiver with a central node and three branches.

One obvious question that we do not address in this paper is the
potential origin of a $U(1)$ factor in a model with non-Higgsable
$SU(3)$.    The simplest approach to including a $U(1)$ factor
would be to tune it by hand, using for example the general $U(1)$
Weierstrass form of \cite{Morrison-Park}.  It would be desirable,
however, in the spirit of this work, to find a more natural way of
including the abelian part of the standard model.
While for some very special (non-toric) bases \cite{Martini-WT,
  Morrison-Park-Taylor} there are non-Higgsable $U(1)$ factors in 6D
F-theory compactifications, it is not clear how frequently such
abelian factors are generic over threefold
bases for 4D F-theory
constructions; this remains an interesting avenue for further investigation.

\begin{figure}[h]
  \centering
  \includegraphics[scale=1]{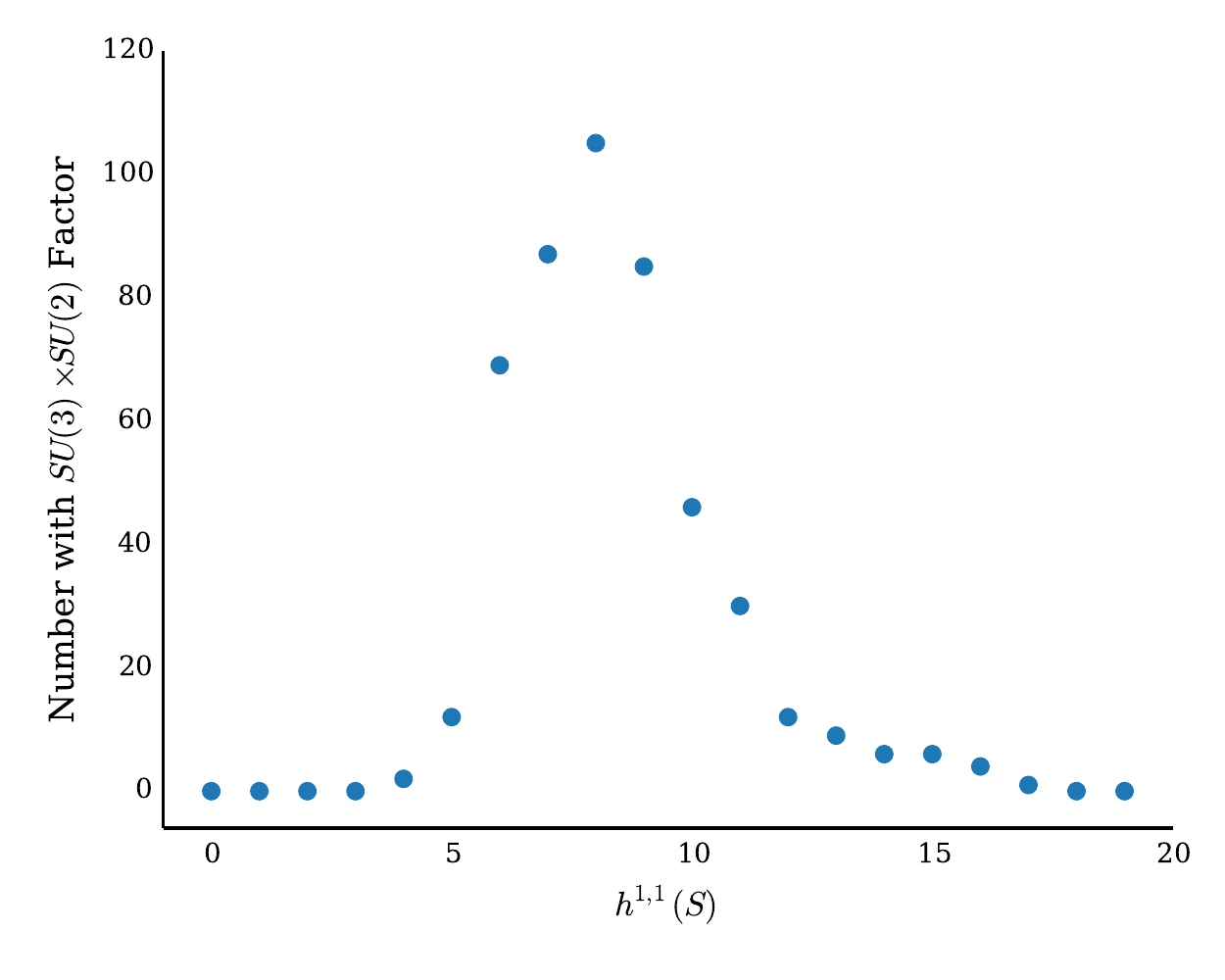}
  \caption{Number of bases $B$ that have a non-Higgsable
    $SU(3)\times SU(2)$ gauge
  factor as a function of $h^{11}(\b2)$.}
\end{figure}

\subsection{Dark matter}
\label{sec:DM}

While the appearance of $SU(3)$ and $SU(2)$ as the only $SU(N)$ gauge
factors that may be non-Higgsable is suggestive for visible sector
phenomenology, there are also at least two distinct ways in which
non-Higgsable clusters may be relevant for dark matter
phenomenology. In this section and the next
we will explore the extent to which
these possibilities arise in our classification.

One possibility (also considered in the F-theory context in \cite{bhv})
is that hidden sector dark matter may arise from gauge
sectors that are topologically disconnected from the visible sector.
Such dark matter would interact with the visible sector
gravitationally but not via gauge forces. Such dark matter can arise
naturally in generic F-theory models since non-Higgsable seven-branes
may be topologically disconnected from those realizing the standard
model. In such a case, the lightest particle charged under the hidden
sector gauge group $G$ is stable, and therefore if cosmologically
produced it will contribute to the dark matter relic abundance.  Even
if no matter is charged under the hidden sector group $G$, uncharged
glueballs may play a role as hidden sector dark matter.

The second possibility that may be relevant is that of exotic weakly
interacting massive particle (WIMP) dark matter. By an exotic WIMP we
mean a a WIMP that arises in an ``exotic'' multiplet, perhaps most
easily defined in an $\mathcal{N}=1$ theory as a chiral multiplet
beyond the MSSM spectrum that contains a WIMP candidate. Such
particles are distinct from the neutralinos of the MSSM, but
nevertheless give rise to interesting dark matter candidates by virtue
of the thermal or non-thermal \cite{NTWimp} WIMP miracle.  Considered
in the context of non-Higgsable clusters such exotic WIMPs arise quite
naturally: if a non-Higgsable $SU(2)_L$ seven-brane intersects another
non-Higgsable seven-brane carrying a non-abelian gauge theory $H$ then
there are exotic multiplets charged under $SU(2)_L\times H$. In fact,
since $H$ arises from a non-Higgsable seven-brane configuration it may only be in
the set $H\in \{G_2,SU(3),SU(2),SO(7)\}$, though there are more
possibilities if $H$ arises on a tuned brane configuration.    It is
also possible for extra matter to arise as codimension two
singularities on a curve within the divisor carrying a gauge group
such as the $SU(2)$ factor in the standard model, even without another
gauge group on another divisor that passes through that curve.  This
could also give rise to interesting exotic WIMP candidates.

In this section we present some results from our classification
related to both of these scenarios, beginning with the first. 
We emphasize from the outset here, however, that particularly in this
set of considerations regarding the structure of multiple and/or multi-factor
non-Higgsable clusters,
the kinds of statistical distributions that we
find are strongly influenced by the specific structure of the
threefold bases we are working with, so the results should be taken
only as very qualitative and suggestive.
As
discussed in section \ref{sec:NH-QCD}, there are $3,092$ examples with
a non-Higgsable $SU(3)$ factor. These are the ones that may realize
the non-Higgsable QCD scenario, and we would like to know whether in
these examples there is at least one gauge factor on a divisor that
does not intersect the divisors supporting the standard model $SU(3)
\times SU(2)$.  Note as mentioned above that we have not explicitly
identified curves on which Higgsable $SU(2)$ factors can be tuned on
all the $SU(3)$ models, so for simplicity we simply look for other
clusters disjoint from the $SU(3)$ to get a qualitative sense of the
possibilities.
In the language of quivers, there must be a quiver
component that is disconnected from the component that contains the
visible sector gauge node.  Given the topology of our $B$'s, if there
is a gauge group on a seven-brane on either of the sections of the
$\P^1$ bundle then there is a single quiver component. Therefore
examples with a non-Higgsable $SU(3)_c$ that have a disconnected
hidden sector gauge group must have no gauge group on either of the
sections; this restricts us to $2,493$ of the possible $3,092$
examples with a non-Higgsable $SU(3)$ factor. Having disconnected
hidden sector gauge groups in these examples is equivalent to having
disconnected quiver components arising from base curves; there are
$2,239$ such examples.

These $2,239$ examples of non-Higgsable QCD have a relatively short
list of disconnected hidden sector gauge groups. The groups of
the disconnected components in these examples, as well as their multiplicities,
are
\begin{align}
SU(2)_{IV}&: \qquad 1753 \nonumber \\
SU(2)_{III}&: \qquad 905 \nonumber \\
F_4&: \qquad 789 \nonumber \\
G_2&: \qquad 762 \nonumber \\
SU(2)_{III}\times G_2&: \qquad 292 \nonumber \\
SO(8)&: \qquad 188 \nonumber \\
SU(2)_{IV}\times G_2&: \qquad 137 \nonumber \\
E_7&: \qquad 91 \nonumber \\
SU(2)_{III}\times SU(2)_{III}&: \qquad 7 \nonumber \\
SU(2)_{III}\times SO(7)\times SU(2)_{III}&: \qquad 3 \nonumber \\
SU(2)_{IV}\times SU(2)_{III}&: \qquad 3
\end{align}
where 
the subscript on each $SU(2)$ indicates whether it is realized on a
Kodaira type III or IV singularity.

For the purposes of studying hidden sector dark matter it is also
useful to examine Table \ref{tab:SU3SU2noexoticsnottb}. This table lists
some of the 85
examples in our set that have a non-Higgsable $SU(3)\times
SU(2)$ sector where the $SU(3)$ and $SU(2)$ seven-branes intersect
each other, but no other non-Higgsable seven-brane, and furthermore
that neither of these seven-branes is on a section. 
Note that in the other 228 cases where there is a disconnected $SU(3)
\times SU(2)$ sector, with one factor on a section, there can be no
hidden sectors in the theory from non-Higgsable seven-brane configurations.  All 85 examples of the type listed in
Table~\ref{tab:SU3SU2noexoticsnottb}, however,
has other non-Higgsable gauge
groups that do not intersect the visible sector. In some cases the
hidden sectors are single factor super Yang-Mills theories, whereas in other cases they
contain product gauge groups with bifundamental matter. We see that
rich dark sectors with no gauge interactions with the visible sector
are common in this subset of models.
It should be emphasized again, however, that the distribution of types of hidden sector
dark matter found in the examples we have considered here is likely
heavily influenced by our choice of bases.  In particular, in general
$E_8$ dark matter sectors are likely much more frequent 
than in our
analysis, where we have only considered strictly
toric bases with no $(4, 6)$ curves on divisors carrying $E_8$
groups.  Also, the non-Higgsable clusters in the hidden sector dark
matter cases described here are heavily influenced by the geometry of
the base surface $S$.  In general we might expect more, and more
complicated hidden sector dark matter structure for typical threefold
bases $B$, which will have larger Hodge numbers and typically a more
complicated topology.
Nonetheless, the sample considered here gives
some general sense of the kinds of possibilities that may arise, and
suggest that $SU(2)$ hidden sector dark matter may be the most common
occurrence in generic F-theory models.

\vspace{.5cm} We now study the possibility of exotic WIMP dark matter
associated with an additional hidden gauge sector; pure matter WIMP
contributions are considered in the next subsection. The simplest
measure of the possibility of WIMP dark matter associated with a
hidden gauge sector is to simply count and categorize the intersection
of non-Higgsable $SU(2)$ seven-branes with other seven-branes.

We present three types of counts, with the first being the simplest,
all presented in Table \ref{table:exotic WIMPs}. The first count is a
count of the number of times that a non-Higgsable seven-brane with a
given gauge group $G$ intersects a non-Higgsable $SU(2)$
seven-brane. In this counting method, we note that $G_2$ is the most
common exotic WIMP hidden sector group, followed by $SO(7)$; this
closely mimics what is expected from six-dimensional
compactifications, in which case these are the \emph{only}
non-Higgsable groups that may intersect a non-Higgsable $SU(2)$. This
is again likely because of the structure of our bases and in
particular because in that context we have allowed for $SU(2)$ factors
that arise on seven-branes on divisors $D_i$ that are always
$\P^1$-bundles over curves in $\b2$, and which dominate the geometry.

The second type of count is to perform the same type of counting, but
to mitigate for the effects of the specific types of base geometry by
including only
only non-Higgsable $SU(2)$ groups that arise on one of the
sections of the $\P^1$-bundle. In this case we see that intersections
with $SO(7)$ \emph{almost never} occur, while intersections with $G_2$
non-Higgsable seven-branes are still the most common. Finally, our
third count is to consider only non-Higgsable $SU(2)$ seven-branes on
one of the sections that also intersect a non-Higgsable $SU(3)$
seven-brane configuration
(other than the instance of the group $G$ being counted, in the case
$G = SU(3)$); that is, these $SU(2)$ configurations are more naturally
identifiable as $SU(2)_L$ given the intersection with an $SU(3)$
seven-brane. The results of these countings are shown in
Figure~\ref{table:exotic WIMPs}.

\begin{table}[h]
  \centering
  \begin{tabular}{cccccc}
    $SU(2)$ Type & $SU(2)_{III}$ & $SU(2)_{IV}$ & $SU(3)$ & $G_2$ & $SO(7)$ \\ \hline
    All & $2322$ & $2148$ & $524$ & $94780$ & $83771$ \\
    Sections & $703$ & $1193$ & $276$ & $1522$ & $1$ \\
    Sections and intersects $SU(3)$& $10$ & $75$ & $9$ & $47$ & $0$ 
  \end{tabular}
  \caption{Frequency of gauge groups under which exotic WIMPs are charged.}
  \label{table:exotic WIMPs}
\end{table}

\subsection{Higgs sector fields}

To realize electroweak symmetry breaking in a supersymmetric theory,
the simplest approach is to have a vector pair of Higgs doublet fields
as in the MSSM.  We briefly investigate here how this might occur in a
scenario where the standard model $SU(3) \times SU(2)$ is realized
through non-Higgsable factors.  Extra matter fields charged under the
$SU(2)$ can arise when the divisor supporting the non-Higgsable
$SU(2)$ factor has an additional codimension two singularity on a
curve where $(f, g)$ vanish to higher order.  When the resulting
matter is in the fundamental representation of $SU(2)$, the fields (if
an appropriate hypercharge is also realized) can give rise to the pair
$H_u$, $H_d$ of the MSSM. Such a Higgs sector could break the $SU(2)$
if radiative corrections due to supersymmetry breaking give the
lightest Higgs field a negative quadratic term so that it is forced to
acquire a vacuum expectation value; {\it i.e.}, if radiative
electroweak symmetry breaking is realized.  

Rather than doing a thorough statistical analysis on the
possibilities, we simply  describe how extra matter
fields can arise from additional codimension two singularities on the
$SU(2)$ divisor, and speculate about how such fields may be stabilized
in a supersymmetric theory.  
Such extra matter fields could be part of a Higgs sector, or
alternatively could play a role as weakly interacting dark matter as
discussed previously.
We conclude with an explicit example in which these kinds of additional $SU(2)$-charged fields arise.

While in any non-Higgsable realization of the product $SU(3) \times
SU(2)$ there is always a codimension two singularity at the
intersection between the $SU(3)$ and $SU(2)$ branes that can carry
matter, additional
 matter curves also arise on a large number of the
$SU(2)$ factors that appear as components of $SU(3) \times SU(2)$
non-Higgsable clusters. In fact, in all the examples in Table
\ref{tab:SU3SU2noexoticsnottb} the residual discriminant intersects
the $SU(2)$ seven-brane along a curve (away from the $SU(3)$ locus)
that is non-trivial in homology.  In future work it may be interesting
to study the detailed structure of these curves.
We describe one explicit example below.

One natural question is how it could be that a massless Higgs pair
$(H_u,H_d)$ that arises in this fashion
would not give rise to a flat direction in the supersymmetric ${\cal
  N} = 1$ F-theory supergravity vacuum, since there is no complex
structure deformation that breaks $SU(2)$, and yet $H_u H_d$ (with
$SU(2)$ indices contracted by $\epsilon_{ij}$) is a gauge invariant
holomorphic function and therefore there is a corresponding D-flat
symmetry breaking direction.  If the obstruction is visible in a
weakly coupled Lagrangian description of this $d=4$ $\mathcal{N}=1$
supergravity theory, the natural possibility is that it is an F-term
obstruction. One possible obstruction may arise due to the possible
appearance of an $SU(3)\times SU(2)$ singlet field $\Phi$ at the same
locus as the Higgs fields $H_u, H_d$.  
The superpotential may then contain a term $\Phi
H_u H_d$; the F-term for $\Phi$ then gives rise in turn to quartic
terms in the potential $(H_u H_d)(H_u^\dagger H_d^{\dagger})$, which
in combination with D-terms could stabilize the Higgs field at quartic
order, providing a physical mechanism that could explain the
non-Higgsable nature of the field in the supersymmetric theory, and
yet remaining compatible with the potential for electroweak symmetry
breaking after radiative corrections push the quadratic term in the
Higgs field negative. Note that such singlets would play the role of the
exotic singlet in the NMSSM extension of the minimal supersymmetric
standard model.   
Other such   terms could potentially
stabilize the squark sector, which would
also need to be fixed in a model with non-Higgsable $SU(3)\times SU(2)$.
Alternatively, it may also be possible that the
obstruction giving rise to non-Higgsability  of the $SU(2)$ gauge
factor
in the supersymmetric ultraviolet theory is due to
non-perturbative physics not captured by a Lagrangian description.

As an explicit example,
consider the base defined by the toric surface with
curves of self-intersection
\begin{equation}
((1, -1, -2, -1, -4, -1, -1))
\end{equation}
and twist
\begin{equation}
T = (0, 0,2, 3, 2, 4, 3) \,.
\end{equation}
This base has a non-Higgsable $SU(2)$ (type III) on $\Sigma_-$ and a
non-Higgsable $SU(3)$ on $D_5$.  There is naturally a codimension two
singularity associated with potential matter on the curve $C
=\Sigma_-\cap D_5$, where $(f, g)$ vanish to orders $(3, 4)$.  There
is also, however a codimension two singularity on the curve $C'
=\Sigma_-\cap D_2$, where $(f, g)$ vanish to orders $(2, 3)$.  This
curve\footnote{ Another way to think of this singularity is that it is
  a curve along which the seven-brane carrying $SU(2)$ intersects
  the residual part of the discriminant, which has an $I_1$ Kodaira
  fiber. } does not intersect with $C$, and is naturally associated
with additional matter charged under the $SU(2)$ gauge group.  
There are also codimension two singularities on the curves $\Sigma_-
\cap D_4, \Sigma_- \cap D_6$ where $(f, g)$ vanish to orders $(2, 2)$.
In
other cases, similar matter curves on an $SU(2)_{III}$ locus can have
other types of increased orders of vanishing of $(f, g)$.  These
codimension two curves with increased vanishing
of $(f, g)$
potentially
give fundamental or other matter representations charged under the
$SU(2)$.  A complete resolution or deformation analysis of any given
model (see, {\it e.g.}, \cite{Katz-Vafa,
mt-singularities, Esole-Yau, Lawrie-sn, Hayashi-ls,
  hlms, Esole-sy, Braun-sn} or \cite{ghs1, ghs2}) would need to be done to
determine the specific matter content over any given curve.  Matter
fields produced in this way, as mentioned above, can potentially
play a role as a Higgs field or as weakly interacting dark matter candidates.

\section{Conclusions}
\label{sec:conclusions}

In this paper we have explored a specific class of complex threefold bases for
F-theory compactifications to ${\cal N} = 1$ supergravity theories in
four space-time dimensions.  We have constructed roughly 100,000 bases
$B$ that have the form of $\P^1$ bundles over toric base surfaces.  We
have studied the non-Higgsable cluster content over each of these
bases and extracted some general lessons from the analysis.

The  general conclusions for which we have found evidence are that:
\vspace*{0.05in}

\noindent
1) Geometrically non-Higgsable gauge groups and matter are ubiquitous
features in the landscape of ${\cal N} = 1$ 4D F-theory vacua.
\vspace*{0.05in}

\noindent
2) The specific group $SU(3)$ and the product $SU(3) \times SU(2)$
appear as non-Higgsable gauge group components
less frequently than a handful of other possibilities, but they are
realized geometrically in a wide range of F-theory vacua.
\vspace*{0.05in}

In addition to analyzing the  statistics of non-Higgsable
clusters on the specific bases we have constructed we have also made
more general arguments, in part by analogy with the better understood
6D story, that support the conclusions that these hypotheses hold in a
wide range of F-theory vacua beyond the particular constructions
considered here.
Note that many of the features found here,
including the genericity of non-Higgsable clusters  when
$h^{1, 1}(B)$  is not small, and the general broad distribution of
non-Higgsable gauge group factors and clusters, are also confirmed from
an analysis of toric bases using a Monte Carlo approach,
which will be presented elsewhere \cite{Wang-WT-MC}.

The work presented here suggests that geometrically non-Higgsable
gauge groups are a fairly universal feature in nonperturbative
F-theory vacua.  These structures may play an important role in better
understanding the global structure of the space of vacua, and in
realizing the observed standard model of particle physics
in F-theory.  The results of this paper, however,
 represent  only a small first step towards a systematic understanding
 of these issues.  While in 6D, there is a growing body of evidence
 \cite{Martini-WT, Johnson-WT, Wang-WT} that known toric
 constructions of elliptic Calabi-Yau threefolds provide a good qualitative
 picture of the space of all elliptic CY threefolds even outside the
 domain of toric geometry, the situation is less clear in 4D.  There
 is no proof that the number of elliptic Calabi-Yau
 fourfolds is finite, and there are base threefolds that are not rational
 that support elliptic CY fourfolds, unlike the case of base
 surfaces.  Nonetheless, it seems feasible that all threefold bases
 that support elliptic Calabi-Yau fourfolds are connected through
 geometric transitions to the set of toric bases, which may be
controllable in a similar fashion to the set of toric bases for
threefolds, which have been completely enumerated \cite{mt-toric}.
Thus, it may be possible to capture general aspects of
the distribution of non-Higgsable clusters in 4D
F-theory models by the analysis
of simple toric constructions.

One particularly noteworthy aspect of the distribution of elliptic CY
fourfolds described here is the increasing probability and number of
non-Higgsable clusters as the Hodge numbers increase.  From the
landscape point of view, the fourfolds with large Hodge numbers are
those that give rise to the greatest number of flux vacua
\cite{Douglas-Kachru, Denef-F-theory}.  This gives additional weight to
the notion that the F-theory landscape is dominated by vacua with
geometrically non-Higgsable gauge groups and matter.  While recent
work has begun to explore some more explicit aspects of the
distribution of 4D F-theory
vacua \cite{bkl, Braun-Watari}, these analyses have
focused primarily on almost-Fano bases, which have no non-Higgsable clusters and
as we have found here represent a small and non-representative sample
of the set of possible bases.  There are many features of the larger
set of vacua available, even just in the toric context, which would be
interesting to study further in the broader class of vacua that
include non-Higgsable structure.

In any event, the constructions considered here provide a rich set of
examples with which to further explore the nature and consequences of
non-Higgsable gauge groups and matter in 4D supersymmetric F-theory
compactifications.  The analysis undertaken here relies completely on
the geometry and complex structure of the F-theory base threefolds. 
A major outstanding challenge for F-theory is to systematically
understand the role of fluxes and additional degrees of freedom on
seven-brane world-volumes, which may among other things enhance or
break parts of the geometrically non-Higgsable gauge group.  A serious
effort to understand the phenomenology of models based on
non-Higgsable $SU(3)$ or $SU(3) \times SU(2)$ groups in global F-theory
compactifications will likely involve
significant advancements in our understanding of these issues.

As a resource for the reader, we have provided the details of the set
of threefold bases constructed in this paper, and information about
the corresponding non-Higgsable structures, in a data file that can be
accessed online \cite{files}.

\vspace*{0.1in}

{\bf Acknowledgements}: We would like to thank Andreas Braun, Lara
Anderson, Antonella Grassi, James Gray, Thomas Grimm, Sam Johnson, David Morrison, Brent Nelson, Julius L. Shaneson, and Yinan Wang for
helpful discussions and Bryan White for time on the Galileo Cluster at
the University of Virginia.  The research of W.T.\ is supported by the
U.S.\ Department of Energy under grant Contract Number DE-SC00012567.
The research of J.H.\ is supported in part by the National Science
Foundation under grant PHY11-25915.

\appendix 

\section{Improved Constraints}
In this paper we have used constraints on twists which are stronger
than those 
given in (\ref{eq:t-0}-\ref{eq:t-big}), which were
derived in \cite{Anderson-Taylor}. In this appendix we
briefly describe   two different approaches to deriving stronger
constraints.  We implemented both of these approaches to give a
complete enumeration of allowed $\P^1$ bundles in the desired class,
and got identically matching results from both approaches.

\subsection{Constraints on twists: I}
\label{sec:method-1}

We begin by reviewing in the toric language the origin of the
constraints  (\ref{eq:t-0}-\ref{eq:t-big}), following similar logic to
that given in 
\cite{Anderson-Taylor}.
Consider a local component of the toric geometry of the $\P^1$
bundle.  We choose coordinates where
\begin{eqnarray}
w_1 & = &  (1, 0, 0)\label{eq:w1}\\
w_2 & = &  (0, -1, 0)\\
s_-& = & (0, 0, 1) \,. \label{eq:s}
\end{eqnarray}
These correspond to the divisors $D_1, D_2, \Sigma_-$.  Now, assume
$d_2$ is a curve of self-intersection $-n$.  Then we have
\begin{equation}
w_3 = (-1, -n, t) \,,
\end{equation}
where $t = \tilde{t}_2 = T \cdot d_2 =t_3$.  If we choose coordinates
$(a, b, c)$ on the dual lattice $N^*$ encoding monomials, the
condition that $(f, g)$ do not vanish to orders $(4, 6)$ 
on $\Sigma_-\cap D_2$
is that there
must be some monomial $(a, b, c) \in{\cal G}$ such that
\begin{equation}
(a, b, c) \cdot (0, -1, 1) =c-b\leq -7 \, \label{eq:abc}
\end{equation}
where $ (a, b, c) \cdot (-1, -n, t) \geq -6$.
We have $a \geq -6, b \leq 6, c \geq -6$ from the definition of ${\cal
  G}$ (\ref{eq:g}) and the rays (\ref{eq:w1}-\ref{eq:s}).
Assuming that $n, t > 0$, we see that the most negative values of
(\ref{eq:abc}) can come from the monomials $(-6, 6, -1)$ or $(-6, 1,
  -6)$, depending on whether $ t > n$ or $n > t$.  If both of these
    monomials are ruled out then there is a (4, 6) vanishing on
    $\Sigma \cap D_2$.
The condition is then
\begin{equation}
-n-6t \geq -12 \;\;\;\;\; {\rm or}\;\;\;\;\;
-6n-t \geq -12 \,.
\end{equation}
One of these must be satisfied to avoid the $(4, 6)$ curve.  It is
straightforward to check that this condition precisely gives the rules
(\ref{eq:t-0}-\ref{eq:t-big}).
A straightforward generalization of this analysis, however, gives
further constraints on the rays $w_i, i > 3$. If any such ray has the form
$w_i = (-m, -n, t)$ with $m, n, t > 0$, then there is a similar
    constraint
\begin{equation}
-n-6t \geq -6(1 + m) \;\;\;\;\; {\rm or} \;\;\;\;\;
-6n-t \geq -6 (1 + m) \,.
\end{equation}
These constraints can be interpreted in terms of constraints on
sequences of twists that can appear over any given base.  For example,
if $d_2$ is a -12 curve, we must have $\tilde{t}_2 = 0$.  We must also
however then have a $-1$ curve $d_3$, so that $w_4 = (-1, -11,
\tilde{t}_3)$, from which the above constraints give $\tilde{t}_3 =
0$.  Similar arguments place very strong constraints on the kinds of twists that are
allowed for a general $\P^1$ bundle base.  By implementing these constraints
for all curves $\Sigma_\pm \cap D_i$ and all $w_i$ with rays that lie
in the appropriate octant in a local coordinate system as above, the
problem of classifying the $\P^1$ bundle bases over the toric surfaces
from \cite{mt-toric} becomes a tractable computational procedure,
which we have implemented.

\subsection{Constraints on twists: II}
\label{sec:method-2}

The second approach is slightly more abstract and based on the
principle of the Zariski decomposition, as used in \cite{mt-toric} to
classify the non-Higgsable clusters for complex surface bases.  This
second approach could be used for base surfaces $S$ that are not
necessarily toric.  The basic idea is that if one has information
about connected curves $C_i$ in a surface, certain curve classes $X$ could
contain higher multiples of the $C_i$ as components than the number of
components that would be derived if one only
used knowledge of a single curve. It is straightforward to see why
this phenomenon may be possible. Consider an effective irreducible
curve $L$ in a surface. We would like to know whether a given
effective class $X$ contains $L$ as a component. A fact from the
algebraic geometry of surfaces, which underlies the Zariski
decomposition on surfaces, is that if $L\cdot L < 0$ and $X \cdot L <
0$, then $X$ contains $L$, i.e.  $X = L + X_1$ for some effective
divisor (curve) $X_1$. Now by the same argument if $X_1 \cdot L < 0$,
$X_1$ contains $L$ and we iterate this process until $X = n L + X_n$
for an effective divisor $X_n$ with $X_n \cdot L \geq 0$. Without
further information about the surface and the curves in it, we have
pulled out as many copies of $L$ from $X$ as we can.

Suppose, though, that there is another curve $R$ with $R \cdot R < 0$ and $X_n\cdot R < 0$.
Then $X=nL + R + X_{n+1}$ and again we recurse until we end up with $X=nL + mR + X_{n+m}$
for some effective divisor $X_{n+m}$ satisfying $X_{n+m}\cdot R\geq 0$. Now note that
\begin{equation}
L \cdot X_{n+m} = L \cdot X - n L\cdot L - n m L \,\cdot R = L \cdot X_n - n m \,L\cdot R
\label{eqn:intxn+m}
\end{equation}
satisfies $L\cdot X_{n+m}\geq 0$ if $L\cdot R=0$, but if $L\cdot R \ne
0$ then $L\cdot X_{n+m}<0$ if $nm L\cdot R > L \cdot X_n\,$. That is,
if $L$ intersects a curve $R$ which is also a component of $X$, then
$X$ may contain additional components of $R$. To determine the full
number of times that a class $X$ contains some negative self
intersection curves $C_i$ in a connected set as components, one simply
recurses over all curves in the set until
\begin{equation}
X = \sum_i n_i C_i + X
\end{equation}
with $X\cdot C_i \geq 0$ for all $i$. This can be quickly implemented
on a computer.  We will henceforth refer to this method of pulling off
as many components as possible based on adjacency data
(self-intersections and twists of adjacent curves) as a
\emph{recursive pull}.

Now consider the relevance for minimality of the Weierstrass equation. Write
\begin{equation}
f = \sum_{k=0}^8 z^k w^{8-k} \, f_k \qquad \qquad \text{and} \qquad \qquad g = \sum_{k=0}^{12} z^k w^{12-k} g_k,
\end{equation}
where $f$ and $g$ are sections of $\cO(-nK_2 -(n-k)T)$ for $n=4,6$, respectively, and define
the class $[n_k]\equiv -nK_2-(n-k)T$. For any genus zero curve $C$ in the twofold base,
$\chi(G) = 2 = \int_C c_1(C)=(-K_2-C)\cdot C$, and therefore
\begin{equation}
[n_k] \cdot C = n\, (2+C\cdot C)-(n-k)\,T\cdot C,
\end{equation}
where $T\cdot C$ is what we have called $\tilde t$ in the toric
context. Though the following twist bounds apply beyond the context
for toric varieties, we will use the $\tilde t$ notation for
simplicity. Note that the only geometric data this depends on is the
self-intersection and the twist, which allows for the analysis of
twist bounds without needing to specify a base.

Building on the previous ideas, we now present a recursive and
relatively efficient algorithm for the classification of allowed
$\bP^1$-bundles over a toric base.
This algorithm could be generalized to a broader class of base surfaces.

Since we are building all allowed $\bP^1$ bundles over the toric
surfaces of \cite{mt-toric}, a natural question is how to break this
difficult combinatoric problem into manageable parts.  In order to
tackle the combinatorics, we would like to note the following facts:
\begin{itemize}
\item The weakest bounds on twists in (\ref{eq:t-0}) are for $C^2=0$
  curves in $\b2$.  However, since there are at most two 0-curves in
  any of the bases of \cite{mt-toric}, the bigger combinatoric penalty
  comes from the $(-1)$-curves, which typically occur with higher
  frequency in the examples we study.
\item Since $(-1)$ curves may potentially give rise to additional
  components when adjacent to other negative self-intersection curves
  with twists over them, the twists over the $(-1)$ curves may be
  constrained beyond the results of \cite{Anderson-Taylor}.
\item While there are sometimes short chains of
  $(-1)$-curves\footnote{The longest such chain occurring in a base of
  \cite{mt-toric} has length 6, and this only occurs in one example,
  $dP_3$.},
  it is typically the case that $(-1)$-curves are separated by
sequences
  of curves with negative self-intersection. These are the
  non-Higgsable clusters of \cite{clusters}; for simplicity we refer to
  such a sequence as a \emph{6NHC} since they correspond to a
  non-Higgsable cluster in the 6d theory obtained from
  compactifying F-theory on a Calabi-Yau elliptic fibration over that
  $\b2$.
\end{itemize}

Instead of the costly algorithm for classifying allowed $\bP^1$
bundles in which one simply applies the constraints
(\ref{eq:t-0}-\ref{eq:t-big}), the algorithm we employ first
classifies the possible allowed twists over each 6NHC where a
recursive pull is used to strengthen the bounds. This can be done
efficiently by taking the chain of curves and iteratively assigning twists,
subject to the condition that the 
recursive pull
on $[n_k]$ does not give rise
to a $(4,6)$ curve. A $(4,6)$ curve arises whenever $[n_k]$ contains $(n-k)$
components of $C$ for all $k\in 0\dots n-1$ and $n=4,6$. This is already a
significant improvement, but it does not improve on the combinatorics
of $(-1)$-curves since there is never a $(-1)$ curve in a 6NHC.

It is often the case, however, in the bases of interest that a $(-1)$
curve is adjacent to a 6NHC. So, given the classification of twists
over 6NHC's, one can put a $(-1)$-curve on either end, or on both
ends, and study the classification of twists over that sequence of
curves --- letting ``o'' denote the presence of a $(-1)$ curve, we
call such sequences o6NHC, 6NHCo, and o6NHCo. Similarly, a recursive
pull can be used to classify the possible twists over such sequences,
and the classification is much smaller than direct application of the
constraints (\ref{eq:t-0}-\ref{eq:t-big}).  For example, the longest
6NHC that appears in the bases of interest
is a chain of 23 consecutive $(-2)$ curves, from which we can
form the o6NHCo
\begin{equation}
[-1,-2,-2,-2,-2,-2,-2,-2,-2,-2,-2,-2,-2,-2,-2,-2,-2,-2,-2,-2,-2,-2,-2,-2,1] \,.
\end{equation} 
In this case the simple constraints
(\ref{eq:t-0}-\ref{eq:t-big}) would allow for about
$13^2\cdot 3^{23} \sim 16$ trillion possible sets of twists, whereas a
recursive pull shows that again there are only
$587$ possibilities. This is the most dramatic combinatoric gain, but large
gains also exist for the other o6NHCo's.

Having classified all allowed twists over 6NHC, o6NHC, 6NHCo, and
o6NHCo sequences our algorithm proceeds to construct the complete
base surfaces,
breaking the bases into building blocks that include these
various different types of sequences. Instead of applying the
constraints curve-by-curve, we proceed building
block by building block, using twists over building blocks which have
been classified. For example, instead of viewing one particular $\b2$
as
\begin{equation}
  [-6, -1, -3, -1, -3, -2, -1, -6, -1, -2, -2, -2, -2, -2, -2, -2, -1]
\end{equation}
the algorithm we wrote (which is efficient but could be made more efficient) views it as
\begin{equation}
  [[-6], [-1], [-3, -1], [-3, -2, -1], [-6, -1], [-2, -2, -2, -2, -2, -2, -2], [-1]]
\end{equation}
where, for example, the allowed twists over $[-6]$ and $[-1]$ were
classified by the simple single-curve twist rules but the method
described above classifies the twists over the remaining ``building
blocks'' which have more than one curve of negative self
interaction. The combinatoric gains for the entire base are compounded
from the combinatoric gains of the building blocks. The algorithm is to fix a base
surface $S$ and break it into building blocks as above. Then one uses the classified
twists over each building block to construct the possible twists over all curves in
$S$, checking for $(4,6)$ divisors and curves after assigning all twists, and
keeping any pair of $S$ and twists that does not have $(4,6)$ divisors or curves.

The results of implementing this algorithm agree perfectly with those
from the more explicit toric method described above.
It is interesting to note that while these constraints are clearly
necessary for the construction of a consistent base, it turns out that
when this set of constraints is implemented completely, these are also
sufficient conditions, which we have confirmed by comparison with the
bases produced by the direct toric algorithm.  This shows that there
are no additional nonlocal constraints on the set of allowed bases.
This observation may be helpful in construction of more general
classes of threefold F-theory bases.

\section{Additional Constraints on $E_8$ Non-Higgsable Seven-branes}
\label{sec:appendix-e8}

Throughout this work we have been careful to check whether the
elliptic fibration contains curves where $(f,g)$ vanish to orders
$(4,6)$ or higher, which often occurs when two non-Higgsable
seven-branes intersect. For non-Higgsable seven-branes that carry an
$E_8$ factor, however, a $(4,6)$ curve can be obtained from a
collision with the $I_1$ locus in a way that is not manifested on a
toric curve.  This is the analogue for 4D F-theory compactifications
of the $-9, -10,$ and $-11$ curves that have similar properties in 6D
F-theory compactifications.  While in the 6D story these are
relatively easy to handle by blowing up points on the base, the
situation is more complicated for curves on threefolds, so we do not
consider threefold bases in this paper that have such $(4, 6)$ curves
on $E_8$ divisors.  In this brief appendix we describe the details of
how we rule out such bases.

Consider a Weierstrass model on a threefold base $B$
that contains a non-Higgsable $E_8$ seven-brane configuration along a 
divisor $Z$.  If the base is toric, we have a homogeneous coordinate $z$ so
that
Z =$\{z = 0\}$, and the coefficients in the Weierstrass model
can always be written in the form
\begin{equation}
  f = z^4\,  \tilde{f}_4 =
z^4\, f_4 + z^5\, f_5 + \cdots \qquad \qquad g = z^5\, \tilde{g}_5 =
z^5\, g_5 + z^6\, g_6 + \cdots
\end{equation}
The discriminant locus takes the form
\begin{equation}
\Delta = 27 z^{10} g_5^2  +{\cal O} (z^{11})
\end{equation}
Thus, a $(4, 6)$ curve arises on the divisor $Z$ whenever $g_5$
generically vanishes along some curve.  In the toric context, $g_5$ is
described by a set of monomials in ${\cal G} \subset N^*$.  If $g_5$
contains more than one monomial then there is always a vanishing
curve, as can be seen in any local coordinate chart, where $g_5$
simply becomes a polynomial in two variables.  If $g_5$ contains only
one monomial, then the only vanishing curves can be toric curves,
which we already check in our analysis.  Thus, to rule out the bases
that have $(4, 6)$ vanishing loci on non-toric curves within $E_8$
divisors, we can simply restrict attention to bases where $g_5$ only
has one monomial in $G$.

From a more general point of view, we see
that the $I_1$ locus intersects the $E_8$ seven-brane if the curve
$C\equiv \{z=g_5=0\}$ exists in $B$. Taking $z \in \cO_{B}(Z)$, we
have $\tilde{g}_5 \in \cO_{B}(-6K_{B} - 5Z)$ and therefore the curve
$C$ is in the class $[C] = (-6K_{B} - 5Z)\cdot Z \in H_2(B,\Z)$. If
$[C]$ is trivial in homology, the intersection doesn't exist and there
is no associated $(4,6)$ curve; this happens if $[g_5]$ is itself
trivial, i.e. $g_5$ is a section of $\cO_{B}$, which is sufficient but
not necessary for $[C]=0$.  From the point of view of $B$, the
necessary and sufficient condition for avoiding a $(4,6)$ curve from
an $E_8$-$I_1$ collision on $Z$ is that $[C]$ is trivial. If $Z$ is
$\P^1$ bundle over a curve in $\b2$, a short calculation shows that
$[C]=0$ only if $\tilde{g}_5\in \cO_{B}$. On the other hand, if the
divisor $Z$ carrying the $E_8$ is one of the sections of the
$\P^1$-bundle, then one might imagine that it could be the case that
$[C]=0$ even if $[g_5]\ne 0$, that is if $\tilde{g}_5$ appears as a
section of a non-trivial line bundle on $B$; one should then
explicitly compute the class of $[C]$, which we have done for each of
the bundles we have constructed, giving precisely the same results as
the monomial analysis above.  In general, though, the question from
the point of view of $B$ is simply whether or not the curve $C$
exists.

The desired condition can also be stated in terms of the local geometry of
$Z$; this perspective, which is that taken in
\cite{4D-NHC}, is particularly useful since it extends to
general base threefolds even when they are not toric.
By the adjunction
formula, when we restrict $g_5$ to $Z$ we have
\begin{equation}
g_5 =   \tilde{g}_5|_{z=0} \in \cO_Z(-6K_Z)\otimes N_{Z/B}.
\end{equation}
The condition that this line bundle have no vanishing sections is that
$N_{Z/B} ={\cal O}_Z (6K_Z)$.  This is a more general way of stating
the condition that an $E_8$ divisor have no $(4, 6)$ curves, which was
also described in \cite{Anderson-Taylor} in the context of F-theory
models with heterotic duals.

\end{document}